 \documentstyle[eqsecnum,aps]{revtex}			
\def\DPS{\displaystyle}

\def\ie{{\it i.e.\/}}
\def\eg{{\it e.g.\/}}
\def\etc{{\it etc\/}}                                                  
\def\viz{{\it viz\/}}                                                  
\def\LocFr{q}		
\def\FTDen{Q}		
\def\possym{\dagger}            
\def\beq{\begin{equation}}
\def\eeq{\end{equation}}
\def\bea{\begin{eqnarray}}
\def\eea{\end{eqnarray}}
\def\bml{\begin{mathletters}}
\def\eml{\end{mathletters}}
\def\density{\nu}               
\def\denfrac{\nu}       
\def\deninve{\nu}       
\def\tlzs{{\tilde{\lambda}}_{0}^{2}}

\def\tlns{{\tilde{\lambda}}_{n}^{2}}

\def\calfars{{\cal F}_{n}}              
\def\calfzrs{{\cal F}^{\rm den}}        
\def\calfgdf{{\cal F}_{n}^{\rm gdf}}    
\def\fgenf{\Phi_{n}}          	
\def\fspa{\tilde{f}^{\rm var}}          
\def\fgdf{\tilde{f}^{\rm gdf}}          
\def\sigz{\sigma_{0}}           

\def\real{{\rm Re\,}}           
\def\imag{{\rm Im\,}}           
\def\dmzrs{{\cal D}_{0}^{\possym}}
\def\dmwrs{{\cal D}_{n}^{\possym}}
\def\dmhrs{{{\overline{\cal D}}^{\possym}}}
\def\dbar{{\mathchar'26\mkern-12mud}}   

\def\lamvec{{\bf c}}            
\def\muvec{{\bf m}}             
\def\lefver{\vert}              
\def\rigver{\vert}              
\def\sumr{\sum_{r=1}^{\infty}}

\def\sumhatr{\sum_{{\hat{k}}_{1},\dots,{\hat{k}}_{r}}}
\def\inthatr{\int\dbar{\hat{k}}_{1}\cdots\dbar{\hat{k}}_{r}}
\def\sumrhor{\sum_{\rho=1}^{r}}
\def\prodrhor{\prod_{\rho=1}^{r}}
\def\sumrhorhopr{\sum_{\rho,\rho^{\prime}=1}^{r}}
\def\sumaln{\sum_{\alpha=0}^{n}}
\def\prodaln{\prod_{\alpha=0}^{n}}
\def\sumin{\sum_{i=1}^{N}}

\def\ltave{\left\{}
\def\rtave{\right\}_{\tau}}
\def\dintr{
\int_{0}^{\infty}
d\tau_{1}p(\tau_{1})
\cdots
d\tau_{r}p(\tau_{r})}           
\def\sintr{
\int_{0}^{1}
ds_{1}{\cdots}ds_{r}}           
\def\sintrPT{
\int_{0}^{1}
ds_{1}{\cdots}ds_{r+2}}         
\def\smatr{{\cal S}_{\rho\rho^{\prime}}}
\def\smatrNI{{\cal S}}  
\def\cmatr{{\cal C}_{\rho\rho^{\prime}}^{(r)}}
\def\cmatrNI{{\cal C}^{(r)}}    
\def\cmatrQi{{\tilde{\cal C}}_{\rho\rho^{\prime}}^{(r)}}        
\def\cmatrQiNI{{\tilde{\cal C}}^{(r)}}  
 \def\rmatr{{\cal R}_{\rho\rho^{\prime}}^{(r)}}
\def\rmatrNI{{\cal R}^{(r)}}    
\def\rmatrinv{\Big({\cal R}^{(r)}\Big)^{-1}\Big\vert_{\rho\rho^{\prime}}}
\def\rmatrinvNI{\Big({\cal R}^{(r)}\Big)^{-1}}                  
\def\rmatrinvNItxt{\big({\cal R}^{(r)}\big)^{-1}}                  
\def\rmatrzero{{\cal R}_{0}^{(r)}\Big\vert_{\rho\rho^{\prime}}}
\def\rmatrzeroNI{{\cal R}_{0}^{(r)}}                        

\def\rmatrzeroinvtxt{\big({\cal R}_{0}^{(r)}\big)^{-1}\big\vert_{\rho\rho^{\prime}}}
\def\rmatrzeroinvNI{\Big({\cal R}_{0}^{(r)}\Big)^{-1}}        
\def\imatr{\delta_{\rho\rho^{\prime}}}
\def\imatrNI{{\cal I}^{(r)}}
\def\uvecr{{\cal U}_{\rho}^{(r)}}
\def\uvecrNI{{\cal U}^{(r)}}    
\def\wscar{{\cal W}^{(r)}}
\def\trar{{\rm   Tr}^{(r)}\,}           
\def\detr{{\rm  Det}^{(r)}\,}           
\def\qdetr{\tilde{{\rm Det}}^{(r)}\,}     
\def\qdetrTXT{{\tilde{{\rm Det}}}^{(r)}\,}  
\def\detn{{\rm Det}_{n}\,}              
\def\imatn{\delta^{\alpha\alpha^{\prime}}}      
\def\imatnNI{I_{n}}                             
\def\somemat{{\cal A}}          
\def\delcomp{\delta_{\rm c}}    
\def\rsb{a}
\def\eqref#1{Eq.~(\ref{#1})}
\def\sumhat#1{\sum_{\hat{#1}}}                  
\def\sumwrsNP#1#2{\sum_{#1=0}^{n}
        {\sum_{{\bf #2}}}^{\prime}}             
\def\sumhrsNP#1{\overline{\sum_{\hat{#1}}}}             
\def\sumzrsHP#1{
        {\sum_{\bf #1}}^{\possym}}                      
\def\sumzrsHPNL#1{
        {\sum}_{\bf #1}^{\possym}}              
\def\sumwrsHP#1#2{\sum_{#1=0}^{n}
        {\sum_{\bf #2}}^{\possym}}                      
\def\sumwrsHPNL#1#2{\sum_{#1=0}^{n}
        {\sum}_{\bf #2}^{\possym}}              
\def\sumhrsHP#1{{\overline{\sum_{\hat{#1}}}}^{\possym}} 
\def\sumhrsHPNL#1{{\overline{\sum}}_{\hat{#1}}^{\possym}} 

\def\Ohat#1{\Omega_{\hat{#1}}}
\def\tilvec#1{\tilde{\bf #1}}
\def\kdelvec#1#2{\delta_{       {\bf #1},{\bf #2}}^{(d)}}       
\def\kdelvecT#1#2{\delta_{\tilde{\bf #1},{\bf #2}}^{(d)}}       

\def\kdelhat#1#2{\delta_{\hat{#1},\hat{#2}}^{(nd+d)}}
\def\kdelhatNS{\delta^{(nd+d)}}                 
\def\dint#1{\int_{0}^{\infty}d#1\,p(#1)}
\def\dinttau#1{\int_{0}^{\infty}d\tau_{#1}\,p(\tau_{#1})}
\def\green#1{{\cal G}_{\bf #1}^{\alpha\alpha^{\prime}}} 
\def\greenNI#1{{\cal G}_{\bf #1}}               
\def\debyeZ#1{g_{0}(\vert{\bf #1}\vert^{2})}     
\def\debyeZbare#1{g_{0}(\vert{#1}\vert^{2})}  	
\def\debyeW#1{g_{1}(\vert{\bf #1}\vert^{2})}     
\def\debyeT#1{g_{2}(\vert{\bf #1}\vert^{2})}     
\def\bigh#1{H_{\bf #1}}
\def\bighz#1{H_{\bf #1}^{(0)}}
\def\bighw#1{H_{\bf #1}^{(1)}}
\def\lith#1{h_{\bf #1}}
\def\lithz#1{h_{\bf #1}^{(0)}}
\def\lithw#1{h_{\bf #1}^{(1)}}
\def\smatrCI#1{{\cal S}_{#1}}                   
\def\disfac{\chi}                               
\def\syfac{\sigma(\disfac)}                     
\def\delfac#1{\Delta_{#1}(\disfac)}             
\def\achfac#1{H_{#1}}                           
\def\trfac#1{{\rm tr}_{#1}\,}                     

\def\nn{\nonumber \\}

\font\smc=cmcsc10

\begin{document}
\preprint{P-95-09-062-iii}
\draft
\title{Randomly Crosslinked Macromolecular Systems: Vulcanisation \\
Transition to and Properties of the Amorphous Solid State}
\author{
Paul M.~Goldbart$^{(a,b,c)}$ 
Horacio E.~Castillo\rlap,$^{(a)}$ and
Annette Zippelius$^{(d)}$  
}
\address{$^{(a)}$Department of Physics, 
$^{(b)}$Materials Research Laboratory and 
$^{(c)}$Beckman Institute, \\
University of Illinois at Urbana-Champaign,
Urbana, Illinois 61801, USA; \\
$^{(d)}$Institut f{\"u}r Theoretische Physik,
Georg August Universit{\"a}t, \\
D 37073 G{\"o}ttingen, Germany}
 \date{September 19, 1995}
\maketitle
\begin{abstract}As Charles Goodyear discovered in 1839, 
when he first vulcanised
rubber, a macromolecular liquid is transformed into a solid when a
sufficient density of permanent crosslinks is introduced at random.  
At this continuous equilibrium phase transition, the liquid state, 
in which all macromolecules are delocalised, is transformed into 
a solid state, in which a nonzero fraction of macromolecules have
spontaneously become localised.  This solid state is a most unusual
one:  localisation occurs about mean positions that are distributed
homogeneously and randomly, and to an extent that varies randomly from
monomer to monomer. Thus, the solid state emerging at the vulcanisation
transition is an {\it equilibrium amorphous solid state\/}: it is
properly viewed as a solid state that bears the same relationship to
the liquid and crystalline states as the spin glass state of certain
magnetic systems bears to the paramagnetic and ferromagnetic states, in
the sense that, like the spin glass state, it is diagnosed by a subtle
order parameter.

In this article we give a detailed exposition of a theoretical approach
to the physical properties of systems of randomly, permanently
crosslinked macromolecules.  Our primary focus is on the equilibrium
properties of such systems, especially in the regime of Goodyear's
vulcanisation transition.  This approach rests firmly on techniques
from the statistical mechanics of disordered systems pioneered by
Edwards and co-workers in the context of macromolecular systems, and
by Edwards and Anderson in the context of magnetic systems.
We begin 
with a review of the semi-microscopic formulation of the
statistical mechanics of randomly crosslinked macromolecular systems
due to Edwards and co-workers, in particular discussing the role of
crosslinks as quenched random variables.
Then we turn to 
the issue of order parameters, and review a version
capable, {\it inter alia\/}, of diagnosing the amorphous solid state.
To develop some intuition, we examine the order parameter in an
idealised situation, which subsequently turns out to be surprisingly
relevant.  
Thus, we are motivated 
to hypothesise an explicit form for
the order parameter in the amorphous solid state that is parametrised
in terms of two physical quantities: the fraction of localised
monomers, and the statistical distribution of localisation lengths of
localised monomers.
Next, we review 
the symmetry properties of the system itself, the liquid
state and the amorphous solid state, and discuss connections with
scattering experiments.
Then, we review 
a representation of the statistical mechanics of
randomly crosslinked macromolecular systems from which the quenched
disorder has been eliminated via an application of the replica
technique.
We transform 
the statistical mechanics into a field-theoretic
representation, which exhibits a close connection with the order
parameter, and analyse this representation at the saddle-point level.
This analysis reveals that sufficient crosslinking causes
an instability of the liquid state, this state giving way to the
amorphous solid state.
To address the properties of the amorphous solid state itself, we solve
the self-consistent equation for the order parameter by adopting the
hypothesis discussed earlier.  Hence, we find that the vulcanisation
transition is marked by the appearance of a nonzero fraction of
localised monomers, which we compute, the dependence of this fraction 
on the crosslink density indicating a connection with random graph theory 
and percolation.
We also compute the distribution of localisation lengths that
characterises the ordered state, which we find to be expressible in
terms of a universal scaling function of a single variable, at least in
the vicinity of the transition.
Finally, we analyse the consequences of incorporating a certain
specific class of correlations associated with the excluded-volume
interaction.
\end{abstract}
\newpage
\tableofcontents
\newpage
{{\obeylines
\hfill {\it 	There is probably no other inert substance, 
\hfill 		the properties of which excite in the human 
\hfill		mind, when first called to examine it, an equal   
\hfill 		amount of curiosity, surprise, and admiration.  
\hfill 		Who can examine, and reflect upon {\rm[{\it the properties of\/}]} 
\hfill 		gum-elastic, without adoring the wisdom of the Creator\/}?
\smallskip
\hfill {\smc Charles Goodyear\/}, {\sl Gum-Elastic and its Varieties, 
\hfill with a Detailed Account of its Applications and Uses,
\hfill and of the Discovery of Vulcanization\/} (1855)}}
\section{Introduction and overview}\label{SEC:Introduce}
In this article we aim to present a theoretical description of the
physical properties of systems of macromolecules that have been randomly
and permanently crosslinked. Our focus will be on the equilibrium properties
of such systems, especially in the regime of the vulcanisation
transition.  By the term vulcanisation transition we mean the sharp
thermodynamic phase transition occurring when the mean density of
crosslinks exceeds a certain critical value.  At this critical
crosslink-density, the equilibrium state of the system undergoes a
continuous transition: for subcritical values the equilibrium state is a
liquid state, in which all the macromolecules are delocalised; for
supercritical values the equilibrium state is an amorphous solid state,
in which a nonzero fraction of macromolecules form a macroscopic network, 
and spontaneously become localised, albeit about certain random 
locations. Thus, our focus will be on the spontaneous emergence of the
equilibrium amorphous solid state at the vulcanisation transition, and
the consequent properties of this unusual state.

It must be emphasised at the outset that the theoretical description of
the vulcanisation transition to and properties of the amorphous solid
state of randomly crosslinked macromolecular networks presented here
represents the confluence of two pioneering contributions to the theory
of condensed matter:  the Deam-Edwards theory of a single crosslinked
macromolecule \cite{REF:DeamEd}, and the Edwards-Anderson theory of spin
glasses \cite{REF:SFEandPWA}.

The basic ingredients of the present approach to the physical properties
of randomly crosslinked macromolecular networks are as follows.
We adopt a semi-microscopic description of the macromolecules, in which
the detailed microscopic chemistry of the macromolecules and solvent (if
any there be) feature only to the extent that they determine the
following effective parameters: the total arclength of each
macromolecule, the persistence length (\ie, the length of the
statistically independent macromolecular segments, which we term
monomers), and the excluded-volume strength (\ie, the parameter that
describes the effective repulsion between monomers).
Thus, we regard the macromolecules as extended, featureless, flexible
linear objects, each capable of exhibiting a large number of
configurations, and use classical equilibrium statistical mechanics to
address the properties of systems composed of a thermodynamically large
number of such macromolecules.
We regard the crosslinks as permanent elements that constrain certain
randomly chosen monomers to remain adjacent to one another.  Thus the
crosslinked macromolecular system is a system with quenched disorder, in
the sense that in addition to the macromolecular freedoms---the
so-called annealed variables, which undergo equilibrium
statistical-mechanical fluctuations---there are additional variables,
those that specify the detailed realisation of the crosslinking, that do
not undergo equilibrium statistical-mechanical fluctuations.
Instead, these variables---the so-called quenched random
variables---vary only between realisations of the physical system.  We
treat these quenched random variables statistically, too, but account
for their quenched nature by invoking the replica technique.

We now give an overview of our basic strategy for determining the
physical properties of randomly crosslinked macromolecular networks.

We characterise the plausible equilibrium states of the system---liquid,
globule, crystalline solid, amorphous solid---in terms of an 
order parameter, designed to discriminate between these states.  This
order parameter is a vastly more intricate object than the order
parameters that arise, say, in the study of ferromagnetism or even spin
glasses, and one must explore enormously larger spaces to find its
equilibrium value.  By investigating a simple caricature of the
amorphous solid state we are, however, able to identify a physically
well-motivated scheme for parametrising the amorphous solid state order
parameter at a manageable level: via a single number---the fraction of
spatially localised monomers---and a normalised probability
distribution---the statistical distribution of localisation lengths of
the localised monomers.

We focus on the free energy and the order parameter for the system of
interacting macromolecular freedoms subject to the crosslinking
constraints.  Application of the replica technique to these quantities
allows for the elimination of the quenched random variables; the price
for this elimination is the introduction of an unusual and rather
complicated effective coupling amongst the replicated macromolecular
freedoms. Our scheme for parametrising the order parameter leaves 
intact the permutation symmetry amongst the replicas.
Next, we introduce a certain stochastic field, the argument of which is
replicated (real or wave-vector) space, via which we transform the
replicated macromolecular description into a field-theoretic one.  In
this representation, the individual macromolecules are coupled to one
another only indirectly, via their coupling to the fluctuations of 
the stochastic field, although the replicas of any given macromolecule
remain directly coupled to each other.  Via the introduction of a
suitable external potential we demonstrate the connection between the
order parameter and the stochastic field.

In order to elucidate the physical properties of the system we adopt a
mean-field approach, which amounts to our approximating, via the
saddle-point method, the averages over the stochastic field in the 
field-theoretic expressions for the free energy and order parameter.
The state of the physical system then follows from the form of the
appropriate stationary value of the stochastic field or, equivalently,
from the form of the self-consistent value of the order parameter.
For mean crosslink densities smaller than a certain critical value, of
order one crosslink per macromolecule, there is only one stationary
value, which is elementary, and the corresponding state is the liquid
state. For supercritical crosslink densities the appropriate stationary
value is most definitely not elementary, and the corresponding state is
the amorphous solid state.  At the critical crosslink density the system
undergoes a continuous vulcanisation transition from the liquid state to
the amorphous solid state.

To determine the properties of the amorphous solid state itself, we
hypothesise that the self-consistent value of the order parameter lies
within the family of order-parameter values reachable via our (severely
restrictive but nevertheless physically plausible) parametrisation.
Quite remarkably, this is indeed the case: our parametrisation does not
merely yield a variational approximation to the amorphous solid state.
Instead, although one has no {\it a priori\/} reason to suppose that it
should, it permits an {\it exact\/} mean-field description of randomly
crosslinked macromolecular networks to be constructed.  What emerges is
an amorphous solid state characterised by a nonzero fraction of
localised monomers.  The precise value of this fraction depends on the
crosslink density, and vanishes continuously at the transition and in
the liquid state. This fraction depends on the crosslink density in a
manner identical to that found in random graph theory and percolation.
The state is further characterised by a crosslink-density--dependent
distribution of localisation lengths, which quantifies the manner in
which the localised monomers have become localised around their random
mean positions.  The typical localisation length diverges continuously
at the transition and in the liquid state.  In the vicinity of the
transition, the distribution of localisation lengths has a scaling form
governed by a universal function, which we compute.  To date, we have
been unable to obtain conclusive results for the distribution of
localisation lengths far from the amorphous solidification transition.
The reason for this is purely technical: at a certain stage in the
development we employ a perturbative calculation, in which the small
parameter is the characteristic inverse localisation length, measured
in units of the radius of gyration of a single, noninteracting
macromolecule, this parameter being zero in the liquid state, and small
in the amorphous solid state only in the vicinity of the transition.

In addition to developing a purely mean-field description of the
transition to and properties of the amorphous solid state, we
incorporate a class of correlations associated with the excluded-volume
interactions. We accomplish this by treating a certain sector of the
stochastic field at the gaussian level, rather than the saddle-point
level.  What emerges from this treatment is merely a particularly simple
finite renormalisation of a specific parameter of the mean-field
theory, at least in three spatial dimensions.

Approaches based on percolation theory \cite{REF:DStauffer}
are sometimes pursued in the context of the modelling of randomly 
crosslinked macromolecular networks; 
see Ref.~\cite{REF:StCoAd}.  However such approaches entail only
a single statistical ensemble, and hence cannot treat the equilibration
of thermal (\ie, annealed) freedoms in the presence of quenched
freedoms (crosslinks, in the present case). If the percolation ensemble
is interpreted as describing the quenched freedoms then it can account
for the appearance of an infinite network at a critical crosslink
density, but it cannot account for the thermal fluctuations, which
determine the physical properties of the liquid and amorphous solid
states. In particular the degree of localisation in the amorphous solid
state and the entropic elasticity of the network cannot be calculated
from first principles in a percolation-based model, which does not
distinguish between thermal and quenched freedoms. If, on the other
hand, the percolation ensemble is interpreted as describing the thermal
fluctuations of the macromolecular freedoms then it may serve as a
model for chemical gelation, in which crosslinks continuously form and
break up, but it cannot account for permanent, quenched crosslinks.

An approach to the properties of randomly permanently crosslinked
macromolecular systems that accounts for both thermal fluctuations and
quenched disorder has been introduced by Edwards and co-workers
\cite{REF:DeamEd,REF:RCBallPaper,REF:RCBallThesis}.  
In addition to formulating the
statistical mechanics of a single randomly crosslinked macromolecule
using the replica technique, Deam and Edwards investigated the elastic
properties of the system by invoking a variational method
\cite{REF:DeamEd}.  Ball and Edwards investigated the incorporation of
correlations, and Ball also studied certain aspects of
many-macromolecule systems \cite{REF:RCBallThesis}.

The point of view that the solid state of randomly crosslinked
macromolecular networks represent an unusual, equilibrium amorphous
solid state of matter was introduced and developed in detail in
Refs.~\cite{REF:prl1987,REF:pra1989,REF:kyoto1988,REF:macro1989}.  In
addition, the order parameter appropriate for detecting this amorphous
solid state was introduced and explained in these papers, and the
instability of the liquid state with respect to the formation of an
amorphous solid, induced by sufficient crosslinking, was also
identified there.  Extending these ideas, the transition to and certain
properties of the amorphous solid state, including its elastic
properties, were obtained within the context of a variational
mean-field approach in Ref.~\cite{REF:PMGandAZprl}.  This variational
approach was improved through an exact mean-field approach, the central
elements of which were reported in Ref.~\cite{REF:CGZjourEPL}.
The present article is intended to provide a fairly complete exposition
of this approach to the statistical mechanics of randomly crosslinked
macromolecular networks, including a detailed account of the work
reported in Refs.~\cite{REF:PMGandAZprl,REF:CGZjourEPL}.

By adopting the Deam-Edwards approach \cite{REF:DeamEd} and the
amorphous solid order parameter picture of 
Refs.~\cite{REF:prl1987,REF:pra1989},
Panyukov has made a number of contributions to the theory of
well-crosslinked macromolecular networks \cite{REF:PanPapers}.  To
accomplish these, he considers a single linear macromolecule,
randomly crosslinked, which is intended to represent a physical network
of many randomly crosslinked macromolecules.  These contributions,
which are reviewed in Ref.~\cite{REF:PanReview}, are based on the
introduction of a well-known free-field representation of a random walk
\cite{REF:ItzDro}, the resulting field being treated at the mean-field
level, along with an additional replica limit.  Focusing, as it does,
on the statistical mechanics of a single, well-crosslinked macromolecule, 
Panyukov's approach is unsuitable for developing a theory of the transition
between the liquid and amorphous solid states, which occurs at
crosslink densities of order one per macromolecule.

The present approach to the physics of randomly crosslinked
macromolecular systems possesses the following distinctive virtues.
First, both thermal freedoms (\ie, the macromolecular positions) and
quenched freedoms (\ie, the crosslink locations) are incorporated, and
handled appropriately, in contrast with percolative pictures.  The
replica technique provides the tool for accomplishing this.  In fact,
the percolative picture emerges from the present approach in the form of
statistical information concerning the presence of localised
macromolecules. However, the present approach is considerably richer,
additionally yielding statistical-mechanical information about the
(thermally fluctuating) macromolecular system.  In particular, it 
allows a unified treatment of liquid and amorphous solid states.
Second, the physical many-macromolecule character of the system is
maintained, in contrast with approaches that consider instead the
properties of a single macromolecule. Especially in the vicinity of the
solidification transition, where the number of physical crosslinks is of
order one per macromolecule, this is particularly significant, and
allows us to develop a theory of the solidification transition.
Third, the present approach leads directly to an order parameter for the
amorphous solid state, which is related to that of spin glass physics.
The order parameter has a natural, physical interpretation, which
facilitates the hypothesising of an appropriate form for it.
Fourth, the physical freedoms, \viz., the macromolecular configurations,
appear directly throughout the development, not being exchanged for some
problem-specific formal representation.  We can, therefore, be confident
that the macromolecular character of the system is retained, especially
when approximations are made. Indeed, the entire approach is very
robust, so that, in addition to being of interest in the context of
vulcanised macromolecular systems, it can readily be extended to address
a wide range of other physical systems, such as crosslinked manifolds
\cite{REF:Roos}, endlinked systems of flexible, semi-flexible and rigid
macromolecules \cite{REF:EndLink}, continuous random network models of
structural glasses \cite{REF:StrGla}, and proteins. In
addition, it should prove possible to extend the present approach to
address issues of dynamics.  Moreover, looking beyond the algebraic
details, one sees a theoretical superstructure that is rather natural,
direct and perhaps even conventional, at least from the point of view of
statistical field theory.
Fifth, the use of the present approach has primarily been restricted to
the mean-field level of approximation.  However, it is not a variational
mean-field theory: instead the relevant saddle-point is determined
exactly. This makes the approach a particularly promising starting point
for future developments, such as the elucidation of the elastic
properties and of the role of thermal fluctuations.  It should be 
mentioned that the full stability of the saddle-point that we have 
determined has not yet been established. 
One current shortcoming of the present approach arises from the
technical difficulty of computing properties in the regime of high
crosslinking (\eg, deep in the amorphous solid state).  This does not
limit the scope of our primary aim, \viz., to explore the vicinity of
the transition to the amorphous solid state. A second shortcoming is 
the inability of the approach, at least in its present formulation, to
respect the interlocking of closed loops of macromolecules (discussed 
in Sec.~\ref{SEC:XLasQRV}) that crosslinking can induce.

The present article is organised as follows. 
In the present, introductory, section we provide an overview of the
article.
In Sec.~\ref{SEC:Model} we discuss the basic elements of the model of
macromolecular systems on which the present approach is based, including
the level of description of macromolecular configurations, the Edwards
measure for their statistical weights, and the notion of crosslinks as
quenched random variables.  We also discuss the partition function, free 
energy and issues of indistinguishability, along with the statistical
characterisation of the crosslinks, and the notion of disorder averages
of certain physically relevant quantities.
In Sec.~\ref{SEC:OPFields} we develop the general subject of order
parameters appropriate for the amorphous solid and other states, discussing
the properties that such order parameters should possess.  We explore
a simple scenario for the amorphous solid state, which
provides physical motivation for a certain specific hypothesis that we
make for the form taken by the amorphous solid order parameter in the
amorphous solid state.  At this stage we introduce the concept of gel 
and sol fractions and the statistical distribution of localisation
lengths associated with localised monomers.  We also analyse the
symmetry properties of the ordered state, and discuss the connection
between the order parameter and the elastic scattering of neutrons.
In Sec.~\ref{SEC:DoaDecd} we address the statistical mechanics of
randomly crosslinked macromolecular networks, invoking the replica
technique in order to eliminate the (quenched random) crosslink
variables.  We also introduce a suitable replica-Helmholtz free energy,
which is dependent on a convenient generalised external potential.
In Sec.~\ref{SEC:HSdcrpf} we reformulate the statistical mechanics of
randomly crosslinked macromolecular networks in field-theoretic terms by
introducing a certain stochastic field, which is closely related to the
amorphous solid state order parameter.
In Sec.~\ref{SEC:StatPtCrit} we explore the properties of the resulting
field theory within the context of a natural mean-field approximation.
We exhibit the instability of the liquid state, and we compute the free
energy and self-consistent order parameter in the vicinity of the
transition. We also describe the characteristics of the amorphous solid
state that emerge from this approach.
In Sec.~\ref{SEC:DenSecOV} we incorporate certain correlations
associated with density-sector fluctuations, and demonstrate that the
results of the previous sections of this article are robust with respect
to the incorporation of these fluctuations.
In Sec.~\ref{SEC:Conclude} we make some concluding remarks.  We have
organised this article so that the main text is, to a large degree,
unencumbered by lengthy mathematical details. Wherever possible such
details have been relegated to one of eleven appendices, in which we
demonstrate in full detail how the results of the main text are
established.

Perhaps the most significant property to emerge at the vulcanisation
transition is rigidity with respect to shear deformations, \ie,
elasticity.  In a forthcoming article \cite{REF:CGZelastic} we shall
report in detail on the extension of the present work to the
investigation of the elastic properties of the amorphous solid state of
randomly crosslinked macromolecular networks.
\section{Model of the macromolecular system}\label{SEC:Model}
\subsection{Macromolecular system prior to crosslinking}\label{SEC:EdwardsHam}
We consider a system consisting of a large number $N$ of long, 
flexible macromolecules, initially identical, and contained in a 
large $d$-dimensional hypercube of volume $V$.  The macromolecules are 
characterised by their common arclength $L$ and 
(weakly temperature-dependent) persistence length 
$\ell(\ll L)$, so that the number of effectively statistically 
independent segments comprising each macromolecule is of order 
$L/\ell\gg 1$.  Semi-microscopic spatial configurations of the system 
are characterised by the collection of spatial configurations 
of the macromolecules $\{{\bf R}_{i}(\sigma)\}_{i=1}^{N}$, 
in which ${\bf R}_{i}(\sigma)$ is the $d$-dimensional position 
vector of the monomer an arclength distance $\sigma$ from a 
specific end of macromolecule $i$, the (discrete) macromolecule 
index $i$ ranging from $1$ to $N$ and the (continuous) arclength 
variable $\sigma$ ranging from $0$ to $L$.

It is convenient to exchange the dimensionful position vector 
${\bf R}$ and arclength $\sigma$ for dimensionless versions 
${\bf c}$ and $s$ via the transformation
\bml
\begin{eqnarray}
{\bf R}_{i}(\sigma)&\equiv&
\sqrt{\ell L/d}\,\,{\bf c}_{i}(s),
\label{EQ:transform_A}
\\
\sigma&\equiv&L\,s.
\label{EQ:transform_B}
\end{eqnarray}
\eml Thus, 
we shall be measuring spatial distances in units of 
$\big(\ell L/d\big)^{1/2}$ 
(\ie, the root mean squared end-to-end distance of a free macromolecule
divided by $\sqrt{d}$), and arclength distances in units of the total
arclength $L$.  We shall measure energies in units such that 
$k_{\rm B}T$ is unity.

At the level of the present semi-microscopic description, and prior to
the incorporation of the effects of either monomer-monomer interactions
or crosslinks, we account for the connectivity of the constituent
macromolecules by adopting the Wiener measure
\cite{REF:SFEdwardsEV,REF:SFEdMDoiBook,REF:OonoReview}, in terms of
which the statistical weight \cite{REF:NoTemp} of the configuration 
of the system $\{{\bf c}_{i}(s)\}_{i=1}^{N}$ is proportional to 
$\exp\big(-H_{1}^{\rm W}\big)$, where
\begin{equation}
H_{1}^{\rm W}
\equiv
{1\over{2}}
\sum_{i=1}^{N}\int_{0}^{1}ds
\bigg\lefver{d\over{ds}}{\bf c}_{i}(s)\bigg\rigver^{2}.
\label{EQ:WienerMeasure}
\end{equation} 
The subscript $1$ on $H_{1}^{\rm W}$ anticipates the introduction of
replicas of the system, which we shall need to make below.  We shall
often need to consider normalised expectation values taken with respect
to the Wiener measure, which we shall denote by the angle-bracket pair
$\langle\cdots\rangle_{1}^{\rm W}$, defined by
\begin{equation}
\Big\langle\cdots\Big\rangle_{1}^{\rm W}
\equiv
{\DPS\int{\cal D}{\bf c}\,\exp\Big({-H_{1}^{\rm W}}\Big)\cdots
\over{\DPS\int{\cal D}{\bf c}\,\exp\Big({-H_{1}^{\rm W}}\Big)}}, 
\label{EQ:wiener_average}
\end{equation} 
where the dots represent an arbitrary function of the 
configuration of the system, and the measure 
\beq
{\cal D}{\bf c}\equiv
\prod_{i=1}^{N}
\prod_{0\le s\le 1}
d{\bf c}_{i}(s)
\label{EQ:FlatMeasure}
\eeq
indicates functional integration over all spatial configurations of the
system, \ie, over all configurations of the $N$ macromolecules.  The
subscript $1$ on $\langle\cdots\rangle_{1}^{\rm W}$ also indicates that
the average is taken only over the configurations of a single copy of
the system, also anticipating the introduction of replicas.

We account for monomer-monomer interactions in a phenomenological
manner, by augmenting the Wiener measure with an additional factor that
has the effect of suppressing the statistical weight of configurations
in which pairs of monomers occupy common regions
of space \cite{REF:SFEdwardsEV,REF:SFEdMDoiBook,REF:OonoReview}.  To
this end, we replace the Wiener measure, Eq.~(\ref{EQ:WienerMeasure}),
by the Edwards measure, in terms of which the statistical weight of the
configuration
$\{{\bf c}_{i}(s)\}_{i=1}^{N}$ 
is proportional to $\exp\big(-H_{1}^{\rm E}\big)$, where 
\begin{equation}
H_{1}^{\rm E}=
\frac{1}{2}\sum_{i=1}^{N}
\int_{0}^{1}ds
\Big\lefver
  {d\over{ds}}{\bf c}_{i}(s)
\Big\rigver^{2}
+\frac{\lambda^{2}}{2}\sum_{i,i^{\prime}=1}^{N}
\int_{0}^{1}ds
\int_{0}^{1}ds^{\prime}\,
\delta^{(d)}
\big(
 {\bf c}_{i}         (s)
-{\bf c}_{i^{\prime}}(s^{\prime})
\big).
\label{EQ:EdwardsMeasure}
\end{equation} 
Here, $\delta^{(d)}\big({\bf c}\big)$ is the $d$-dimensional Dirac
$\delta$-function, and the dimensionless (real, positive) parameter
$\lambda^{2}$ characterises the strength of the suppression of
statistical weight due to the (repulsive) ex{\-}cluded-volume
interaction between monomers
\cite{REF:SFEdwardsEV,REF:SFEdMDoiBook,REF:OonoReview}.  
The excluded-volume interaction is suitably modified so as to exclude 
interactions between adjacent monomers on a common macromolecule 
(\ie, monomers for which $\vert{s-s^{\prime}}\vert<\ell/L$).
The expression for the Edwards measure in terms of dimensionful
variables is given in footnote~\cite{REF:DimensionForm}.
The system can be regarded as a melt of macromolecules, in which case
the interaction parameter $\lambda^{2}$ is intended to account for the
monomer-monomer interaction.  Alternatively, it can be regarded as a
solution of macromolecules dissolved in a good solvent, in which case
$\lambda^{2}$ is intended to represent the effective monomer-monomer
interaction (\ie, the bare interaction renormalised by the
monomer-solvent and solvent-solvent interactions, the solvent degrees
of freedom having been integrated out). 
In both cases, $\lambda^{2}$ is weakly temperature-dependent.
Even at the level of mean-field theory, the excluded-volume interaction
plays a crucial role:  it partially compensates the effective
monomer-monomer attraction due to the crosslinks in just such a fashion
as to maintain the macroscopic homogeneity of the system whilst
allowing for the possibility of transition from the liquid to the
amorphous solid state.

We shall need to consider normalised expectation values 
taken with respect to the Edwards measure, which we shall denote 
by the angle-bracket pair $\langle\cdots\rangle_{1}^{\rm E}$, 
defined by 
\begin{equation}
\Big\langle\cdots\Big\rangle_{1}^{\rm E}
\equiv
{\DPS\int{\cal D}{\bf c}\,\exp\Big({-H_{1}^{\rm E}}\Big)\cdots
\over{\DPS\int{\cal D}{\bf c}\,\exp\Big({-H_{1}^{\rm E}}\Big)}}, 
\label{EQ:EdwardsAverage}
\end{equation} 
where the dots represent an arbitrary function of the configuration of
the system, and once again ${\cal D}{\bf c}$ indicates functional
integration over all configurations of the system.  Again, the
subscripts $1$ on $H_{1}^{\rm E}$ and 
$\langle\cdots\rangle_{1}^{\rm E}$ 
anticipate the introduction of replicas.

It should be noted that neither the Wiener measure nor the Edwards
measure explicitly breaks translational or rotational symmetry:  the
statistical weight of a configuration remains unchanged if all the
monomers are simultaneously translated through a common amount or
rotated through a common angle about a common axis.
\subsection{Crosslinks as quenched random variables}\label{SEC:XLasQRV}
Our aim is to address the statistical mechanics of thermodynamically
large systems of macromolecules into which a large number of crosslinks
have been permanently introduced at random.  Each crosslink has the
effect of constraining two randomly selected monomers, the locations of
which were kinematically independent prior to the introduction of the
crosslink, to occupy a common spatial location.  Thus the effect of the
crosslinks is to eliminate from the ensemble of configurations of the
system all configurations that do not obey the entire set of random
constraints enforced by the crosslinks.  Our task is therefore to
address the statistical mechanics of macromolecular systems in the
presence of a large number of random constraints.

A specific realisation of the crosslinking is fully described by
specifying which randomly selected pairs of monomers are connected by
each crosslink, \ie, that the crosslink labelled by the index $e$
serves to connect the monomer at arclength $s_{e}$ on macromolecule
$i_{e}$ to the monomer at arclength $s_{e}^{\prime}$ on macromolecule
$i_{e}^{\prime}$, for $e=1,\dots,M$, with $M$ being the total number of
crosslinks. Thus, only those configurations that satisfy the
constraints
\begin{equation}
{\bf c}_{i_{e}         }(s_{e}         )=
{\bf c}_{i_{e}^{\prime}}(s_{e}^{\prime}), 
\qquad
({\rm with}\,\,\,e=1,\dots,M)
\label{EQ:constraints}
\end{equation}
are retained in the ensemble. It should be noted that these constraints
do not explicitly break translational symmetry.

In principle, of course, neither the crosslinks nor the integrity of
the macromolecules are truly permanent. However, in many physical
realisations of crosslinked macromolecular systems there is a very wide
separation between the time-scale required for the crosslink-constrained
macromolecular system to relax to a state of thermodynamic equilibrium
and the much longer time-scale required for either the crosslinks or the
macromolecules to break.  For such systems, and it is such systems that
we have in mind, the crosslinks and the macromolecules should be
regarded as permanent, so that the number and identity of the monomers
participating in crosslinks,
$\{i_{e},s_{e};i_{e}^{\prime},s_{e}^{\prime}\}_{e=1}^{M}$, should be
treated as nonequilibrating (\ie, quenched) random variables.  The
unconstrained macromolecular freedoms are regarded as reaching
equilibrium in the presence of fixed values of the quenched variables.
Thus, it is a meaningful task to address the equilibrium statistical
mechanics of permanently crosslinked macromolecular systems.

It should be remarked that the relative statistical weights of the
configurations that do satisfy the crosslinking constraints are
hypothesised, at least {\it a priori\/}, to be unaffected by the
introduction of crosslinks.  That is, the statistical weights are
proportional to $\exp\big({-H_{1}^{\rm E}}\big)$ for configurations
satisfying the crosslinks and zero otherwise.  However, as we shall see
in detail below, for a sufficiently large density of crosslinks the
translational and rotational symmetry of the equilibrium state of the
system is spontaneously broken.  That is, in a given (pure) state the
statistical weights of configurations that are translations and
rotations of one another are no longer identical, and thus localisation
can arise. Indeed, only one member of a family of translated and
rotated configurations has a nonzero weight in a given (pure) state.
The associated transition to an amorphous solid state is precisely the
transition on which we are focusing.  We remark that in the present
context of amorphous solidification, translational and rotational
symmetry are spontaneously broken in an unusual sense, in that they
remain fully intact at the macroscopic level.

A second mechanism that leads to the violation of the hypothesis
mentioned in the previous paragraph arises because sufficient
crosslinking is liable to give a topological character to the system of
macromolecules, at least in three spatial dimensions, in the sense that
for a given set of crosslinks there will be families of configurations
allowed by the crosslinks that are mutually inaccessible.  We mean by
this that, because of the possibility of interlocking closed loops
formed by macromolecules, there will be families of configurations
between which the system cannot continuously evolve without the
necessity either of the breaking of at least one crosslink or the
passage of one monomer through another.  We distinguish between
constraints arising indirectly from crosslinking via the interlocking of
closed loops and constraints arising directly from the crosslinks
themselves by referring to the former as {\it anholonomic\/} constraints
and the latter as {\it holonomic\/} constraints.  In principle, a
statistical-mechanical approach should incorporate, at most, those
configurations that are mutually accessible, \ie, should respect both
holonomic and anholonomic constraints. The theory presented here treats
the holonomic constraints as quenched but the anholonomic constraints
as annealed, therefore not incorporating the latter.  We know of no
explicit semi-microscopic strategy that is capable of handling the 
anholonomic constraints.
\subsection{Partition function}\label{SEC:PartFunction}
We define the statistical-mechanical partition function 
$\tilde{Z}\big(\{i_{e},s_{e};i_{e}^{\prime},s_{e}^{\prime}\}_{e=1}^{M}\big)$ 
that characterises the crosslinked system via
\begin{equation}
\tilde{Z}\big(\{i_{e},s_{e};i_{e}^{\prime},s_{e}^{\prime}\}_{e=1}^{M}\big)
\equiv
\Big\langle
\prod\limits_{e=1}^{M}\delta^{(d)}
        \Big(
     {\bf c}_{i_{e}         }(s_{e}         )
    -{\bf c}_{i_{e}^{\prime}}(s_{e}^{\prime})
        \Big)
\Big\rangle_{1}^{\rm E}.
\label{EQ:partition}
\end{equation}
The product of Dirac $\delta$-functions serves to remove from the sum
over configurations implicit in the
angle brackets~(\ref{EQ:EdwardsAverage}) any configuration that fails
to satisfy the constraints~(\ref{EQ:constraints}) \cite{REF:NoDelta}, the 
remaining configurations contributing with weights given by the Edwards
measure~(\ref{EQ:EdwardsMeasure}).  With this definition,
$\tilde{Z}\big(\{i_{e},s_{e};i_{e}^{\prime},s_{e}^{\prime}\}_{e=1}^{M}\big)$
is normalised relative to the uncrosslinked system. Consequently, the
free energy derived from this partition function will in fact be the
increase in free energy that arises upon crosslinking.  This allows us
to focus on the implications of crosslinking rather than the properties
of the uncrosslinked system.

At first sight, the quantity
$\tilde{Z}\big(\{i_{e},s_{e};i_{e}^{\prime},s_{e}^{\prime}\}_{e=1}^{M}\big)$
in Eq.~(\ref{EQ:partition}), which we are calling the partition
function, appears to be precisely the physical partition function of
the crosslinked system, at least relative to that of the uncrosslinked
system.  However, for a straightforward reason associated with the
notion of indistinguishability, a reason that we discuss in
Secs.~\ref{SEC:indisti} and \ref{SEC:avSym},
$\tilde{Z}\big(\{i_{e},s_{e};i_{e}^{\prime},s_{e}^{\prime}\}_{e=1}^{M}\big)$
as defined in Eq.~(\ref{EQ:partition}) is not quite the correct
definition of the physical partition function.  However, as we shall
see, the partition function
$\tilde{Z}\big(\{i_{e},s_{e};i_{e}^{\prime},s_{e}^{\prime}\}_{e=1}^{M}\big)$
will turn out to be adequate for our purposes.
\subsection{Indistinguishability}\label{SEC:indisti}
As first pointed out by Gibbs \cite{REF:JWGibbs}, the (configurational
aspect of the) physical partition function for systems involving one or
more species of identical constituents is to be found by summing over
all configurations of the system whilst ignoring the issue of the
distinguishability of the constituents, and subsequently dividing by an
appropriate combinatorial factor to account for the indistinguishability
of the constituents.  This strategy compensates for the over-counting
of configurations that has arisen from the neglect of
indistinguishability.

What are the implications of indistinguishability in the present context?
For the case of the system of $N$ identical uncrosslinked macromolecules, 
the appropriate factor is $N!$, and thus the physical partition function is 
given by 
\begin{equation}
\frac{1}{N!}
\int{\cal D}{\bf c}\,\exp\left(-H_{1}^{\rm E}\right).
\label{EQ:pf_not_linked}
\end{equation} 
If, for the case of the crosslinked system, the appropriate factor were also 
$N!$ (which it is not) then its physical partition function would be given by
\begin{equation}
\frac{1}{N!}
\int{\cal D}{\bf c}\,\exp\left(-H_{1}^{\rm E}\right)
\prod_{e=1}^{M}
\delta^{(d)}
        \big(
     {\bf c}_{i_{e}         }(s_{e}         )
    -{\bf c}_{i_{e}^{\prime}}(s_{e}^{\prime})
        \big),
\label{EQ:pf_yes_linked_wrong}
\end{equation}
and thus the relative physical partition function would indeed be given 
by Eq.~(\ref{EQ:partition}), the factors of $1/N!$ in the 
numerator and denominator cancelling. 

However, the process of crosslinking alters the system from one that
comprises $N$ copies of a single species of identical elements.
Instead, the crosslinked system will contain a variety of species, such
as macromolecules that do not participate in any crosslinks, as well
as clusters of macromolecules of many types.  By clusters we mean
assemblages of macromolecules that are (directly or indirectly)
connected by crosslinks or interlockings, and therefore cannot be
separated by arbitrary distances.  Examples of clusters include pairs
of macromolecules that participate in a single crosslink, 
that crosslink being located between some specific pair of arclength
locations (say $(s,s^{\prime})=(0.12,0.57)$), 
single macromolecules crosslinked to
themselves at some specific pair of arclength locations, triplets of
macromolecules connected by two specifically located crosslinks, as
well as more complicated species such as pairs of self-crosslinked
macromolecules interlocking one another.
 
Let us label the various possible cluster species by the index 
$a=1,2,3,\ldots$, and let $a=0$ label the uncrosslinked 
macromolecule species. Then, for a specific realisation of the 
disorder (\ie, the crosslinks and the interlockings)
let the number of uncrosslinked macromolecules be $\nu_{0}$, and the 
number of clusters of species $a$ be $\nu_{a}$ \cite{REF:discrete}.  
Then the incorrect 
combinatorial factor of $N!$ should be replaced by the correct factor 
of $\prod_{a}\nu_{a}!$, this factor varying across disorder realisations.  
The physical partition function for a given realisation of the 
system is then given by 
\beq
\frac{1}{\prod_{a}\nu_{a}!}
\int{\cal D}{\bf c}\,\exp\left(-H_{1}^{\rm E}\right)
\prod_{e=1}^{M}\delta^{(d)}
        \big(
     {\bf c}_{i_{e}         }(s_{e}         )
    -{\bf c}_{i_{e}^{\prime}}(s_{e}^{\prime})
        \big).
\label{EQ:PFyesLinkedBetter}
\eeq
This correction of the combinatorial factor is mirrored by the absence,
due to the constraints, in the summation over system configurations of
those configuration in which macromolecules participating in a cluster
are widely separated, which results in the loss of volume factors.
Together, the corrected combinatorial factor and the loss of volume
factors conspire to yield a thermodynamic free energy that is properly
extensive.

In common with much work on the physics of disordered systems, we shall 
not focus on the statistical mechanics of a system having a particular 
realisation of the disorder.  Instead we shall take a probabilistic 
approach, focusing on the typical properties of randomly crosslinked 
macromolecular systems. To do this, we shall need to consider the 
statistical distribution of crosslink locations.  In fact, we shall 
also allow the number of crosslinks to fluctuate across realisations. 
\subsection{Deam-Edwards crosslink distribution}\label{SEC:DECrossDist}
To compute physical quantities characterising the system of randomly 
crosslinked macromolecules for a specific realisation of the large set 
of quenched random variables 
$\{i_{e},s_{e};i_{e}^{\prime},s_{e}^{\prime}\}_{e=1}^{M}$ is, of course, 
neither possible nor particularly useful. Instead we shall focus on 
{\it typical\/} values of physical quantities, constructed by suitably 
averaging them over the quenched random variables.  To perform this 
averaging we shall need to choose a probability distribution that 
assigns a sensible statistical weight 
${\cal P}_{M}
\big(
        \{i_{e},         s_{e};
          i_{e}^{\prime},s_{e}^{\prime}
        \}_{e=1}^{M}
\big)$ 
to each possible realisation of the number $M$ and location 
$\{i_{e},s_{e};i_{e}^{\prime},s_{e}^{\prime}\}_{e=1}^{M}$ of the 
crosslinks. Following an elegant strategy due to Deam and Edwards 
\cite{REF:DeamEd}, we assume that the normalised crosslink 
distribution is given by
\begin{equation}
{\cal P}_{M}
\big(
        \{i_{e},         s_{e};
          i_{e}^{\prime},s_{e}^{\prime}
        \}_{e=1}^{M}
\big) 
=
{\DPS
\big(\mu^{2}V/2N\big)^{M}
\,
\tilde{Z}\big(\{i_{e},s_{e};i_{e}^{\prime},s_{e}^{\prime}\}_{e=1}^{M}\big)
        \over{\DPS
M!\,
\Big\langle
\exp
        \Big( 
\frac{\mu^{2}V}{2N}
\sum\limits_{i,i^{\prime}=1}^{N}
\int\nolimits_{0}^{1}ds
\int\nolimits_{0}^{1}ds^{\prime}\,
\delta^{(d)}
\big(
     {\bf c}_{i         }(s         )
    -{\bf c}_{i^{\prime}}(s^{\prime})
\big)
        \Big)
\Big\rangle_{1}^{\rm E}
        }
}, 
\label{EQ:distribute}
\end{equation}
where 
$\tilde{Z}\big(\{i_{e},s_{e};i_{e}^{\prime},s_{e}^{\prime}\}_{e=1}^{M}\big)$
is given by Eq.~(\ref{EQ:partition}), and can be regarded as probing
the equilibrium correlations of the underlying uncrosslinked liquid 
\cite{REF:TDlimit}.
Such correlations were omitted from the crosslink distribution in certain 
previous works 
\cite{REF:prl1987,REF:pra1989,REF:kyoto1988,REF:macro1989,REF:AZPGNG}, 
which led to difficulties in obtaining a quantitative description of the 
amorphous solid state.  It is not, at present, clear whether this 
omission is significant for the liquid state.

The Deam-Edwards distribution can be envisaged as arising from a realistic 
vulcanisation process, in which crosslinks are introduced simultaneously 
and instantaneously into the liquid state in equilibrium 
\cite{REF:XLkinetics}.  Specifically, it incorporates the notion that 
all pairs of monomers that happen (at some particular instant) to be 
nearby are, with a certain probability controlled by the crosslink density 
parameter $\mu^{2}$, crosslinked.  Thus, the correlations of the crosslink 
distribution reflect the correlations of the uncrosslinked liquid, and it 
follows that realisations of crosslinks only acquire an appreciable 
statistical weight if they are compatible with some reasonably probable 
configuration of the uncrosslinked liquid.  
This good feature of the Deam-Edwards distribution is compatible 
with the random, space-filling, {\it frozen liquid\/}, nature of the 
equilibrium amorphous state that is achieved upon sufficient crosslinking.

We allow the number of crosslinks to fluctuate in a quasi-Poisson
manner, controlled by the parameter $\mu^{2}$.  All that we shall need
to know about $\mu^{2}$ is that the mean number of crosslinks per
macromolecule, which we denote by $[M]/N$, is a smooth,
monotonically-increasing function of $\mu^{2}$ that can, in principle,
be determined using the distribution ${\cal P}$ \cite{REF:Rough}.
We remark that the control parameter $\mu^{2}$ appears in
\eqref{EQ:distribute} divided by $N/V$. This factor is simply the
(dimensionless) density of macromolecules, which is an intensive
quantity. As we shall see, this choice leads to an equation of state
that does not depend on the density of macromolecules, at least at the
level of mean-field theory.  We also remark that no delicate scaling of
the control parameter is needed to achieve a good thermodynamic limit,
in contrast with the case of the Sherrington-Kirkpatrick spin-glass
model \cite{REF:SKmodel}.

As discussed in Sec.~\ref{SEC:XLasQRV}, at least in three dimensions
crosslinking confers anholonomic topological constraints on the
network as well as holonomic ones.  Thus, the statistical-mechanical
tool for constructing the crosslink distribution is not entirely
correct.  In principle, crosslink-realisations should be labelled not
only by $\{i_{e},s_{e};i_{e}^{\prime},s_{e}^{\prime}\}_{e=1}^{M}$, \ie,
by the number and arclength locations of the crosslinks, but also by
the precise topology of the realisation, \ie, by the manner in which
the macromolecules thread through the closed loops made by one
another.  Then the statistical weight attributed to a
crosslink-and-topology realisation would be better modelled as arising
from those configurations of the underlying equilibrium liquid that not
only satisfy the holonomic constraints but also the anholonomic ones.
As remarked in Sec.~\ref{SEC:XLasQRV}, no mathematical tool yet exists
for accomplishing this refinement analytically.  In other words, we are
treating the random topology of the system as annealed rather than
quenched.

One should pause to notice the striking feature that at the heart
of the Deam-Edwards crosslink distribution is the partition function
$\tilde{Z}\big(\{i_{e},s_{e};i_{e}^{\prime},s_{e}^{\prime}\}_{e=1}^{M}\big)$
of the crosslinked system, \ie, the crosslink distribution is itself
proportional to the partition function, the logarithm of which it is to
be used to average.  This fact gives the development a structure that
is rather appealing, at least from the point of view of form.  This
will become especially apparent in Sec.~\ref{SEC:DoaDecd} in the
context of the replica technique, in which this distribution is
generated via an additional (\ie, zeroth) replica, the permutation
aspect of the symmetry of the theory thereby being enlarged from the
permutation group ${\cal S}_{n}$ to ${\cal S}_{n+1}$, where $n$ is the
number of replicas \cite{REF:GoZiJoPhyA,REF:Neural}.  There is,
however, no physical basis for restricting attention {\it solely\/} to
crosslink distributions generated by the partition function identical
to that of the crosslinked system.  For example, one might imagine
crosslinking at a different temperature or solvent quality, which would
break the symmetry between the crosslink distribution and the partition
function of the crosslinked system; then, in the context of the replica
technique, the permutation aspect of the symmetry of the theory would
remain ${\cal S}_{n}$.
\subsection{Disorder averages and symmetry factors}\label{SEC:avSym}
How are we to use the Deam-Edwards crosslink distribution?  As is well
known, it is generally inappropriate in disordered systems to average
the partition function itself over the quenched random variables, as
this would amount to treating the quenched random variables as annealed
variables (\ie, equilibrated variables having the same status as the
variables describing the configurations of the system that can be
accessed during equilibrium fluctuations).  Rather, it is
thermodynamically extensive or intensive quantities, such as the free
energy or the order parameter, that should be averaged over the
quenched random variables \cite{REF:MPVbook}.  To illustrate this
point, consider the free energy relative to that of the uncrosslinked
system, $-\ln\tilde{Z}$. (Recall that we are measuring energies in
units such that $k_{\rm B}T=1$.)\thinspace\ Then the disordered-average
of the free energy per macromolecule (relative to the free energy of
the uncrosslinked system) per space dimension, which we denote
$\tilde{f}$, is given by
\begin{equation}
-dN\tilde{f}=
\left[\ln \tilde{Z}
\big(
        \{i_{e},s_{e};i_{e}^{\prime},s_{e}^{\prime}
        \}_{e=1}^{M}
\big)
\right], 
\label{EQ:logAverage}
\end{equation}
where the square brackets indicate a disorder-average, \viz., 
\bea
&&
\left[
{\cal O}_{M}
\big(
        \{i_{e},         s_{e};
       i_{e}^{\prime},s_{e}^{\prime}
        \}_{e=1}^{M}
\big)
\right]
\equiv
{\cal P}_{0}
{\cal O}_{0}+
\sum_{M=1}^{\infty}
\int\nolimits_{0}^{1}ds_{1}         \cdots ds_{M}
\int\nolimits_{0}^{1}ds_{1}^{\prime}\cdots ds_{M}^{\prime}
\nonumber
\\
&&
\qquad\qquad
\sum_{i_{1}=1}^{N}
\cdots
\sum_{i_{M}=1}^{N}
\sum_{i_{1}^{\prime}=1}^{N}
\cdots
\sum_{i_{M}^{\prime}=1}^{N}
{\cal P}_{M}
\big(
        \{i_{e},         s_{e};
       i_{e}^{\prime},s_{e}^{\prime}
        \}_{e=1}^{M}
\big)
\,
{\cal O}_{M}
\big(
        \{i_{e},         s_{e};
       i_{e}^{\prime},s_{e}^{\prime}
        \}_{e=1}^{M}
\big),
\label{EQ:AnyAverage}
\eea where 
${\cal O}_{M}\big(
        \{i_{e},         s_{e};
       i_{e}^{\prime},s_{e}^{\prime}
        \}_{e=1}^{M}\big)$ 
is an arbitrary function of the realisation of crosslinks.  
The average over the locations of the crosslinks excludes
realisations of the disorder in which two positions on  
the same macromolecule located closer than a persistence length 
participate in crosslinks.  This can be accomplished
by suitably cutting off the arclength integrations.
Inasmuch as $\tilde{Z}$ is not strictly speaking the physical partition
function, $\tilde{f}$ is not strictly speaking the physical free
energy.  In fact, from \eqref{EQ:PFyesLinkedBetter} we know that the
physical partition function (normalised with respect to that of the
uncrosslinked system) is given by
\beq 
Z\big(\{i_{e},s_{e};
i_{e}^{\prime},s_{e}^{\prime}\}_{e=1}^{M}\big)
=
{N!\over{\prod_{a}\nu_{a}!}}\,
\tilde{Z}\big(\{i_{e},s_{e};
i_{e}^{\prime},s_{e}^{\prime}\}_{e=1}^{M}\big).
\label{EQ:PartFuncPhys}
\eeq 
Thus, for the disorder-averaged physical free energy $f$, which is given by 
\beq 
-dNf\equiv
\left[\ln Z\big(\{i_{e},s_{e};
i_{e}^{\prime},s_{e}^{\prime}\}_{e=1}^{M}\big)\right],
\label{EQ:FrEnDefn}
\eeq 
we obtain 
\beq 
-dNf=-dN\tilde{f}+\ln N!-\Big[\ln\prod_{a}\nu_{a}!\Big].
\label{EQ:FrEnWiSyFac}
\eeq 
Now, $f$ is an intensive quantity.  However, $\Delta f$, defined by 
\beq 
\Delta f\equiv
f-\tilde{f}
=
-{1\over{dN}}\left[
\ln\left({\prod_{a}\nu_{a}!\over{N!}}\right)\right], 
\label{EQ:FrEnDelta}
\eeq 
is in general proportional to $\ln N$ (for large $N$). The constant of 
proportionality is, in general, difficult to compute: it will, however, 
be a small number for the case of lightly crosslinked systems, 
increasing to $d^{-1}$ in the high-crosslinking limit, for which all
macromolecules are connected to a single cluster. 
(For the uncrosslinked system $\Delta f=0$.)\thinspace\ 
Thus, $\tilde{f}$ ($=f-\Delta f$) contains a
term proportional to the logarithm of the size of the system, \ie, is
not intensive.  Despite this unphysical feature of $\tilde{f}$, it is
$\tilde{f}$ that we shall be computing, rather than $f$, because our
inability to compute $\big[\ln\prod_{a}\nu_{a}!\big]$ precludes us from
computing $f$.  However, the physical properties of the system, such as
the order parameter, \eg,
are determined by certain disorder-averaged quantities that we shall
show to be insensitive to the indistinguishability factor
$\prod_{a}\nu_{a}!$, 
and which can thus be computed in the present approach.
\section{Order parameter for the amorphous solid state}\label{SEC:OPFields}
\subsection{General properties of the order parameter}\label{SEC:GeneralProps}
We now discuss a certain order parameter constructed with the intention
of distinguishing between equilibrium states that are liquid (in which
the monomers are all delocalised), crystalline solid (in which a nonzero
fraction are localised in a spatially periodic fashion), globular (in
which the monomers have condensed within a spatial subvolume of the
system), and amorphous solid (in which a nonzero fraction are localised
in a spatially random fashion) \cite{REF:prl1987,REF:pra1989,REF:kyoto1988}.  
This order parameter
is an extension, {\it mutatis mutandis\/}, of the order parameter
introduced by Edwards and Anderson in the context of a class of
amorphous magnetic systems known as spin glasses
\cite{REF:SFEandPWA,REF:MPVbook}.

For a specific realisation of the crosslinks (\ie, prior to disorder
averaging), the appropriate order parameter is given by
\beq
\frac{1}{N}\sumin\int_{0}^{1}ds\,
\langle\exp\big(i{\bf k}^{1}\cdot{\bf c}_{i}(s)\big)\rangle_{\disfac}
\langle\exp\big(i{\bf k}^{2}\cdot{\bf c}_{i}(s)\big)\rangle_{\disfac}
\cdots
\langle\exp\big(i{\bf k}^{g}\cdot{\bf c}_{i}(s)\big)\rangle_{\disfac},
\label{EQ:OPdefNDA}
\eeq
for $g=1,2,3,\ldots$, none of the $d$-dimensional wave vectors 
$\{{\bf k}^{1},\ldots,{\bf k}^{g}\}$ being zero.  The angle brackets
$\langle\cdots\rangle_{\disfac}$ indicate an average over the
equilibrium state in question for a particular realisation of the
disorder, indicated by the subscript $\disfac$.  Such equilibrium states
may correspond to situations in which the translational symmetry of the
system is spontaneously broken, in which case they are not ergodic.
However, we shall not dwell here on the possibility of further
ergodicity-breaking (\eg, of the type commonly associated with the
concept of replica-symmetry breaking; see Ref.~\cite{REF:MPVbook}). This
restriction is consistent with the results presented below. For a
discussion of ergodicity-breaking in systems of crosslinked
macromolecular networks, see 
Refs.~\cite{REF:prl1987,REF:pra1989,REF:kyoto1988,REF:macro1989,REF:GoZiJoPhyA}.

The disorder-averaged order parameter is denoted by 
\beq
\left[
\frac{1}{N}\sumin\int_{0}^{1}ds\,
\langle\exp\big(i{\bf k}^{1}\cdot{\bf c}_{i}(s)\big)\rangle_{\disfac}
\langle\exp\big(i{\bf k}^{2}\cdot{\bf c}_{i}(s)\big)\rangle_{\disfac}
\cdots
\langle\exp\big(i{\bf k}^{g}\cdot{\bf c}_{i}(s)\big)\rangle_{\disfac}
\right].
\label{EQ:opDefinition}
\eeq
For any particular positive integer $g$, this order parameter may be
regarded as the $g^{\rm th}$ moment of the distribution of random static
density fluctuations ${\cal N}\big(\{\rho_{\bf k}\}\big)$ (see
Ref.~\cite{REF:AZPGNG}), which is defined by
\beq
{\cal N}\big(\{\rho_{\bf k}\}\big)
\equiv
\left[
\frac{1}{N}\sumin\int_{0}^{1}ds\,
{\prod_{{\bf k}}}^{\possym}
\delcomp\Big(
\rho_{\bf k}-
\langle\exp\big(i{\bf k}\cdot{\bf c}_{i}(s)\big)\rangle_{\disfac}\Big)
\right],
\label{EQ:stat_den_dist}
\eeq
where $\prod\nolimits_{{\bf k}}^{\possym}$ denotes the product over all
$d$-vectors ${\bf k}$ in the half-space given by the condition 
${\bf k}\cdot{\bf n}>0$ for a suitable unit $d$-vector ${\bf n}$, 
and the Dirac $\delta$-function of complex argument $\delcomp(z)$ is 
defined by $\delcomp(z)\equiv\delta(\real z)\,\delta(\imag z)$, where 
$\real z$ and $\imag z$ respectively denote the real and imaginary parts of 
the complex number $z$.  Thus,
\bea
&&
\int{\prod_{{\bf k}}}^{\possym}
d\,\real\rho_{\bf k}\,d\,\imag\rho_{\bf k}\,\,
{\cal N}\big(\{\rho_{\bf k}\}\big)\,\,
\rho_{{\bf k}^{1}}\rho_{{\bf k}^{2}}\cdots\rho_{{\bf k}^{g}}
\nonumber
\\
&&
\qquad
=
\left[
\frac{1}{N}\sumin\int_{0}^{1}ds\,
\langle\exp\big(i{\bf k}^{1}\cdot{\bf c}_{i}(s)\big)\rangle_{\disfac}
\langle\exp\big(i{\bf k}^{2}\cdot{\bf c}_{i}(s)\big)\rangle_{\disfac}
\cdots
\langle\exp\big(i{\bf k}^{g}\cdot{\bf c}_{i}(s)\big)\rangle_{\disfac}
\right].
\label{EQ:moment_stat_den}
\eea 

To see why formula~(\ref{EQ:OPdefNDA}) is indeed an order parameter
appropriate for distinguishing between liquid, crystalline, globular and
amorphous solid states, let us examine its qualitative properties.
First, suppose that the state is liquid.  Then each monomer $(i,s)$ is
to be found, with equal probability, in the vicinity of any location in
the container. Consequently, the equilibrium expectation value of its density 
$\langle\delta^{(d)}\big({\bf r}-{\bf c}_{i}(s)\big)\rangle_{\disfac}$ is the constant $V^{-1}$, and the Fourier transform 
$\langle\exp\big(i{\bf k}\cdot{\bf c}_{i}(s)\big)\rangle_{\disfac}$ vanishes (except for the trivial case of ${\bf k}={\bf 0}$).  Thus, for a liquid
state the order parameter~(\ref{EQ:OPdefNDA}) vanishes, all terms in
the summation over monomers vanishing identically.  This corresponds to
a state having full translational and rotational symmetry.

Next, consider the case when a nonzero fraction of monomers are
localised in the vicinity of specific points in space, albeit exhibiting
thermal fluctuations about these points.  In this case, for many
monomers $(i,s)$ the quantity $\langle\delta^{(d)}\big({\bf r}-{\bf
c}_{i}(s)\big)\rangle_{\disfac}$ will be more or less sharply peaked at
some point in space and, correspondingly,
$\langle\exp\big(i{\bf k}\cdot{\bf c}_{i}(s)\big)\rangle_{\disfac}$ 
will not vanish identically, instead varying with ${\bf k}$ so as to 
reflect the spatial localisation of monomer $(i,s)$.  Then
\beq
\big\langle\exp\big(i{\bf k}\cdot{\bf c}_{i}(s)\big)\big\rangle_{\disfac}
=
\exp\big(i{\bf k}\cdot{\bf b}_{i}(s)\big)
\,\wp_{(i,s)}\big({\bf k}\big)
\label{EQ:xtal_den_dist}
\eeq
where ${\bf b}_{i}(s)$ is the site about which monomer $(i,s)$ is localised, 
and $\wp_{(i,s)}\big({\bf k}\big)$, which does not vanish identically, 
is the Fourier transform of the density 
of a monomer localised at the origin. In such a state, translational 
invariance is broken at the microscopic level.  However, the symmetry of 
the state of the system at the macroscopic level is not settled without 
further information.  

What possibilities present themselves in the situation in which a
nonzero fraction of the monomers are localised?  If the mean locations
$\{{\bf b}_{i}(s)\}$ of the localised monomers are distributed randomly 
and homogeneously over the volume of the system then the state is said 
to be macroscopically translationally invariant (MTI), the inclusion of
rotational invariance being understood.  We mean by this that there is
no periodicity, or any other macroscopic feature capable of
distinguishing one equilibrium state from any global translation or
rotation of it.  We refer to such states as (equilibrium) amorphous
solid states. On the other hand, if the mean locations $\{{\bf
b}_{i}(s)\}$ of the localised monomers are distributed inhomogeneously
over the volume of the system then the state is said to break
translational invariance macroscopically.  Examples of such states are
the globular state \cite{REF:LIFandKOKH}, in which the monomers have
condensed (in space) within a subvolume of the system, and the
crystalline state, in which the mean locations of the monomers are
arranged in a periodic lattice (and the monomers may be regarded as
having condensed in wave vector space).

How are the various possible states diagnosed by the order parameter?
As we showed above, the order parameter is zero for all 
$\{{\bf k}^{1},\ldots,{\bf k}^{g}\}$ in a state that is translationally
invariant at the microscopic level (\ie, a liquid).  On the other hand,
it will take at least some nonzero values for any state in which
translational invariance is broken at the microscopic level.  By using
\eqref{EQ:xtal_den_dist} we see that in such a state the order
parameter~(\ref{EQ:OPdefNDA}) becomes
\beq
\frac{1}{N}\sumin\int_{0}^{1}ds\,
\wp_{(i,s)}\big({\bf k}^{1}\big)
\wp_{(i,s)}\big({\bf k}^{2}\big)
\cdots
\wp_{(i,s)}\big({\bf k}^{g}\big)
\,
\exp
\Big(
i\big(
{\bf k}^{1}+
{\bf k}^{2}+\cdots+
{\bf k}^{g}
\big)
\cdot{\bf b}_{i}(s)
\Big).
\label{EQ:op_xtal}
\eeq
This order parameter also provides a way to distinguish between
nonliquid states that are MTI and those that are not.  In the case of an
MTI state the summation of complex phase factors will totally
destructively interfere unless the wave vectors happen to sum to zero,
the random locations of the mean monomer-positions otherwise leading to
random phase cancellations.  Hence, the order parameter will only fail
to vanish for values of the wave vectors
$\{{\bf k}^{1},{\bf k}^{2},\dots,{\bf k}^{g}\}$ 
that sum to zero.  This property of being MTI is a fundamental
characteristic of the amorphous solid state.  In the non-MTI case total
destructive interference is avoided not only if the wave vectors sum to
zero but also under other circumstances.  Hence, in this case the order
parameter will also fail to vanish for certain values of the wave
vectors $\{{\bf k}^{1},{\bf k}^{2},\dots,{\bf k}^{g}\}$ that do not sum
to zero. To establish this, consider how formula~(\ref{EQ:op_xtal})
transforms under a global translation by an arbitrary vector ${\bf a}$:
\bea
&&
\frac{1}{N}\sumin\int_{0}^{1}ds\,
\wp_{(i,s)}\big({\bf k}^{1}\big)
\wp_{(i,s)}\big({\bf k}^{2}\big)
\cdots
\wp_{(i,s)}\big({\bf k}^{g}\big)
\,
\exp
\Big(
i\big(
{\bf k}^{1}+
{\bf k}^{2}+\cdots+
{\bf k}^{g}
\big)
\cdot{\bf b}_{i}(s)
\Big)
\nn
\noalign{\medskip}
&&
\qquad
\rightarrow
\exp
\Big(
i\big(
{\bf k}^{1}+
{\bf k}^{2}+\cdots+
{\bf k}^{g}
\big)
\cdot{\bf a}
\Big)
\nn
\noalign{\medskip}
&&
\qquad
\times
\frac{1}{N}\sumin\int_{0}^{1}ds\,
\wp_{(i,s)}\big({\bf k}^{1}\big)
\wp_{(i,s)}\big({\bf k}^{2}\big)
\cdots
\wp_{(i,s)}\big({\bf k}^{g}\big)
\,
\exp
\Big(
i\big(
{\bf k}^{1}+
{\bf k}^{2}+\cdots+
{\bf k}^{g}
\big)
\cdot{\bf b}_{i}(s)
\Big).
\label{EQ:translate}
\eea
In situations of  MTI, this transformation must leave the order
parameter unchanged for all vectors ${\bf a}$.  This enforces the
condition that for MTI situations the order parameter must vanish unless
${\bf k}^{1}+{\bf k}^{2}+\cdots+{\bf k}^{g}={\bf 0}$, and thus the order
parameter becomes
\beq
\delta_{{\bf 0},{\bf k}^{1}+\cdots+{\bf k}^{g}}\,
\frac{1}{N}\sumin\int_{0}^{1}ds\,
\wp_{(i,s)}\big({\bf k}^{1}\big)
\cdots
\wp_{(i,s)}\big({\bf k}^{g}\big),
\label{EQ:OPisMTI}
\eeq
where $\delta_{{\bf p}^{1},{\bf p}^{2}}$ is a $d$-dimensional Kronecker
$\delta$-factor, which is nonzero only if the $d$-vectors ${\bf p}^{1}$
and ${\bf p}^{2}$ have all components equal, in which case it has the
value unity.  In this state, in contrast with the crystalline state,
there is no periodicity associated with the spatial pattern of localised
monomers, and thus there will not be a collection of reciprocal lattice
vectors for which the order parameter fails to vanish.  In particular,
the order parameter vanishes for $g=1$.  (One may equivalently regard
the amorphous solid state as the special case of the crystalline state
in which the unit cell of the crystal is the entire sample, \ie, a
realisation of Schr{\"o}dinger's \lq\lq aperiodic
solid\rlap.\rq\rq\ \cite{REF:ESgap})\thinspace\ The equilibrium
amorphous solid state is characterised by the presence of random (\ie,
nonperiodic) static density fluctuations, which spontaneously break
translational symmetry at the microscopic (but not the macroscopic)
level.

In the non-MTI case, either the transformation~(\ref{EQ:translate}) must
leave the order parameter unchanged for a discrete lattice of vectors
${\bf a}$, or it need not leave the order parameter unchanged for any
value of ${\bf a}$. When there is invariance for a discrete lattice of
vectors (\ie, in the crystalline state), the order parameter must vanish
unless ${\bf k}^{1}+{\bf k}^{2}+\cdots+{\bf k}^{g}={\bf G}$, where 
${\bf G}$ is any reciprocal lattice vector of the crystal (including the 
zero reciprocal lattice vector). When there is no vector for which the
invariance holds (\ie, in the globular state) the order parameter need
not vanish on symmetry grounds for any values of the wave vectors
$\{{\bf k}^{1},\ldots,{\bf k}^{g}\}$.

To summarise, the values of the order parameter for the various values
of $g$ and the wave vectors 
$\{{\bf k}^{1},{\bf k}^{2},\dots,{\bf k}^{g}\}$ serve to distinguish 
between liquid, crystalline, globular and amorphous solid states: for
liquid states the order parameter vanishes for $g=1,2,3,\dots$; for
amorphous solid states it vanishes for all wave vectors that do not sum
to zero (and thus vanishes for $g=1$); for crystalline states it only
vanishes for wave vectors that fail to sum to a reciprocal lattice
vector (and therefore is nonzero for some 
$\{{\bf k}^{1},{\bf k}^{2},\dots,{\bf k}^{g}\}$, even when $g=1$); and 
for globular states it need not vanish in symmetry grounds for any values 
of $\{{\bf k}^{1},{\bf k}^{2},\dots,{\bf k}^{g}\}$.
\subsection{A simple idealisation: 
	generalised Einstein model}\label{SEC:EinsteinMTI}
To illustrate the general properties of the order parameter, and to
motivate the specific hypothesis for the form of the order parameter
described in Sec.~\ref{SEC:OPhypoth} and applied in
Sec.~\ref{SEC:StatPtCrit}, we examine a simple caricature of the
amorphous solid state.  We refer to this caricature as a generalised
Einstein model, by analogy with the Einstein model of a crystalline
solid adopted for the computation of the specific heat, in which it is
assumed that every atom is independently localised by an identical
harmonic potential \cite{REF:AlEinstein}.  In the context of amorphous
solidification, the caricature is obtained by asserting that a fraction
$(1-{\LocFr})$ of the monomers (the so-called sol fraction) are
delocalised, with each monomer $(i,s)$ of the remaining fraction
${\LocFr}$ (the so-called gel fraction) being localised near a random
mean position ${\bf b}_{i}(s)$, its location exhibiting thermal
fluctuations about that mean position.  We emphasise that our usage of
the terms gel and sol in this article refers solely to the issue of
whether or not a monomer is localised.  Ultimately, however, we shall
see that the gel fraction defined in this way coincides with the more
common architectural definition, in the sense that localisation will be
seen to occur only for crosslink densities for which the network spans
the entire system.  It is further asserted that the probability
distribution for the fluctuations in location of each localised monomer
$(i,s)$ about its mean position is gaussian and isotropic, and
characterised by an inverse square localisation length $\tau_{i}(s)$.
Then if $(i,s)$ is a localised monomer its Fourier-transformed density
would be given by
\beq
\big\langle
\exp\big(i{\bf k}\cdot{\bf c}_{i}(s)\big)
\big\rangle_{\disfac}
=\exp\big(i{\bf k}\cdot{\bf b}_{i}(s)\big)
 \exp\big(-k^{2}/2\tau_{i}(s)\big),
\label{EQ:gauss_dense}
\eeq
so that the order parameter (prior to disorder-averaging) becomes
\beq
(1-{\LocFr})\prod_{a=1}^{g}\kdelvec{{\bf 0}}{{\bf k}^{\mit a}}
+\frac{1}{N}\sumin\int_{0}^{1}ds\,
\exp
\Big(
i{\bf b}_{i}(s)\cdot\sum_{a=1}^{g}{\bf k}^{a}
\Big)\,
\exp
\Big(
-\sum_{a=1}^{g}
\vert{\bf k}^{a}\vert^{2}/2\tau_{i}(s)
\Big),
\label{EQ:op_alpha}
\eeq
where it is understood that the summation in the second term 
only includes localised monomers.

To obtain the disorder-average of the order parameter we assume that 
for each monomer $(i,s)$ the random variables ${\bf b}_{i}(s)$ and 
$\tau_{i}(s)$ are uncorrelated.  Furthermore we assume that ${\bf b}_{i}(s)$ 
is unformly distributed over the volume $V$ and that the inverse square 
localisation length $\tau_{i}(s)$ has the probability distribution $p(\tau)$. 
(The notion that the state could be characterised by a statistical 
distribution of localisation lengths was introduced in \cite{REF:kyoto1988} 
and implemented in \cite{REF:CGZjourEPL}.)\thinspace\  
In this case the order parameter becomes
\beq
(1-{\LocFr})\prod_{a=1}^{g}\kdelvec{{\bf 0}}{{\bf k}^{\mit a}}
\,+\,
{\LocFr}\,\kdelvec{\bf 0}{\sum_{\mit a=1}^{\mit g}{\bf k}^{\mit a}}
\int_{0}^{\infty}d\tau\,p(\tau)
\exp\Big(-\sum\limits_{a=1}^{g}
\vert{\bf k}^{a}\vert^{2}/2\tau\Big).
\label{EQ:op_form}
\eeq
The first term accounts for the delocalised monomers, and the second
term accounts for the localised monomers.  If ${\LocFr}=0$ then the state
described by this order parameter is the liquid state.  If ${\LocFr}\ne 0$ then
it describes an amorphous solid state. The Kronecker $\delta$ factor in
front of the second term is a reflection of the MTI that characterises
the amorphous solid state.  This hypothesis is a refinement of the
gaussian hypothesis used in a number of contexts \cite{REF:GaussHypo}.
It is useful to observe that the gel fraction ${\LocFr}$ can be extracted from
the order parameter~\eqref{EQ:OPdefNDA} by taking the limit of the
order parameter as 
$\{{\bf k}^{1},\ldots,{\bf k}^{g}\}
\rightarrow\{{\bf 0},\ldots,{\bf 0}\}$ 
through a sequence for which 
$\sum_{\mit a=1}^{\mit g}{\bf k}^{\mit a}={\bf 0}$.
\subsection{Replica--order-parameter hypothesis: gel fraction and 
	distribution of localisation lengths}\label{SEC:OPhypoth}
Having discussed the physical order parameter capable of diagnosing
equilibrium amorphous solidification, we now anticipate the development
of the replica approach by describing the particular form that we shall
hypothesise for the replica order parameter, \ie, the order parameter
that emerges from the application of the replica technique and
represents, in the replica approach, the physical order parameter
discussed in the previous two subsections of the present section.  This
form is motivated by the general characterisation of amorphous
solidification in terms of the gel fraction ${\LocFr}$ and the distribution of
inverse square localisation lengths $p(\tau)$ given in
Sec.~\ref{SEC:EinsteinMTI}.  Below, in Sec.~\ref{SEC:StatPtCrit}, we
shall show that within the context of a certain model of randomly
crosslinked macromolecular networks the form that we now hypothesise for
the replica order parameter is sufficiently broad to allow us to provide
an exact and physically appealing mean-field--level description of the
transition to and properties of the equilibrium amorphous solid state of
randomly crosslinked macromolecular networks.

As we shall see in detail in Sec.~\ref{SEC:DoaDecd}, the replica 
representation of the physical order parameter is given by
\beq
\Big\langle
\frac{1}{N}\sum_{i=1}^{N}
\int_{0}^{1}ds\,
\exp\big(
i\sumaln{\bf k}^{\alpha}\cdot{\bf c}_{i}^{\alpha}(s)\big)
\Big\rangle_{n+1}^{\rm P},
\label{EQ:ReplicaOrder}
\eeq
in the replica limit, $n\rightarrow 0$. As we shall also see there,
$\langle\cdots\rangle_{n+1}^{\rm P}$ denotes an expectation value for a
pure (\ie, quenched-disorder--free) system of $n+1$ coupled replicas of
the original macromolecular system.  Note the inclusion of degrees of
freedom associated with a replica labelled by $\alpha=0$. For the sake
of notational convenience we introduce hatted vectors (\eg, ${\hat k}$
or ${\hat c}$), which are $(n+1)d$-component vectors comprising
$(n+1)$-fold replicated sets of $d$-component vectors 
(\eg,  the wave vectors $\{{\bf k}^{0},{\bf k}^{1},\ldots,{\bf k}^{n}\}$ 
or the position vectors $\{{\bf c}^{0},{\bf c}^{1},\ldots,{\bf c}^{n}\}$).  
We define the extended scalar product ${\hat k}\cdot{\hat c}$ by 
${\hat k}\cdot{\hat c}
\equiv
\sum_{\alpha=0}^{n}
{\bf k}^{\alpha}
\cdot{\bf c}^{\alpha}$, 
having the special cases
${\hat k}^{2}\equiv{\hat k}\cdot{\hat k}$ and 
${\hat c}^{2}\equiv{\hat c}\cdot{\hat c}$. 
In terms of this notation, the order parameter becomes
\beq
\Big\langle
\frac{1}{N}\sum_{i=1}^{N}
\int_{0}^{1}ds\,
\exp\big(
i\hat{k}\cdot\hat{c}_{i}(s)\big)
\Big\rangle_{n+1}^{\rm P}.
\label{EQ:ReplicaOrdHat}
\eeq

By translating formula~(\ref{EQ:op_form}) into the replica language,
through the use of \eqref{EQ:OPsource}, we shall assume that the replica
order parameter takes on values expressible in the following form:
\beq
\left(1-{\LocFr}\right)\kdelhat{k}{0}+
{\LocFr}\,\kdelvecT{k}{0}\dint{\tau}\,
\exp(-\hat{k}^{2}/2\tau), 
\label{EQ:OPhypothesis}
\eeq
where $\tilvec{k}\equiv\sum_{\alpha=0}^{n}{\bf k}^{\alpha}$ is a
permutation-invariant $d$-vector built by summing the elements of the
replicated vector $\hat{k}$, and
$\kdelhat{p}{q}\equiv\prod_{\alpha=0}^{n} \kdelvec{{\bf
p}^{\alpha}}{{\bf q}^{\alpha}}$.  Thus, we parametrise the order
parameter in terms of the gel fraction ${\LocFr}$ and the distribution of
(inverse square) localisation lengths $p(\tau)$ \cite{REF:condition}.
This parametrisation is severely restrictive but physically plausible.
In order for $p(\tau)$ to be interpreted as a probability distribution
it must be non-negative. This condition is not imposed {\it a priori\/},
but emerges from the stationarity condition.  The ranging of $\tau$ only
over positive values reflects the fact that inverse square localisation
lengths are positive. Moreover, delocalised monomers are accounted for
by the term proportional to $(1-{\LocFr})$, so that $p(\tau)$ must not contain
a Dirac $\delta$-function--like piece at $\tau=0$.  The factor
$\kdelvecT{k}{0}$ incorporates the MTI property into the hypothesised
form.  It should be emphasised that the hypothesised form is invariant
under the permutation of the replicated vectors 
$\{{\bf k}^{0},{\bf k}^{1},\ldots,{\bf k}^{n}\}$, 
which is a manifestation of its
replica-symmetric character.
\subsection{Symmetry properties of the order parameter 
hypothesis}\label{SEC:SyPropsOPhyp}
We now state explicitly the symmetry properties of the order parameter
hypothesis~(\ref{EQ:OPhypothesis}) that we shall use throughout the
remainder of this article.  As we shall see in Sec.~\ref{SEC:DoaDecd},
the effective hamiltonian of the 
replica theory turns out to have the following symmetries:
(i)~independent translations or rotation of the replicas, and 
(ii)~permutations of the replicas. 
In the liquid state the order parameter retains all these symmetries. 
In the amorphous solid state the symmetry of the order parameter is reduced. 
By invoking our hypothesis for the order parameter we are assuming 
that in the amorphous solid state the residual symmetries are:
(i$^{\prime}$)~common translations and rotations of the replicas, and 
(ii)~permutations of the replicas. 
In other words, in the transition to the amorphous solid state the 
symmetry of independent translations and rotations of the replicas 
is spontaneously broken.  As a consequence of the spontaneous breaking 
of certain symmetries there is a manifold of symmetry-related 
values of the order parameter that describe the solid state.

In this article we have restricted our attention to order-parameter
hypotheses that are invariant under the permutations of all $(n+1)$
replicas (\ie, that are replica-symmetric).  This mathematical
restriction is equivalent to the physical condition that, upon
amorphous solidification, the (overwhelming fraction of the) system
must exhibit one member of a unique family of equilibrium states (\ie,
statistical arrangements of the macromolecules), this unique family of
states being related by global translations and rotations.  Whilst the
occurrence of a unique family would not be an unreasonable consequence
of crosslinking, especially in view of our exclusion of the anholonomic
contraints that crosslinking introduces into the physical system, one
might anticipate that crosslinking would cause the full physical system
to exhibit many families of states (\ie, there would be states that are
not related by global translations and rotations). Such an occurrence
would be signalled by an order parameter that is no longer invariant
under permutations of the replicas (\ie, for which replica symmetry is
spontaneously broken).  For discussions of these matters, see 
Ref.~\cite{REF:GoZiJoPhyA} as well as 
Refs.~\cite{REF:prl1987,REF:pra1989,REF:kyoto1988,REF:macro1989}. 
The issue of whether or not there exists a more accurate treatment of
randomly crosslinked macromolecular systems that would indicate the
exitence of many un--symmetry-related states is a matter of ongoing
research \cite{REF:RSB}.

It must, however, be emphasised that, regardless of the issue of the
intactness of permutation symmetry, the primary physical phenomemon at
hand in the formation of the equilibrium amorphous solid state is the
spontaneous breaking of translation symmetry (\viz., the spontaneous
random localisation of macromolecules). The issue of
replica-symmetry--breaking is not an alternative to
translational-symmetry--breaking: it simply addresses whether or not a
system with given realisation of crosslinking possesses one or many
un--symmetry-related ways for the macromolecules to be randomly
localised.  To allow for the possibility that
replica-symmetry--breaking accompanies translational-symmetry--breaking
is to explore a more general class of behaviours of the system.
\subsection{Connection with scattering experiments}\label{SEC:scatter}
The order parameter that we have been addressing in the present section
is, in principle, accessible via neutron scattering experiments
\cite{REF:Mezei,REF:Sinha}, at least for the case $g=2$. In fact, the
elastic part of the differential scattering cross-section (per atom) can
be written as
\bea
{1\over{N}}
{d^{2}\sigma\over{d\Omega}}
&=&
S^{\rm el}({\bf q})
=
\lim_{t\to\infty}
\bigg(
{\vert b_{\rm coh}\vert^{2}\over{N}}
\Big\langle
\sum_{i,i^{\prime}=1}^{N}
\int_{0}^{1}ds\,ds^{\prime}
\exp\big(-i{\bf q}\cdot{\bf c}_{i}(s;0)\big)\,
\exp\big(i{\bf q}\cdot{\bf c}_{i^{\prime}}(s^{\prime};t)\big)
\Big\rangle
\nonumber
\\
&&
\qquad\qquad\qquad
+
{\vert b_{\rm incoh}\vert^{2}\over{N}}
\Big\langle
\sum_{i=1}^{N}
\int_{0}^{1}ds
\exp\big(-i{\bf q}\cdot{\bf c}_{i}(s;0)\big)\,
\exp\big(i{\bf q}\cdot{\bf c}_{i}(s;t)\big)
\Big\rangle
\bigg),
\label{EQ:DSXS}
\eea
where $b_{\rm coh}$ is the average scattering length, $b_{\rm incoh}$
is the variance of the scattering length, ${\bf c}_{i}(s;t)$ is the
position of monomer $s$ on macromolecule $i$ at time $t$, and
$\langle\cdots\rangle$ indicates a time-dependent equilibrium
expectation value.  The second part on the right hand side is 
the {\it incoherent\/} contribution, and can be extracted 
in some cases.  By using the fact that the connected correlators 
vanish for $t\rightarrow\infty$, we see that this second part 
reduces to 
\beq
{\vert b_{\rm incoh}\vert^{2}\over{N}}
\sum_{i=1}^{N}
\int_{0}^{1}ds
\Big\langle
\exp\left(-i{\bf q}\cdot{\bf c}_{i}(s)\right)
\Big\rangle\,
\Big\langle
\exp\left(i{\bf q}\cdot{\bf c}_{i}(s)\right)
\Big\rangle,
\label{EQ:Reduce}
\eeq
\ie, formula~(\ref{EQ:OPdefNDA}) evaluated for the special case of 
$\{{\bf  k}^{1},{\bf k}^{2},    \ldots,{\bf k}^{g}\}=
\{ {\bf -q},    {\bf q},{\bf 0},\ldots,{\bf 0}    \}$. 
Thus, the order parameter for $g=2$ is proportional to the 
incoherent part of the elastic neutron scattering cross-section.

Oeser {\it et al.\/} \cite{REF:Oeser} have measured the time persistent
part of the incoherent scattering function in neutron-spin-echo
experiments. They fit their data, which are taken in the high
crosslinking limit, to a gaussian in wave vector space characterised by a
typical length scale $l$, which turns out to be comparable to the radius
of gyration. A potential critique of neutron scattering experiments
results from the available time scales, of order $10\,{\rm ns}$, which
make it difficult to extrapolate to infinite time in order to extract
the time-persistent part of the autocorrelation.  This may not be a
severe problem in the high crosslinking limit, in which one expects
rather small time scales associated with small distances between
crosslinks.  However, it may become prohibitive for weakly crosslinked
systems, which barely sustain an infinite cluster.

Pulsed field gradient NMR (see, \eg, Ref.~\cite{REF:Krager}) is another
experimental technique for measuring the intermediate-time incoherent
scattering function with a spatial and temporal resolution that is
complementary to neutron scattering experiments. Typical time scales in
NMR experiments are of order milliseconds, and length scales are
restricted to be greater than $10\,{\rm nm}$. An example of such
measurements are the detection of spatial fluctuations in swollen
networks in Ref.~\cite{REF:Skirda}.
\section{Replica approach for disorder-averaged quantities}\label{SEC:DoaDecd}
Having prepared the way by discussing the model and the construction of
a suitable probability distribution for the disorder, and defining an
order parameter capable of diagnosing the possible states of the
system, we now turn to the computation of disorder-averages of
important physical quantities, such as the free energy, order parameter
and certain correlators.  A direct assault on this task, as it stands,
seems prohibitively difficult, but it can be rendered tractable by the
use of the replica technique \cite{REF:MPVbook}, pioneered in the
context of macromolecular networks by Deam and Edwards
\cite{REF:DeamEd}.  In this approach, we do not consider just the
original degrees of freedom but, instead, a system comprising $n+1$
interacting copies (\ie, replicas) of it that will be labelled by the
superscript $\alpha=0,1,\ldots,n$.  In this new system, the quenched
randomness disappears from the formulation, at the price of introducing
an inter-replica interaction.
\subsection{Replica-Helmholtz free energy}\label{SEC:RtRpf}
In order to compute quantities such as the disorder-averaged free
energy, the order parameter and the elastic free energy
\cite{REF:CGZforth} in a unified way, it is particularly convenient
to introduce a certain generating functional $\fgenf(\mu^{2},\{U\})$, 
which we refer to as the replica-Helmholtz free energy and define as
follows:
\begin{equation}
\exp\big(-ndN\fgenf(\mu^{2},\{U\})\big)
\equiv 
{\DPS           
{\DPS           
{\DPS           
\Big\langle
\exp
        \Big(-
\frac{\lambda^{2}}{2}
\sum_{i,i^{\prime}=1}^{N}
\int\nolimits_{0}^{1}ds
\int\nolimits_{0}^{1}ds^{\prime}\,
\sum_{\alpha=0}^{n}
\delta^{(d)}
\big(
 {\bf c}_{i}^{\alpha}(s)
-{\bf c}_{i^{\prime}}^{\alpha}(s^{\prime})
\big)
        \Big)
\hfill
\atop{\DPS
\hfill
\quad
\times
\exp
        \Big(
\frac{\mu^{2}V}{2N}
\sum_{i,i^{\prime}=1}^{N}
\int\nolimits_{0}^{1}ds
\int\nolimits_{0}^{1}ds^{\prime}\,
\prod_{\alpha=0}^{n}
\delta^{(d)}
\big(
 {\bf c}_{i}^{\alpha}(s)
-{\bf c}_{i^{\prime}}^{\alpha}(s^{\prime})
\big)
        \Big)
}}	
\hfill\atop{\DPS
\hfill\times
\exp
       \Big(
-\frac{V^{n+1}}{N}
\sum_{i=1}^{N}
\int\nolimits_{0}^{1}ds\,
U\big({\hat c}_{i}(s)\big)
        \Big)
\Big\rangle_{n+1}^{\rm W}
}}      
\over{\DPS
{\DPS   
\Big\langle
\exp
        \Big(-
\frac{\lambda^{2}}{2}
\sum_{i,i^{\prime}=1}^{N}
\int\nolimits_{0}^{1}ds
\int\nolimits_{0}^{1}ds^{\prime}\,
\delta^{(d)}
\big(
 {\bf c}_{i}(s)
-{\bf c}_{i^{\prime}}(s^{\prime})
\big)
        \Big)
\hfill
\atop{\DPS
\hfill
\qquad
\times
\exp
        \Big(
\frac{\mu^{2}V}{2N}
\sum_{i,i^{\prime}=1}^{N}
\int\nolimits_{0}^{1}ds
\int\nolimits_{0}^{1}ds^{\prime}\,
\delta^{(d)}
\big(
 {\bf c}_{i}(s)
-{\bf c}_{i^{\prime}}(s^{\prime})
\big)
        \Big)
\Big\rangle_{1}^{\rm W}
}}      
}}.     
\label{EQ:newPartition}
\end{equation}
The expectation value $\langle\cdots\rangle_{n+1}^{\rm W}$ 
is defined to be
\begin{equation}
\Big\langle\cdots\Big\rangle_{n+1}^{\rm W}
\equiv
{\DPS
\int\prod_{\alpha=0}^{n}{\cal D}{\bf c}^{\alpha}\,
\exp\left({-H_{n+1}^{\rm W}}\right)\cdots
\over{\DPS
\int\prod_{\alpha=0}^{n}{\cal D}{\bf c}^{\alpha}\,
\exp\left({-H_{n+1}^{\rm W}}\right)}}, 
\label{EQ:wienaverrep}
\end{equation} 
\ie, an average with respect to the $(n+1)$-fold replicated Wiener
measure, which is proportional to $\exp\big(-H_{n+1}^{\rm W}\big)$,
where
\begin{equation}
H_{n+1}^{\rm W}=
\frac{1}{2}
\sumaln
\sum_{i=1}^{N}
\int\nolimits_{0}^{1}ds
\Big\lefver
  {d\over{ds}}{\bf c}_{i}^{\alpha}(s)
\Big\rigver^{2}.
\label{EQ:WienerMeasureReplica}
\end{equation} 
In analogy with magnetic systems, the properties of which one can probe
by computing the free energy as a function of an external,
position-dependent, magnetic field, it is useful in the present context
to introduce the external potential $U$ on which $\fgenf$ depends.  The
potential $U$, which takes as its argument an $(n+1)$-fold replicated
$d$-vector $\hat{x}$, acts simultaneously on replicas of all monomers,
has zero average (over ${\hat x}$), and plays the role of a source
field.  As we shall now show, knowledge of
$\fgenf(\mu^{2},\{U\})\big\vert_{U=0}$ allows us to compute the
disorder-averaged free energy per macromolecule $d\tilde{f}$.
Subsequently, we shall see that by (functional) differentiation of
$\fgenf(\mu^{2},\{U\})$ with respect to $U$ we shall be able to
compute the order parameter.  Furthermore, it is possible to
investigate the elastic properties of the amorphous solid state of
randomly crosslinked macromolecular networks via making a Legendre
transformation with respect to $U$ (see \cite{REF:CGZforth}).

The quantity ${\FTDen}_{\hat{k}}$, defined by 
\beq
{\FTDen}_{\hat{k}}
\equiv
\frac{1}{N}\sum_{i=1}^{N}
\int_{0}^{1}ds\,
\exp\Big(
i\hat{k}\cdot\hat{c}_{i}(s)\Big),
\label{EQ:density}
\eeq
is the Fourier transform of the replicated monomer density.  
As we shall see in Sec.~\ref{SEC:compAv}, a certain average of
${\FTDen}_{\hat{k}}$ turns out to be the replica representation 
of the order parameter.  In order to generate ${\FTDen}_{\hat{k}}$ 
it is useful to introduce the Fourier representation of the
external potential $U$, \viz.,
\beq
U(\hat{c})
=
{1\over{V^{n+1}}}
\sum_{{\hat k}}
U_{\hat{k}}
\exp\big(i\hat{k}\cdot\hat{c}\big), 
\label{EQ:FourURep}
\eeq
which leads to the replacement
\beq
\frac{V^{n+1}}{N}
\sum_{i=1}^{N}
\int\nolimits_{0}^{1}ds\,
U\big({\hat c}_{i}(s)\big)
\rightarrow
\sumhat{k}
U_{\hat{k}}\,
{\FTDen}_{\hat{k}},
\eeq
in \eqref{EQ:newPartition}, so that $U_{\hat{k}}$ is a source for
${\FTDen}_{\hat{k}}$.  We shall, in addition, make use of expectation
values involving the weight that features in the replica-Helmholtz free
energy, \eqref{EQ:newPartition}:
\beq
\Big\langle\cdots\Big\rangle_{n+1,U}^{\rm P}
\equiv
{\DPS           
{\DPS           
{\DPS           
\Big\langle
\cdots\,
\exp
        \Big(-
\frac{\lambda^{2}}{2}
\sum_{i,i^{\prime}=1}^{N}
\int\nolimits_{0}^{1}ds
\int\nolimits_{0}^{1}ds^{\prime}\,
\sum_{\alpha=0}^{n}
\delta^{(d)}
\big(
 {\bf c}_{i}^{\alpha}(s)
-{\bf c}_{i^{\prime}}^{\alpha}(s^{\prime})
\big)
        \Big)
\hfill
\atop{\DPS
\hfill\quad
\times
\exp
        \Big(
\frac{\mu^{2}V}{2N}
\sum_{i,i^{\prime}=1}^{N}
\int\nolimits_{0}^{1}ds
\int\nolimits_{0}^{1}ds^{\prime}\,
\prod_{\alpha=0}^{n}
\delta^{(d)}
\big(
 {\bf c}_{i}^{\alpha}(s)
-{\bf c}_{i^{\prime}}^{\alpha}(s^{\prime})
\big)
        \Big)
}}	
\hfill\atop{\DPS\hfill\times
\exp
       \Big(
-\sum\nolimits_{\hat{k}}
U_{\hat{k}}
{\FTDen}_{\hat{k}}
        \Big)
\Big\rangle_{n+1}^{\rm W}
}}      
\over{
{\DPS           
{\DPS           
\Big\langle
\exp
        \Big(-
\frac{\lambda^{2}}{2}
\sum_{i,i^{\prime}=1}^{N}
\int\nolimits_{0}^{1}ds
\int\nolimits_{0}^{1}ds^{\prime}\,
\sum_{\alpha=0}^{n}
\delta^{(d)}
\big(
 {\bf c}_{i}^{\alpha}(s)
-{\bf c}_{i^{\prime}}^{\alpha}(s^{\prime})
\big)
        \Big)
\hfill
\atop{\DPS
\hfill\quad
\times
\exp
        \Big(
\frac{\mu^{2}V}{2N}
\sum_{i,i^{\prime}=1}^{N}
\int\nolimits_{0}^{1}ds
\int\nolimits_{0}^{1}ds^{\prime}\,
\prod_{\alpha=0}^{n}
\delta^{(d)}
\big(
 {\bf c}_{i}^{\alpha}(s)
-{\bf c}_{i^{\prime}}^{\alpha}(s^{\prime})
\big)
        \Big)
}}	
\hfill\atop{\DPS\hfill\times
\exp
       \Big(
-\sum\nolimits_{\hat{k}}
U_{\hat{k}}
{\FTDen}_{\hat{k}}
        \Big)
\Big\rangle_{n+1}^{\rm W}
}}
}}.
\label{EQ:giant}
\eeq
The superscript ${\rm P}$ indicates that the average is taken over a
pure system; the subscript $n+1$ indicates that the system comprises
$n+1$ coupled replicas.  In fact we shall also find ourselves making
use of the zero-potential expectation value
$\langle\cdots\rangle_{n+1,U}^{\rm P}\big\vert_{U=0}$, 
for which we introduce the special notation
$\big\langle\cdots\big\rangle_{n+1}^{\rm P}$. 

The definitions given in the present subsection give the basic
mathematical objects that we shall manipulate in our formulation. As
was anticipated above, the quenched randomness associated with the
random crosslinks does not appear explicitly in these definitions, but
instead a term proportional to $\mu^{2}$ appears, which introduces a
coupling between the replicas.  In contrast with the case of
conventional spin glass models \cite{REF:MPVbook}, the replica
interaction couples all the replicas simultaneously, rather than
pairwise.  This feature is responsible for the occurrence of an order
parameter involving a product over all the replicas, rather than pairs
of replicas.  For a similar feature in the context of dilute 
spin glasses, see Ref.~\cite{REF:Viana}
\subsection{Application of the replica-Helmholtz free energy 
	    to the free energy $\tilde{f}$}\label{SEC:ReThPhiFE}
We now show how knowledge of $\fgenf(\mu^{2},\{0\})$ allows us 
to obtain the disorder-averaged free energy per macromolecule $d\tilde{f}$, 
defined in \eqref{EQ:logAverage}. To see this, consider the quotient 
\beq
\exp\big({-ndN\fgenf(\mu^{2},\{0\})}\big)
\Big/
\exp\big({-ndN\fgenf(0,\{0\})}\big).
\label{EQ:FirstQuo}
\eeq
By using \eqref{EQ:newPartition} we see that this quotient is given by
\begin{equation}
{
\exp\left({-ndN\fgenf\left(\mu^{2},\{0\}\right)}\right)
\over{
\exp\left({-ndN\fgenf\left(0,\{0\}\right)}\right)
}}
=
{\DPS           
\Big\langle
\exp
        \Big(
\frac{\mu^{2}V}{2N}
\sum_{i,i^{\prime}=1}^{N}
\int\nolimits_{0}^{1}ds
\int\nolimits_{0}^{1}ds^{\prime}\,
\prod\limits_{\alpha=0}^{n}
\delta^{(d)}
\big(
 {\bf c}_{i}^{\alpha}(s)
-{\bf c}_{i^{\prime}}^{\alpha}(s^{\prime})
\big)
        \Big)
\Big\rangle_{n+1}^{\rm E}
\over{\DPS
\Big\langle
\exp
        \Big(
\frac{\mu^{2}V}{2N}
\sum\limits_{i,i^{\prime}=1}^{N}
\int\nolimits_{0}^{1}ds
\int\nolimits_{0}^{1}ds^{\prime}\,
\delta^{(d)}
\big(
 {\bf c}_{i}(s)
-{\bf c}_{i^{\prime}}(s^{\prime})
\big)
        \Big)
\Big\rangle_{1}^{\rm E}
}}.     
\label{EQ:TwoQuo}
\end{equation}
Here, we have transformed the expectation values from ones weighted by
the replicated Wiener measure, \eqref{EQ:WienerMeasureReplica}, to ones
weighted by the replicated Edwards measure, which is proportional to
$\exp\big(-H_{n+1}^{\rm E}\big)$, in which 
\begin{equation}
H_{n+1}^{\rm E}
\equiv
\frac{1}{2}
\sumaln
\sum_{i=1}^{N}
\int\nolimits_{0}^{1}ds
\Big\lefver
  {d\over{ds}}{\bf c}_{i}^{\alpha}(s)
\Big\rigver^{2}
+\frac{\lambda^{2}}{2}
\sum_{\alpha=0}^{n}
\sum_{i,i^{\prime}=1}^{N}
\int\nolimits_{0}^{1}ds
\int\nolimits_{0}^{1}ds^{\prime}\,
\delta^{(d)}
\big(
 {\bf c}_{i}^{\alpha}(s)
-{\bf c}_{i^{\prime}}^{\alpha}(s^{\prime})
\big).
\label{EQ:edwards_measure_replica}
\end{equation} 
We indicate such expectation values by $\langle\cdots\rangle_{n+1}^{\rm
E}$, the subscript $n+1$ indicating the presence of replicas.  Next, we
replace the exponential function in the numerator of \eqref{EQ:TwoQuo}
by its power series expansion, thus obtaining
\bea
&&
{
\exp\left({-ndN\fgenf\left(\mu^{2},\{0\}\right)}\right)
\over{
\exp\left({-ndN\fgenf\left(0,\{0\}\right)}\right)
}}
\nonumber
\\
\noalign{\medskip}
&&
\qquad\qquad
=
{\DPS           
\Big\langle
\sum_{M=0}^{\infty}
{1\over{M!}}
\left({\mu^{2}V\over{2N}}\right)^{M}
\!\!
        \Big(
\sum\limits_{i,i^{\prime}=1}^{N}
\int\nolimits_{0}^{1}ds
\int\nolimits_{0}^{1}ds^{\prime}\,
\prod\limits_{\alpha=0}^{n}
\delta^{(d)}
\big(
 {\bf c}_{i}^{\alpha}(s)
-{\bf c}_{i^{\prime}}^{\alpha}(s^{\prime})
\big)
        \Big)^{M}
\Big\rangle_{n+1}^{\rm E}
\over{\DPS
\Big\langle
\exp
        \Big(
\frac{\mu^{2}V}{2N}
\sum\limits_{i,i^{\prime}=1}^{N}
\int\nolimits_{0}^{1}ds
\int\nolimits_{0}^{1}ds^{\prime}\,
\delta^{(d)}
\big(
 {\bf c}_{i}(s)
-{\bf c}_{i^{\prime}}(s^{\prime})
\big)
        \Big)
\Big\rangle_{1}^{\rm E}
}}.     
\eea
By expanding the $M^{\rm th}$ power as a multiple summation and integral 
over products we obtain
\begin{equation}
{
\exp\left({-ndN\fgenf\left(\mu^{2},\{0\}\right)}\right)
\over{
\exp\left({-ndN\fgenf\left(0,\{0\}\right)}\right)
}}
=
{\DPS           
{\DPS           
\Big\langle
\sum_{M=0}^{\infty}
{1\over{M!}}
\left({\mu^{2}V\over{2N}}\right)^{M}
\sum\limits_{i_{1},i_{1}^{\prime}=1}^{N}
\cdots
\sum\limits_{i_{M},i_{M}^{\prime}=1}^{N}
\int\nolimits_{0}^{1}ds_{1}
\int\nolimits_{0}^{1}ds_{1}^{\prime}
\cdots
\hfill\atop{\DPS\hfill\qquad
\cdots
\int\nolimits_{0}^{1}ds_{M}
\int\nolimits_{0}^{1}ds_{M}^{\prime}
\prod\limits_{\alpha=0}^{n}
\prod\limits_{e=1}^{M}
\delta^{(d)}
\big(
 {\bf c}_{i_{e}}^{\alpha}(s_{e})
-{\bf c}_{i_{e}^{\prime}}^{\alpha}(s_{e}^{\prime})
\big)
\Big\rangle_{n+1}^{\rm E}
}}
\over{\DPS
\Big\langle
\exp
        \Big(
\frac{\mu^{2}V}{2N}
\sum\limits_{i,i^{\prime}=1}^{N}
\int\nolimits_{0}^{1}ds
\int\nolimits_{0}^{1}ds^{\prime}\,
\delta^{(d)}
\big(
 {\bf c}_{i}(s)
-{\bf c}_{i^{\prime}}(s^{\prime})
\big)
        \Big)
\Big\rangle_{1}^{\rm E}
}}.     
\end{equation}
We now recognise that the expectation value in the numerator factorises
on the replica index to give $n+1$ factors of the partition function
$\tilde{Z}\big(\{i_{e},s_{e};i_{e}^{\prime},s_{e}^{\prime}\}_{e=1}^{M}\big)$,
\eqref{EQ:partition}.  One of these factors, when taken together with
the denominator and remaining terms (\ie, factors other than $n$
partition functions), reconstructs the Deam-Edwards probability
distribution \eqref{EQ:distribute}.  Thus, by using
\eqref{EQ:AnyAverage} we obtain
\bml\bea
{
\exp\left({-ndN\fgenf\left(\mu^{2},\{0\}\right)}\right)
\over{
\exp\left({-ndN\fgenf\left(0,\{0\}\right)}\right)
}}
&=&
\sum_{M=0}^{\infty}
\sum\limits_{i_{1},i_{1}^{\prime}=1}^{N}
\cdots
\sum\limits_{i_{M},i_{M}^{\prime}=1}^{N}
\int\nolimits_{0}^{1}ds_{1}
\int\nolimits_{0}^{1}ds_{1}^{\prime}
\cdots
\int\nolimits_{0}^{1}ds_{M}
\int\nolimits_{0}^{1}ds_{M}^{\prime}
\nonumber
\\
&&
\qquad\qquad
{\cal P}_{M}
\big(
        \{i_{e},         s_{e};
       i_{e}^{\prime},s_{e}^{\prime}
        \}_{e=1}^{M}
\big)
\,
\tilde{Z}
\big(
        \{i_{e},         s_{e};
       i_{e}^{\prime},s_{e}^{\prime}
        \}_{e=1}^{M}
\big)^{n},
\\
\noalign{\medskip}
&=&
\left[
\tilde{Z}
\big(
        \{i_{e},         s_{e};
       i_{e}^{\prime},s_{e}^{\prime}
        \}_{e=1}^{M}
\big)^{n}
\right].
\eea
\eml
By taking the logarithm of both sides, and using the expansions
\bml
\bea
z^{n}
=
{\rm e}^{n\ln z}
&=&
1+n\ln z+{\cal O}\big(n^{2}\big),
\label{EQ:limitA}
\\
\ln(1+nz)
&=&
nz+{\cal O}\big(n^{2}\big),
\label{EQ:limitB}
\eea
\eml
valid for small $n$, we arrive at the relationship between 
the replica-Helmholtz free energy $\fgenf(\mu^{2},\{U\})$ 
and the disorder-averaged free energy per macromolecule $d\tilde{f}$ 
of \eqref{EQ:logAverage}:
\beq
\lim_{n\to 0}
\left(
\fgenf(\mu^{2},\{0\})-\fgenf(0,\{0\})
\right)
=\tilde{f}.
\label{EQ:ReplLim}
\eeq
\subsection{Application of the replica-Helmholtz free energy 
	to disorder-averaged observables}\label{SEC:compAv}
In the previous subsection we discussed the application of the
replica technique to the computation of the disorder-averaged free energy.
We now turn to the issue of the computation of disorder-averages of
equilibrium expectation values of physical observables as well as of
disorder-averages of sums of products of equilibrium expectation values
of physical observables.  As we have seen in Sec.~\ref{SEC:OPFields}, 
the latter type of quantity arises in the context of the order parameter 
for equilibrium amorphous solidification.  This computation is 
accomplished by using the connection between the replica-Helmholtz free 
energy $\fgenf(\mu^{2},\{U\})$, \eqref{EQ:newPartition}, 
and expectation values of 
${\FTDen}_{\hat{k}}$, \eqref{EQ:density}:
\beq
{\delta\over{\delta U_{\hat{k}}}}
ndN\fgenf(\mu^{2},\{U\})
=
\big\langle {\FTDen}_{\hat{k}}\big\rangle_{n+1,U}^{\rm P}
\label{EQ:FirstOrder}
\eeq
More generally, by (functional) differentiation of
$\fgenf(\mu^{2},\{U\})$ with respect to $U$ we can obtain
average values for powers of the order parameter:
\bea
&&
ndN\fgenf(\mu^{2},\{U\})
\nonumber
\\
&&
\qquad\qquad
=
ndN\fgenf(\mu^{2},\{0\})
+\sum_{r=1}^{\infty}
\frac{(-1)^{r+1}}{r!}
{\sum_{{\hat{k}}_{1}}}
\cdots
{\sum_{{\hat{k}}_{r}}}\,
U_{\hat{k}_1}
\cdots 
U_{\hat{k}_r}\,
\big\langle
{\FTDen}_{\hat{k}_1}
\cdots
{\FTDen}_{\hat{k}_r} 
\big\rangle_{n+1,{\rm c}}^{\rm P}.
\label{EQ:fSourcesQ}
\eea 
The subscript ${\rm c}$ indicates that the correlators are 
connected ones \cite{REF:Amit,REF:Zinn}.

Now, Eqs.~(\ref{EQ:FirstOrder}) and (\ref{EQ:fSourcesQ}) exhibit
relationships between quantities defined in the framework of pure
systems of replicated degrees of freedom.  On the other hand, as
discussed in Sec.~\ref{SEC:OPFields}, the state of the system is
appropriately diagnosed in terms of an order parameter built from the
disorder-average of products of expectation values taken in the
(unreplicated) physical system.  The connection between the former
(replica) quantities and the later (nonreplica) quantities is contained
in the following formula, established in App.~\ref{APP:RepAndObs},
which is valid in replica-symmetric states \cite{REF:MPVbook}:
\beq
[\langle{\cal O}_{0}\rangle_{\disfac}
\langle{\cal O}_{1}\rangle_{\disfac}\ldots\langle{\cal O}_{g}\rangle_{\disfac}]
=
\lim_{n\to 0}
\Big\langle
{\cal O}_{0}(\{{\bf c}_{i}^{0}(s)\})
\cdots
{\cal O}_{g}(\{{\bf c}_{i}^{g}(s)\})
\Big\rangle_{n+1}^{\rm P}.
\label{EQ:verysource}
\eeq
The most
important application of this connection is the case of the order
parameter. For this case, as we also show in App.~\ref{APP:RepAndObs},
\eqref{EQ:verysource} reduces to:
\bea
&&
\lim_{n\to 0}
{\delta\over{\delta U_{\hat{l}}}}
ndN\fgenf(\mu^{2},\{U\})\bigg\vert_{U=0}
=
\lim_{n\to 0}
\big\langle
{\FTDen}_{\hat{l}}
\big\rangle_{n+1}^{\rm P}
\nn
&&
\quad
=
\left[
\frac{1}{N}\sumin\int_{0}^{1}ds\,
\big\langle\exp\big(i{\bf k}^{0}\cdot{\bf c}_{i}(s)\big)\big\rangle_{\disfac}
\big\langle\exp\big(i{\bf k}^{1}\cdot{\bf c}_{i}(s)\big)\big\rangle_{\disfac}
\cdots
\big\langle\exp\big(i{\bf k}^{g}\cdot{\bf c}_{i}(s)\big)\big\rangle_{\disfac}
\right], 
\label{EQ:OPsource}
\eea
where $\hat{l}=\{
{\bf k}^{0},{\bf k}^{1},\ldots,
{\bf k}^{g},{\bf 0},    \ldots,{\bf 0}\}$, 
and $\{{\bf k}^{0},{\bf k}^{1},\ldots,{\bf k}^{g}\}$ are kept fixed as
the limit $n\rightarrow 0$ is taken \cite{REF:invariant}.  Similar
replica-limit expressions for the disorder-average of functions of the
quantity~(\ref{EQ:OPdefNDA}) can also be obtained.
\section{Field-theoretic representation}\label{SEC:HSdcrpf}
The purpose of the present section is two-fold. First, in
Secs.~\ref{SEC:FouRepInt} and \ref{SEC:ApplFourRep} we transform our
representation of the physical problem from one expressed in terms of
the semi-microscopic replicated macromolecular coordinates to one
expressed in terms of generalised monomer-densities.  These densities
are closely related to the order parameter that, as we have seen,
diagnoses the various physical states of the system.  Then, in
Secs.~\ref{SEC:HSDSatZeroP} and \ref{SEC:HSDSatNonzeroP}, we obtain a
field-theoretic representation by applying a sequence of
Hubbard-Stratonovich decoupling transformations
\cite{REF:prl1987,REF:pra1989} to the replica-Helmholtz free
energy~\eqref{EQ:symmPartition} This type of decoupling strategy was
first used in the context of crosslinked macromolecular systems by Ball
and Edwards \cite{REF:RCBallPaper,REF:RCBallThesis}; see also
\cite{REF:PanNew}.

The motivation for making these transformations is that they will
provide us with a suitable starting point for developing a mean-field
description of the transition to and properties of the amorphous solid
state.  Furthermore, they provide a starting point for the analysis of
fluctuations and for the investigation of the elastic properties of the
amorphous solid state (see Ref.~\cite{REF:CGZforth}).
\subsection{Fourier representation of interactions}\label{SEC:FouRepInt}
At this stage it is useful to introduce the following Fourier
representations of the Dirac $\delta$-functions:
\bml
\bea
\delta^{(d)}({\bf c})
&=&
{1\over{V}}
\sum_{\bf p}
\exp\left(i{\bf p}\cdot{\bf c}\right),
\label{EQ:delta_unrep}
\\
\kdelhatNS\big(\hat{c}\big)\equiv
\prod_{\alpha=0}^{n}
\delta^{(d)}({\bf c}^{\alpha})
&=&
{1\over{V^{n+1}}}
\sum_{{\bf p}^{0},\dots,{\bf p}^{n}}
\exp
\Big(
i\sum_{\alpha=0}^{n}{\bf p}^{\alpha}\cdot{\bf c}^{\alpha}
\Big)
\equiv
{1\over{V^{n+1}}}
\sum_{{\hat p}}
\exp
\Big(
i\hat{p}\cdot\hat{c}
\Big), 
\label{EQ:delta_rep}
\eea
\eml where bold-face wave vectors (such as ${\bf p}^{\alpha}$) are
$d$-component vectors having components taking on the usual values
associated with periodic boundary conditions, \ie, all positive and
negative integral multiples of $2\pi/V^{1/d}$ (because $V^{1/d}$ is the
length of each side of the $d$-dimensional cubic container of volume
$V$) \cite{REF:TDlimit}.

We now introduce a particularly convenient decomposition of the terms in
a summation over a replicated wave vector $\hat{p}$, such as that
appearing in \eqref{EQ:delta_rep}.  Consider a generic replicated vector
$\hat{p}\equiv
\{{\bf p}^{0},{\bf p}^{1},\ldots,{\bf p}^{n}\}$.  
Of the $n+1$ component $d$-vectors, establish the number $h$ that are
nonzero $d$-vectors.  Then we say that the replicated vector $\hat{p}$
resides in the $h$-replica sector.  For example, if
$\hat{p}\equiv
\{{\bf 0},{\bf 0},{\bf q}^{2},{\bf 0},
{\bf q}^{4},{\bf 0},{\bf 0},\ldots,{\bf 0}\}$ 
with ${\bf q}^{2}$ and ${\bf q}^{4}$ both nonzero $d$-vectors then
$h=2$, and we say that $\hat{p}$ resides in the $2$-replica sector.  The
decomposition that we are introducing amounts to separating from  the
summation over $\hat{p}$ the term in the $0$-replica sector (\ie, the
term corresponding to
$\hat{p}=\hat{0}\equiv
\{{\bf 0},{\bf 0},\ldots,{\bf 0}\}$), 
and also separating the terms in the $1$-replica sector (\ie, terms
corresponding to those values of $\hat{p}$ in which exactly one
$d$-vector is nonzero).  Thus we shall decompose summations over
$\hat{p}$ into contributions from:
(i)~the $0$-replica sector; 
(ii)~the $1$-replica sector; and 
(iii)~the remainder, which we refer to as the higher-replica 
sector, and which contains the $h$-replica sectors for $2\le h\le n+1$.
Schematically, the decomposition can be expressed in the following way:
\beq
\sum_{\hat{p}}{\cal Q}_{\hat{p}}
\equiv
{\cal Q}_{\hat{0}}+
\sumwrsNP{\alpha}{k}{\cal Q}_{\bf k}^{\alpha}+
\sumhrsNP{k}{\cal Q}_{\hat{k}}
\label{EQ:RepSecDec}
\eeq
where $\sum\nolimits_{\bf k}^{\prime}$ denotes a summation over all
values of the $d$-vector ${\bf k}$ except that the ${\bf k}={\bf 0}$
term is omitted (\ie, it comprises terms in the $1$-replica sector),
${\cal Q}_{\bf k}^{\alpha}$ denotes the value of ${\cal Q}_{\hat{k}}$
when ${\hat{k}}$ is in the $1$-replica sector (\ie, the 
$\alpha^{\rm th}$ $d$-vector entry in ${\hat{k}}$ is nonzero, all other
entries being zero), and ${\overline{\sum}}_{\hat{k}}$ denotes a
summation over replicated vectors $\hat{k}$ residing in the higher
replica sector.  It will turn out to be useful for us to consider $V$ to
be large but (at least initially) finite, in which case it is
straightforward to implement the replica-sector decomposition,
\eqref{EQ:RepSecDec}.

With a view to subsequent decoupling transformations, it is useful to
use the Fourier representations of the Dirac $\delta$-functions and the
replica-sector decomposition in order to re-express the Dirac
$\delta$-function interactions, \ie, the non-Wiener measure terms that
couple the replicated degrees of freedom.  Thus, we see that the
interaction terms in \eqref{EQ:newPartition} can be written as
\bml
\begin{eqnarray}
&&
\sum_{i,i^{\prime}=1}^{N}
\int\nolimits_{0}^{1}ds
\int\nolimits_{0}^{1}ds^{\prime}\,
\sum_{\alpha=0}^{n}
\delta^{(d)}
\big(
 {\bf c}_{i}^{\alpha}(s)
-{\bf c}_{i^{\prime}}^{\alpha}(s^{\prime})
\big)
\nn
&&
\qquad\qquad
=
\frac{N^{2}}{V}
\sum_{\alpha=0}^{n}
\sum_{\bf k}
\Big\lefver
\frac{1}{N}
\sum_{i=1}^{N}
\int\nolimits_{0}^{1}ds
\exp
\left(i{\bf k}\cdot{\bf c}_{i}^{\alpha}(s)\right)
\Big\rigver^{2}
\label{EQ:one_lower_int}
\\
&&
\qquad\qquad
=
\frac{N^{2}}{V}(n+1)
+
\frac{N^{2}}{V}
\sumwrsNP{\alpha}{k}
\Big\lefver
\frac{1}{N}
\sum_{i=1}^{N}
\int\nolimits_{0}^{1}ds
\exp
\left(i{\bf k}\cdot{\bf c}_{i}^{\alpha}(s)\right)
\Big\rigver^{2}
\label{EQ:two_lower_int}
\\
&&
\qquad\qquad
=
\frac{N^{2}}{V}(n+1)
+
\frac{2N^{2}}{V}
\sumwrsHP{\alpha}{k}
\Big\lefver
\frac{1}{N}
\sum_{i=1}^{N}
\int\nolimits_{0}^{1}ds
\exp
\left(i{\bf k}\cdot{\bf c}_{i}^{\alpha}(s)\right)
\Big\rigver^{2}
\label{EQ:three_lower_int}
\\
&&
\qquad\qquad
=
\frac{N^{2}}{V}(n+1)
+
\frac{2N^{2}}{V}
\sumwrsHP{\alpha}{k}
\Big\lefver
{\FTDen}_{\bf k}^{\alpha}
\Big\rigver^{2};
\label{EQ:four_lower_int}
\\
&&
\sum_{i,i^{\prime}=1}^{N}
\int\nolimits_{0}^{1}ds
\int\nolimits_{0}^{1}ds^{\prime}\,
\prod_{\alpha=0}^{n}
\delta^{(d)}
\big(
 {\bf c}_{i}^{\alpha}(s)
-{\bf c}_{i^{\prime}}^{\alpha}(s^{\prime})
\big)
\nn
&&
\qquad\qquad
=
\frac{N^{2}}{V^{n+1}}
\sum_{{\hat k}}
\Big\lefver
\frac{1}{N}
\sum_{i=1}^{N}
\int\nolimits_{0}^{1}ds
\exp
\big(
i\hat{k}\cdot{\hat{c}}_{i}(s)
\big)
\Big\rigver^{2}
\label{EQ:one_upper_int}
\\
&&
\qquad\qquad
=
\frac{N^{2}}{V^{n+1}}
+
\frac{N^{2}}{V^{n+1}}
\sumwrsNP{\alpha}{k}
\Big\lefver
\frac{1}{N}
\sum_{i=1}^{N}
\int\nolimits_{0}^{1}ds
\exp
\left(i{\bf k}\cdot{\bf c}_{i}^{\alpha}(s)\right)
\Big\rigver^{2}
\nn
&&
\qquad\qquad\qquad\qquad\qquad\qquad
+
\frac{N^{2}}{V^{n+1}}
\sumhrsNP{k}
\Big\lefver
\frac{1}{N}
\sum_{i=1}^{N}
\int\nolimits_{0}^{1}ds
\exp
\big(
i\hat{k}\cdot{\hat{c}}_{i}(s)
\big)
\Big\rigver^{2}
\label{EQ:two_upper_int}
\\
&&
\qquad\qquad
=
\frac{N^{2}}{V^{n+1}}
+
\frac{2N^{2}}{V^{n+1}}
\sumwrsHP{\alpha}{k}
\Big\lefver
\frac{1}{N}
\sum_{i=1}^{N}
\int\nolimits_{0}^{1}ds
\exp
\left(i{\bf k}\cdot{\bf c}_{i}^{\alpha}(s)\right)
\Big\rigver^{2}
\nn
&&
\qquad\qquad\qquad\qquad\qquad\qquad
+
\frac{2N^{2}}{V^{n+1}}
\sumhrsHP{k}
\Big\lefver
\frac{1}{N}
\sum_{i=1}^{N}
\int\nolimits_{0}^{1}ds
\exp
\big(
i\hat{k}\cdot{\hat{c}}_{i}(s)
\big)
\Big\rigver^{2}
\label{EQ:three_upper_int}
\\
&&
\qquad\qquad
=
\frac{N^{2}}{V^{n+1}}
+
\frac{2N^{2}}{V^{n+1}}
\sumwrsHP{\alpha}{k}
\Big\lefver
{\FTDen}_{\bf k}^{\alpha}
\Big\rigver^{2}
+
\frac{2N^{2}}{V^{n+1}}
\sumhrsHP{k}
\Big\lefver
{\FTDen}_{\hat{k}}
\Big\rigver^{2}.
\label{EQ:four_upper_int}
\end{eqnarray}
\eml 
For each of the left hand sides we have performed four steps.  In the
first step we have used Eqs.~(\ref{EQ:delta_unrep}) and
(\ref{EQ:delta_rep}) to re-express the Dirac $\delta$-functions.  In the
second step we have performed the replica-sector decomposition,
according to \eqref{EQ:RepSecDec}.  In the third step, we have
recognised that the summands in the summations over wave vectors of the
second step are even functions of the relevant wave vector.
Furthermore, none of the summations includes a zero wave vector.  Thus,
in each case the summation can be restricted to half of the relevant
wave vector space, provided a factor of two is included to compensate.
To represent this, we have introduce the notation
$\sum\nolimits_{\bf k}^{\possym}$ to denote
$\sum\nolimits_{\bf k}^{\prime}$
but with ${\bf k}$ restricted to the half space via the additional
condition ${\bf k}\cdot{\bf n}>0$ for a suitable unit $d$-vector 
${\bf n}$, and 
${\overline{\sum}}_{\hat{k}}^{\possym}$ to denote
${\overline{\sum}}_{\hat{k}}$ 
but with $\hat{k}$ restricted to the half space via the additional
condition $\hat{k}\cdot\hat{n}>0$ for a suitable unit $(n+1)d$-vector
$\hat{n}$.  The virtue of this procedure is that in our subsequent
development it will enable us to avoid the introduction of kinematically
non-independent fields. In the fourth and final step we have used the
definition of ${\FTDen}_{\hat{k}}$ given in \eqref{EQ:density}, which in the
context of the replica-sector decomposition becomes
\bml
\bea
{\FTDen}_{\bf k}^{\alpha}
\equiv&
{\DPS
\frac{1}{N}\sum_{i=1}^{N}
\int_{0}^{1}ds\,
\exp\Big(
i{\bf k}\cdot{\bf c}_{i}^{\alpha}(s)\Big),
}
&\quad\mbox{(1-replica sector)},
\label{EQ:NEWdensityWRS}
\\
{\FTDen}_{\hat{k}}
\equiv&
{\DPS
\frac{1}{N}\sum_{i=1}^{N}
\int_{0}^{1}ds\,
\exp\Big(
i\hat{k}\cdot\hat{c}_{i}(s)\Big),
}
&\quad\mbox{(higher-replica sector)}.
\label{EQ:NEWdensityHRS}
\eea
\eml
\subsection{Applications of the Fourier representation}\label{SEC:ApplFourRep}
By applying the results of Sec.~\ref{SEC:FouRepInt} to 
\eqref{EQ:newPartition} we obtain
\bea
&&
\exp\big(-ndN\fgenf(\mu^{2},\{U\})\big)
=
\nn
\noalign{\medskip}
&&
\qquad\qquad\qquad
{\DPS
{\cal B}_{n}\,
\Big\langle
\exp
        \Big(
-\frac{\tlns N^{2}}{V}
\sumwrsHP{\alpha}{k}
\Big\lefver
	{\FTDen}_{\bf k}^{\alpha}
\Big\rigver^{2}
+\frac{\mu^{2}N}{V^{n}}
{\overline{\sum\limits_{\hat{k}}}}^{\possym}
\Big\lefver
	{\FTDen}_{\hat{k}}
\Big\rigver^{2}
-\sum_{\hat{k}}
U_{\hat{k}}
	\,{\FTDen}_{\hat{k}}
        \Big)
\Big\rangle_{n+1}^{\rm W}
\over{\DPS
\Big\langle
\exp
        \Big(
-\frac{\tlzs N^{2}}{V}
{\sum_{\bf k}}^{\possym}
\Big\lefver
	{\FTDen}_{\bf k}^{0}
\Big\rigver^{2}
        \Big)
\Big\rangle_{1}^{\rm W}
}},
\label{EQ:symmPartition}
\eea
where we have introduced the effective excluded-volume parameter 
\beq
\tlns\equiv\lambda^{2}-\frac{\mu^{2}V}{NV^{n}},
\label{EQ:renln}
\eeq
\ie, the bare excluded-volume parameter $\lambda^{2}$ renormalised to a
smaller value by a correction term proportional to the crosslink
density parameter $\mu^{2}$.  The prefactor ${\cal B}_{n}$ is an
unimportant constant, which arises from terms in the $0$-replica
sector, and is given by
\beq
{\cal B}_{n}
\equiv
\exp\Big(\frac{\mu^{2}N}{2V^{n}}
        -\frac{\lambda^{2}N^{2}n}{2V}\Big)
\Big/
\exp\Big(\frac{\mu^{2}N}{2}\Big).
\label{EQ:DullConstant}
\eeq
We see from Eqs.~(\ref{EQ:symmPartition}) and (\ref{EQ:renln}) that in
the 1-replica sector there is a competition between the excluded-volume
interaction $\lambda^{2}$ and the effect of the crosslinking,
represented by $\mu^{2}$. If $\tlns$ is positive then configurations
having nonzero ${\FTDen}_{\bf k}^{\alpha}$ are disfavoured (and therefore MTI
is favoured), whilst if $\tlns$ is negative then configurations having
nonzero ${\FTDen}_{\bf k}^{\alpha}$ are favoured (and therefore there is a
tendency to violate MTI).  As we are primarily concerned with
investigating the liquid and the amorphous solid states, both of which
are MTI, we focus our attention on the regime $\tlns>0$. On the other
hand, the (higher-replica sector) component of the term due to
crosslinking increases the statistical weight of configurations in which
${\FTDen}_{\hat{k}}\ne 0$ (for $\hat{k}$ in the higher-replica sector).
As we shall see in Sec.~\ref{SEC:InstFlPH}, the coefficient of this
term, $\mu^{2}$, is the control parameter governing the transition from
the liquid state (for small $\mu^{2}$) to the amorphous solid state
(for large $\mu^{2}$), as characterised by the values of the order
parameters shown in Table~\ref{TAB:phases}.
\subsection{Hubbard-Stratonovich decoupling scheme: replica-Helmholtz 
	free energy at zero external potential}\label{SEC:HSDSatZeroP}
We now obtain a field-theoretic representation by applying a sequence of
Hubbard-Stratonovich decoupling transformations to the replica-Helmholtz
free energy, \eqref{EQ:symmPartition}.  We focus on the case $U=0$ in
the present subsection, and in the following subsection present results
that allow for nonzero $U$.  The motivation for these transformations is
that via them all interactions between different macromolecules are
eliminated at the expense of introducing a certain additional stochastic
field to which the monomers are coupled.  This strategy has the
following virtues.  First, the task of summing over the configurations
of the system of $N$ replicated macromolecules is reduced to the task of
summing over the configurations of a {\it single\/} replicated macromolecule,
albeit one that is coupled to the stochastic field.  The monomers that
constitute this replicated macromolecule remain coupled to each other
via the Wiener measure, \eqref{EQ:WienerMeasure}, and by the stochastic
field to which they are coupled.  To the extent that this summation can be
performed, \eqref{EQ:onePartition} gives the replica-Helmholtz free
energy in terms of functional integrals over the stochastic fields. 
Second, the stochastic field itself has a natural physical interpretation: 
as we shall see explicitly in Sec.~\ref{SEC:HSDSatNonzeroP} it is 
related in a direct manner to the order parameter.

The appropriate Hubbard-Stratonovich decoupling transformations are predicated 
on the multiple use of the following pair of integrals:
\bml
\bea
\exp
        \Big(
-a\big\lefver{w}\big\rigver^{2}
        \Big)
&=&
\left(a/\pi\right)
\int
d\real z\,\,
d\imag z
\exp
        \Big(
-a\big\lefver{z}\big\rigver^{2}
        \Big)
\exp
        \Big(
2ia\,\real zw^{\ast}
        \Big),
\label{EQ:HuStExVo}
\\
\exp
        \Big(
+a\big\lefver{w}\big\rigver^{2}
        \Big)
&=&
\left(a/\pi\right)
\int
d\real z\,\,d\imag z
\exp
        \Big(
-a\big\lefver{z}\big\rigver^{2}
        \Big)
\exp
        \Big(
2a\,\real zw^{\ast}
        \Big)
\label{EQ:HuStCrLi},
\eea
\eml where $w$ is an arbitrary complex number, $a$ is a real and
positive (but otherwise arbitrary) number, and the integrals are taken
over the entire complex $z$ plane. We transform each exponential term
inside an expectation value in \eqref{EQ:symmPartition} (with $U=0$) by
using these integrals, those having the coefficient $\tlns$ or $\tlzs$
with \eqref{EQ:HuStExVo} and  those having the coefficient $\mu^{2}$
with \eqref{EQ:HuStCrLi}.

We define the measures $\dmwrs$, $\dmhrs$ and $\dmzrs$ via 
\bml
\bea 
\dmwrs\Omega
&\equiv&
 \prod_{\alpha=0}^{n}
{\prod_{{\bf k}}}^{\possym}
\frac{\tlns N^{2}}{\pi V}
d\real\Omega_{{\bf k}}^{\alpha}\,\,
d\imag\Omega_{{\bf k}}^{\alpha},
\label{EQ:measure_wrs}
\\
\dmhrs\Omega
&\equiv&
\overline{\prod_{\hat{k}}}^{\possym}
\frac{\mu^{2}N}{\pi V^{n}}
d\real\Omega_{\hat{k}}\,\,
d\imag\Omega_{\hat{k}},
\label{EQ:measureHRS}
\\
\dmzrs\omega
&\equiv&
{\prod_{{\bf k}}}^{\possym}
\frac{\tlzs N^{2}}{\pi V}
d\real\omega_{{\bf k}}\,\,
d\imag\omega_{{\bf k}},
\label{EQ:measure_zrs}
\eea
\eml where $\prod\nolimits_{{\bf k}}^{\possym}$ denotes the product over
all $d$-vectors ${\bf k}$ in the half-space given by the condition 
${\bf k}\cdot{\bf n}>0$ for a suitable unit $d$-vector ${\bf n}$, and
${\overline{\prod}}^{\possym}_{\hat{k}}$ denotes the product over all
$(n+1)d$-vectors $\hat{k}$ in the half-space given by the condition
$\hat{k}\cdot\hat{n}>0$ for a suitable unit $(n+1)d$-vector $\hat{n}$.
These definitions have convenient normalisation properties \cite{REF:FieldScale}:
\bml
\bea 
\int\dmwrs\Omega
\exp
\Big(
-{\tlns N^{2}}{V^{-1}}
\sum\nolimits_{\alpha=0}^{n}
\sum\nolimits_{\bf k}^{\possym}
\big\lefver
\Omega_{{\bf k}}^{\alpha}
\big\rigver^{2}
\Big)
&=&
1,
\label{EQ:normWRS}
\\
\int\dmhrs\Omega
\exp
\Big(
-{\mu^{2}N}{V^{-n}}
{\overline{\sum}}_{\hat{k}}^{\possym}
\big\lefver
\Omega_{{\hat k}}
\big\rigver^{2}
\Big)
&=&
1,
\label{EQ:norm_hrs}
\\
\int\dmzrs\omega 
\exp
\Big(
-{\tlzs N^{2}}{V^{-1}}
{\sum}_{\bf k}^{\possym}
\big\lefver
\omega_{{\bf k}}
\big\rigver^{2}
\Big)
&=&
1.
\label{EQ:normZRS}
\eea
\eml By using this sequence of Hubbard-Stratonovich transformations
\eqref{EQ:symmPartition} becomes
\bea
&&
\exp\big(-ndN\fgenf(\mu^{2},\{0\})\big)
=
\nn
\noalign{\medskip}
&&
{\DPS   
{\DPS   
{\cal B}_{n}\,
\int\dmwrs\Omega
\exp
\Big(
-{\tlns N^{2}}{V^{-1}}
 \sumwrsHPNL{\alpha}{k}
\big\lefver
\Omega_{{\bf k}}^{\alpha}
\big\rigver^{2}
\Big)
\int\dmhrs\Omega
\exp
\Big(
-{\mu^{2}N}{V^{-n}}
\sumhrsHPNL{k}  
\big\lefver
\Omega_{{\hat k}}
\big\rigver^{2}
\Big)
\hfill
\atop{\DPS  
\quad
\hfill
\times
\Big\langle
\exp
\Big(
{2i\tlns N^{2}}{V^{-1}}
\sumwrsHPNL{\alpha}{k}
\real
\Omega_{\bf k}^{\alpha\ast}
	\,{\FTDen}_{\bf k}^{\alpha}
+{2\mu^{2}N}{V^{-n}}
\sumhrsHPNL{k}
\real
\Omega_{\hat{k}}^{\ast}
	\,{\FTDen}_{\hat{k}}
\Big)
\Big\rangle_{n+1}^{\rm W}
}}      
\over{\DPS
\int\dmzrs\omega
\exp
\Big(
-{\tlzs N^{2}}{V^{-1}}
\sumzrsHPNL{k}
\big\lefver
\omega_{{\bf k}}
\big\rigver^{2}
\Big)
\Big\langle
\exp
\Big(
{2i\tlzs N^{2}}{V^{-1}}
\sumzrsHPNL{k}
\real
\omega_{\bf k}^{\ast}
	\,{\FTDen}_{\bf k}^{0}
\Big)
\Big\rangle_{1}^{\rm W}
}}      
\label{EQ:funPartition}
\eea

By examining the expectation values $\langle\cdots\rangle_{n+1}^{\rm W}$
and $\langle\cdots\rangle_{1}^{\rm W}$ in \eqref{EQ:funPartition} we see
that indeed the sequence of Hubbard-Stratonovich transformations has led
to the decoupling of the $N$ (replicated) macromolecules from each
other. Moreover, as these expectation values are products of identical
factors, one for each replicated macromolecule, we may replace them by the
$N^{\rm th}$ power of an expectation value involving a single replicated
macromolecule. Thus, we see that \eqref{EQ:symmPartition} is given by
the quotient of partition functions of stochastic fields,
\bml
\beq
\exp\big(-ndN\fgenf(\mu^{2},\{0\})\big)
=
{\cal B}_{n}\,
{\DPS
\int\dmwrs\Omega\,
\int\dmhrs\Omega
\exp
\left(
-ndN\calfars
\big(\{\Omega_{\bf k}^{\alpha},\Omega_{\hat k}\}\big)
\right)
\over{\DPS
\int\dmzrs\omega
\exp
\left(
-ndN\calfzrs\big(\{\omega_{{\bf k}}\}\big)
\right)
}}, 
\label{EQ:onePartition}
\eeq
governed by the effective hamiltonians 
$\calfars\big(\{\Omega_{{\bf k}}^{\alpha},\Omega_{{\hat k}}\}\big)$ 
and 
$\calfzrs\big(\{\omega_{{\bf k}}\}\big)$, where
\bea
&&
nd\calfars\big(\{\Omega_{{\bf k}}^{\alpha},\Omega_{{\hat k}}\}\big)
\equiv
{\tlns N}{V^{-1}}
\sumwrsHPNL{\alpha}{k}
\big\lefver
\Omega_{{\bf k}}^{\alpha}
\big\rigver^{2}
+
{\mu^{2}}{V^{-n}}
\sumhrsHPNL{k}
\big\lefver
\Omega_{{\hat k}}
\big\rigver^{2}
\nonumber
\\
&&
\qquad\qquad\qquad
-\ln
\Big\langle
\exp
\Big(
{2i\tlns N}{V^{-1}}
\sumwrsHPNL{\alpha}{k}
\real
\Omega_{\bf k}^{\alpha\ast}
\int\nolimits_{0}^{1}ds\,
\exp
\left(
i{\bf k}\cdot{\bf c}^{\alpha}(s)
\right)
\Big)
\nonumber
\\
&&
\qquad\qquad\qquad\qquad
\qquad\qquad
\times
\exp
\Big(
{2\mu^{2}}{V^{-n}}
\sumhrsHPNL{k}
\real
\Omega_{\hat{k}}^{\ast}
\int\nolimits_{0}^{1}ds\,
\exp
\big(
i{\hat{k}}\cdot{\hat{c}}(s)
\big)
\Big)
\Big\rangle_{n+1}^{\rm W}\,,
\label{EQ:hamnumer}
\\
\noalign{\smallskip}
&&
nd\calfzrs\big(\{\omega_{{\bf k}}\}\big)
\!\equiv\!
\tlzs NV^{-1}
\sumzrsHPNL{k}
\big\lefver
\omega_{{\bf k}}
\big\rigver^{2}
\!-\!\ln\!
\Big\langle\!
\exp
\Big(
2i
\tlzs NV^{-1}
\sumzrsHPNL{k}
\real
\omega_{\bf k}^{\ast}
\int\nolimits_{0}^{1}\!\!ds\,\!
\exp
\left(i{\bf k}\cdot{\bf c}(s)
\right)
\Big)
\Big\rangle_{1}^{\rm W}.
\label{EQ:hamdenom}
\eea
\eml 
We can compute 
$\calfars\big(\{\Omega_{{\bf k}}^{\alpha},\Omega_{{\hat k}}\}\big)$ and 
$\calfzrs\big(\{\omega_{{\bf k}}\}\big)$ 
perturbatively \cite{REF:LGWform}, order by order, 
in powers of $\left\{\Omega_{{\bf k}}^{\alpha},\Omega_{{\hat k}}\right\}$ 
and $\left\{\omega_{{\bf k}}\right\}$.
This perturbative construction of the Landau-Ginzburg-Wilson effective 
hamiltonian is equivalent to that arising in many other contexts in 
statistical physics.  
\subsection{Hubbard-Stratonovich decoupling scheme: replica-Helmholtz 
	free energy at nonzero external potential}\label{SEC:HSDSatNonzeroP}
We now turn to the general case of the replica-Helmholtz free energy, in
which $U$ need not vanish. As with the $U=0$ case, it will be convenient
to perform a Hubbard-Stratonovich decoupling transformation on
$\fgenf(\mu^{2},\{U\})$. This transformation is predicated on
the multiple use of the following pair of integrals, which are
generalisations of those given in Eqs.~(\ref{EQ:HuStExVo}) and
(\ref{EQ:HuStCrLi}):
\bml
\bea
&&
\exp
        	\Big(
-a	\big(
\big\lefver{w}\big\rigver^{2}
-jw-\bar{\jmath}w^{\ast}
	\big)
        	\Big)
=
\left(a/\pi\right)
\exp\big(aj\bar{\jmath}\big)
\nonumber
\\
&&
\qquad\qquad
\times
\int
d\real z\,\,
d\imag z
\exp
        	\Big(
-a\big\lefver{z}\big\rigver^{2}
+2ia\,\real zw^{\ast}
- ia\,\big(zj+z^{\ast}\bar{\jmath}\big)
        	\Big),
\label{EQ:HSsourceA}
\\
&&
\exp
        	\Big(
 a	\big(
\big\lefver{w}\big\rigver^{2}
-jw-\bar{\jmath}w^{\ast}
	\big)
        	\Big)
=
\left(a/\pi\right)
\exp\big(-aj\bar{\jmath}\big)
\nonumber
\\
&&
\qquad\qquad
\times
\int
d\real z\,\,
d\imag z
\exp
        	\Big(
-a\big\lefver{z}\big\rigver^{2}
+2 a\,\real zw^{\ast}
-  a\,\big(zj+z^{\ast}\bar{\jmath}\big)
        	\Big),
\label{EQ:HSsourceB}
\eea
\eml where $w$, $j$ and $\bar{\jmath}$ are arbitrary complex numbers, 
$a$ is a real and positive (but otherwise arbitrary) number, 
and the integrals are taken over the entire complex $z$ plane. 
Following a strategy analogous to that used above for the $U=0$ case, 
we find
\bea
&&
ndN\fgenf(\mu^{2},\{U\})
=
-\ln{\cal B}_{n}
+\ln\int\dmzrs\omega
\exp\left(-ndN\calfzrs\big(\{\omega_{{\bf k}}\}\big)\right)
\nn
&&
\qquad\qquad\qquad
-\frac{V}{N^{2}\tlns}
\sumwrsHPNL{\alpha}{k}
U_{\bf k}^{\alpha}
U_{-{\bf k}}^{\alpha}
+\frac{V^{n}}{N\mu^{2}}
\sumhrsHPNL{k}
U_{ \hat{k}}
U_{-\hat{k}}
\nn
&&
\qquad
-\ln
\int
\dmwrs\Omega\,
\dmhrs\Omega
\exp
	\Big(
-ndN\calfars\big(\{\Omega_{\bf k}^{\alpha},\Omega_{\hat k}\}\big)
+i\sumwrsNP{\alpha}{k}
U_{\bf k}^{\alpha}
\Omega_{\bf k}^{\alpha}
-\sumhrsNP{k}
U_{\hat{k}}\,\Omega_{\hat{k}}
	\Big),
\label{EQ:FusHS}
\eea
where the effective hamiltonian $\calfars$ is given in \eqref{EQ:hamnumer}. 
Thus we have transformed our description into an effective one for the
fields $\{\Omega_{\bf k}^{\alpha},\Omega_{\hat k}\}$.  It should be
noted that the coefficients of the powers of 
$\{U_{\bf k}^{\alpha},U_{\hat k}\}$  in the functional Taylor series for
$\fgenf(\mu^{2},\{U\})$ are simply related to the connected correlators
of $\{\Omega_{\bf k}^{\alpha},\Omega_{\hat k}\}$:
\bml
\bea
&&
ndN\fgenf(\mu^{2},\{U\})=
ndN\fgenf(\mu^{2},\{0\})
-\frac{V}{2N^{2}\tlns}
\sumwrsNP{\alpha}{k}
U_{  \bf k }^{\alpha}
U_{-{\bf k}}^{\alpha}
+
\frac{V^{n}}{2N\mu^{2}} 
\overline{\sum}_{\hat{k}} 
U_{ \hat{k}}
U_{-\hat{k}}
\nn
&&
\qquad\quad
+
\sum_{{\scriptstyle r,s=0
\atop{\scriptstyle (r,s)\ne(0,0)}}}^{\infty}
\frac{i^{r}(-1)^{s+1}}{r!\,s!}
\sumwrsNP{\alpha_{1}}{k_{\mit 1}}
\cdots
\sumwrsNP{\alpha_{r}}{k_{\mit r}}\,\,
\overline{\sum_{{\hat{k}}_{1}}}
\cdots
\overline{\sum_{{\hat{k}}_{s}}}
U_{{\bf k}_1}^{\alpha_1}
\cdots
U_{{\bf k}_r}^{\alpha_r}\,
U_{\hat{k}_1}\cdots 
U_{\hat{k}_s}\,
\nn
&&
\qquad\qquad\qquad
\qquad\qquad\qquad
\qquad\qquad\qquad
\times
\Big\langle
\Omega_{{\bf k}_1}^{\alpha_1}
\cdots
\Omega_{{\bf k}_r}^{\alpha_r}
\Omega_{\hat{k}_1}
\cdots
\Omega_{\hat{k}_s} 
\Big\rangle_{n+1,{\rm c}}^{\cal F},
\label{EQ:FrVarUpot}
\eea
where the expectation value $\langle\cdots\rangle_{n+1}^{\cal F}$ 
is defined via
\beq
\Big\langle
\cdots
\Big\rangle_{n+1}^{\cal F}
\equiv
{\DPS
\int
\dmwrs\Omega\,
\dmhrs\Omega
\,\cdots\,
\exp
	\Big(
-ndN\calfars\big(\{\Omega_{\bf k}^{\alpha},\Omega_{\hat k}\}\big)
	\Big)
\over{\DPS
\int
\dmwrs\Omega\,
\dmhrs\Omega\,
\exp
	\Big(
-ndN\calfars\big(\{\Omega_{\bf k}^{\alpha},\Omega_{\hat k}\}\big)
	\Big)
}}.
\eeq
\eml

The virtue of the present development is that it allows us to construct
quantities of physical interest, which typically involve the densities
$\{{\FTDen}_{\bf k}^{\alpha},{\FTDen}_{\hat k}\}$ in terms of more 
readily computable quantities, which involve the stochastic fields 
$\{\Omega_{\bf k}^{\alpha},\Omega_{\hat k}\}$.  Indeed, by using
Eqs.~(\ref{EQ:fSourcesQ}) and (\ref{EQ:FrVarUpot}) we see that
\bml
\bea
\big\langle
{\FTDen}_{\bf k}^{\alpha}
\big\rangle_{n+1}^{\rm P}
&=&
-i\big\langle
\Omega_{\bf k}^{\alpha}
\big\rangle_{n+1}^{\cal F},
\label{EQ:ExEqWRS}
\\
\big\langle
{\FTDen}_{\hat{k}}
\big\rangle_{n+1}^{\rm P}
&=&
\phantom{-i}\big\langle
\Omega_{\hat{k}}
\big\rangle_{n+1}^{\cal F}.
\label{EQ:ExEqHRS}
\eea
\eml Thus, by using \eqref{EQ:OPsource} we see that we can 
relate the order parameter to the expectation value of the 
stochastic field $\Omega_{\hat{k}}$: 
\beq
\left[
\frac{1}{N}\sumin\int_{0}^{1}ds\,
\big\langle\exp\big(i{\bf k}^{0}\cdot{\bf c}_{i}(s)\big)\big\rangle_{\disfac}
\big\langle\exp\big(i{\bf k}^{1}\cdot{\bf c}_{i}(s)\big)\big\rangle_{\disfac}
\cdots
\langle\exp\Big(i{\bf k}^{g}\cdot{\bf c}_{i}(s)\Big)\rangle_{\disfac}
\right]
=
\lim_{n\to 0}
\big\langle
\Omega_{\hat{l}}
\big\rangle_{n+1}^{\cal F},
\label{EQ:newOPsource}
\eeq
where $\hat{l}=\{
{\bf k}^{0},{\bf k}^{1},\ldots,
{\bf k}^{g},{\bf 0},    \ldots,{\bf 0}\}$.
In the following section we make explicit use of this development 
in order to compute the order parameter.
\section{Saddle-point approximation in the critical regime}\label{SEC:StatPtCrit}
In the preceding sections we have developed an exact, formal,
field-theoretic representation of the statistical mechanics of randomly
crosslinked macromolecular networks.  In the present section we shall
explore the properties of such systems, focusing our attention on the
regime of crosslink densities near to the equilibrium phase transition
from the liquid state to the amorphous solid state that sufficient
crosslinking causes.  We shall do this by analysing the field-theoretic
representation at the level of mean-field theory, considering in detail
expressions for the free energy and the order parameter.  Following
this, in Sec.~\ref{SEC:DenSecOV}, we shall consider
the implications of a certain, physically important class of fluctuations.
In order to streamline the presentation, a considerable amount of
technical detail has been relegated to appendices. The main results of 
this section have been briefly reported in Ref.~\cite{REF:CGZjourEPL}. 
\subsection{Approximation strategy}\label{SEC:AppStrat}
The mean-field level of approximation follows from computing the
functional integral in the numerator of Eq.~(\ref{EQ:onePartition}) 
by using the saddle-point method. This amounts to replacing the
functional integral by the value of its integrand that is stationary
with respect to variations of 
$\{{\Omega}_{\bf k}^{\alpha},{\Omega}_{\hat k}\}$, so that, omitting 
unimportant constants, we obtain the following approximations
for $\fgenf(\mu^{2},\{0\})$, 
for $\langle\Omega_{\bf k}^{\alpha}\rangle_{n+1}^{\cal F}$, 
for $\langle\Omega_{\hat k}        \rangle_{n+1}^{\cal F}$ and 
for $\langle{\FTDen}\rangle_{n+1}^{\rm P}$: 
\bml
\bea
\fgenf(\mu^{2},\{0\})
&=&
\calfars\big(\{\bar{\Omega}_{\bf k}^{\alpha},\bar{\Omega}_{\hat k}\}\big),
\label{EQ:BasicMFT}
\\
\noalign{\medskip}
i
\langle {\FTDen}_{\bf k}^{\alpha}\rangle_{n+1}^{\rm P}
&=&
\langle\Omega_{\bf k}^{\alpha}\rangle_{n+1}^{\cal F}
=
\bar{\Omega}_{\bf k}^{\alpha},
\\
\noalign{\medskip}
\langle {\FTDen}_{\hat k}\rangle_{n+1}^{\rm P}
&=&
\langle\Omega_{\hat k}\rangle_{n+1}^{\cal F}
=
\bar{\Omega}_{\hat k},
\eea
\eml
where $\bar{\Omega}_{\bf k}^{\alpha}$ and $\bar{\Omega}_{\hat k}$ make 
$\calfars\big(\{\Omega_{\bf k}^{\alpha},\Omega_{\hat k}\}\big)$ 
stationary, \ie, satisfy the stationarity conditions
\bml
\bea
{\delta\calfars
\over{
\delta\Omega_{\bf q}^{\alpha\,\ast}
}}
\bigg\vert_{\{\bar{\Omega}_{\bf k}^{\alpha},\bar{\Omega}_{\hat k}\}}
&=&0,
\\
\noalign{\bigskip}
{\delta\calfars
\over{
\delta\Omega_{\hat q}^{\ast}
}}
\bigg\vert_{\{\bar{\Omega}_{\bf k}^{\alpha},\bar{\Omega}_{\hat k}\}}
&=&0.
\label{EQ:NewStat}
\eea
\eml By using \eqref{EQ:hamnumer} the stationarity conditions become
\bml
\bea
\bar{\Omega}_{\bf q}^{\alpha}
&=&
{\DPS
{\DPS
i\Big\langle
\int_{0}^{1}\!\!dt\,\!
\exp\Big(i{\bf q}\cdot{\bf c}^{\alpha}(t)\Big)\,
\exp
\Big(
{2i\tlns N}{V^{-1}}\!
\sumwrsHPNL{\alpha}{k}
\real
\bar{\Omega}_{\bf k}^{\alpha\ast}
\int\nolimits_{0}^{1}\!\!ds\,
\exp
\left(
i{\bf k}\cdot{\bf c}^{\alpha}(s)
\right)
\Big)
\hfill\atop{\DPS\qquad\hfill\times
\exp
\Big(
{2\mu^{2}}{V^{-n}}
\sumhrsHPNL{k}
\real
\bar{\Omega}_{\hat{k}}^{\ast}
\int\nolimits_{0}^{1}ds\,
\exp
\big(
i{\hat{k}}\cdot{\hat{c}}(s)
\big)
\Big)
\Big\rangle_{n+1}^{\rm W}
}}
\over{\DPS
{\DPS
\Big\langle
\exp
\Big(
{2i\tlns N}{V^{-1}}
\sumwrsHPNL{\alpha}{k}
\real
\bar{\Omega}_{\bf k}^{\alpha\ast}
\int\nolimits_{0}^{1}ds\,
\exp
\left(
i{\bf k}\cdot{\bf c}^{\alpha}(s)
\right)
\Big)
\hfill\atop{\DPS\qquad\hfill\times
\exp
\Big(
{2\mu^{2}}{V^{-n}}
\sumhrsHPNL{k}
\real
\bar{\Omega}_{\hat{k}}^{\ast}
\int\nolimits_{0}^{1}ds\,
\exp
\big(
i{\hat{k}}\cdot{\hat{c}}(s)
\big)
\Big)
\Big\rangle_{n+1}^{\rm W}
}}
}},
\label{EQ:selfconseqWRS}
\\
\bar{\Omega}_{\hat{q}}
&=&
{\DPS
{\DPS
\Big\langle
\int_{0}^{1}\!\!dt\,\!
\exp\Big(i\hat{q}\cdot\hat{c}(t)\Big)\,
\exp
\Big(
{2i\tlns N}{V^{-1}}\!
\sumwrsHPNL{\alpha}{k}
\real
\bar{\Omega}_{\bf k}^{\alpha\ast}
\int\nolimits_{0}^{1}\!\!ds\,\!
\exp
\left(
i{\bf k}\cdot{\bf c}^{\alpha}(s)
\right)
\Big)
\hfill\atop{\DPS\qquad\hfill\times
\exp
\Big(
{2\mu^{2}}{V^{-n}}
\sumhrsHPNL{k}
\real
\bar{\Omega}_{\hat{k}}^{\ast}
\int\nolimits_{0}^{1}ds\,
\exp
\big(
i{\hat{k}}\cdot{\hat{c}}(s)
\big)
\Big)
\Big\rangle_{n+1}^{\rm W}
}}
\over{\DPS
{\DPS
\Big\langle
\exp
\Big(
{2i\tlns N}{V^{-1}}
\sumwrsHPNL{\alpha}{k}
\real
\bar{\Omega}_{\bf k}^{\alpha\ast}
\int\nolimits_{0}^{1}ds\,
\exp
\left(
i{\bf k}\cdot{\bf c}^{\alpha}(s)
\right)
\Big)
\hfill\atop{\DPS\qquad\hfill\times
\exp
\Big(
{2\mu^{2}}{V^{-n}}
\sumhrsHPNL{k}
\real
\bar{\Omega}_{\hat{k}}^{\ast}
\int\nolimits_{0}^{1}ds\,
\exp
\big(
i{\hat{k}}\cdot{\hat{c}}(s)
\big)
\Big)
\Big\rangle_{n+1}^{\rm W}
}}
}}.
\label{EQ:selfconseqHRS}
\eea
\eml

\subsection{Instability of the liquid state}\label{SEC:InstFlPH}
In the context of the mean-field approximation, 
the liquid state corresponds to 
$\bar{\Omega}_{\bf k}^{\alpha}=\bar{\Omega}_{\hat{k}}=0$, 
which can readily be checked to solve Eqs.~(\ref{EQ:selfconseqWRS}) 
and (\ref{EQ:selfconseqHRS}).  
To address the stability of this state we examine
$\calfars\big(\{\Omega_{{\bf k}}^{\alpha},\Omega_{{\hat k}}\}\big)$,
Eq.~(\ref{EQ:hamnumer}), and expand perturbatively about 
$\Omega_{\bf k}^{\alpha}=\Omega_{\hat{k}}=0$ to second order in 
$\Omega_{\bf k}^{\alpha}$ and $\Omega_{\hat{k}}$.  This gives
\bea
nd\calfars\big(\{\Omega_{{\bf k}}^{\alpha},\Omega_{{\hat k}}\}\big)
&=&
{\tlns N}{V^{-1}}
\sumwrsHPNL{\alpha}{k}
\left(1+{\tlns N}{V^{-1}}
\debyeZ{k}\right)
\big\lefver
\Omega_{{\bf k}}^{\alpha}
\big\rigver^{2}
\nonumber
\\
&&
\qquad\qquad\qquad\qquad
+{\mu^{2}}{V^{-n}}
\sumhrsHPNL{k}
\left(1-
{\mu^{2}}{V^{-n}}
\debyeZbare{{\hat{k}}}
\right)
\big\lefver
\Omega_{{\hat k}}
\big\rigver^{2}
+\cdots.
\label{EQ:hampert}
\eea
The correlators necessary to calculate the terms in this expansion are
computed in App.~\ref{APP:WMC}, and the function $\debyeZ{k}$ resulting
from the subsequent arclength integrations is defined in
App.~\ref{APP:debye} and has the value

\beq
\debyeZ{k} =
{
{\rm e}^{-k^{2}/2}-\Big(1-\frac{1}{2}k^{2}\Big)
\over{
\frac{1}{2}\left(\frac{1}{2}k^{2}\right)^{2}
}}
\sim
\cases{1 - k^{2}/6,     &if $k^{2}\ll 1$;\cr
       4/k^{2}&if $k^{2}\gg 1$.\cr}
\eeq

As we anticipated at the end of Sec.~\ref{SEC:ApplFourRep}, the
stability of the 1-replica sector is controlled by the coefficient of
the $\big\lefver\Omega_{{\bf k}}^{\alpha}\big\rigver^{2}$ term in this
expansion.  This coefficient, together with the positive-definiteness of
$\debyeZ{k}$, show that provided the crosslink-renormalised
excluded-volume parameter $\tlns$, \eqref{EQ:renln}, is positive, the
1-replica sector is locally stable.   Thus, the saddle-point
value $\bar{\Omega}_{{\bf k}}^{\alpha}$ is zero.  The
positive-definiteness of $\tlns$ requires that
\beq
\lambda^{2}>\frac{\mu^{2}V}{NV^{n}},
\label{EQ:majorWRS}
\eeq
\ie, that the repulsive character of the physical excluded-volume
parameter $\lambda^{2}$ is sufficiently strong to enable the system to
withstand the effective tendency towards collapse afforded by the
crosslinking.  Thus, we see that even at the level of mean-field theory
it is only as a consequence of the presence of the excluded-volume
interaction that the system can, at the same time, be stable with
respect to collapse to the (inhomogeneous) globular state and yet
unstable with respect to the formation of the (macroscopically
homogeneous) amorphous solid state.

The stability of the higher-replica sector is controlled by the
coefficient of the $\big\lefver\Omega_{\hat{k}}\big\rigver^{2}$ term in
the expansion, \eqref{EQ:hampert}, \ie, by
\beq
1-{\mu^{2}}{V^{-n}}\debyeZbare{{\hat{k}}}
\label{EQ:majorHRS}
\eeq
(considering, as we do, $\mu^{2}\ge 0$). The two contributions to this
coefficient enter with competing signs, owing to the attractive nature
of the effective term arising from crosslinking, and thus provide the
opportunity for the loss of positivity of this coefficient.  Indeed,
the coefficient indicates that the liquid state will be stable for
${\mu^{2}}<1$ and unstable for ${\mu^{2}}>1$, \ie, stable only for
sufficiently small crosslink density, the factor of $V^{-n}$ in
\eqref{EQ:majorHRS} being eliminated by first taking the limit
$n\rightarrow 0$, and subsequently taking the thermodynamic limit
($V\rightarrow\infty$, $N\rightarrow\infty$, $N/V$ fixed, $\mu^{2}$
fixed) \cite{REF:cautionA}.  The least stable modes correspond to long
wavelengths, $\hat{k}^{2}\rightarrow 0$, for which
$\debyeZbare{{\hat{k}}}\rightarrow 1$ from below \cite{REF:cautionB}.

The linear stability analysis of the present subsection indicates that
the liquid state, as characterised by the order parameter discussed in
Sec.~\ref{SEC:OPFields}, is stable when the mean number of crosslinks
per macromolecule $[M]/N$ is smaller than a certain critical value
$M_{\rm c}/N$, \ie, those mean crosslink densities corresponding to
$\mu^{2}<1$.  However, for larger crosslink densities, 
$([M]/N)>(M_{\rm c}/N)$, \ie, $\mu^{2}>1$, the liquid state is unstable
\cite{REF:HighNumber}, being replaced by an alternative state which, as
we shall see in the following two subsections, is an amorphous solid
state, characterised by $\Omega_{{\hat k}}\ne 0$ but 
$\Omega_{{\bf k}}^{\alpha}=0$.  In fact, the state that replaces the
liquid state will turn out to have the property of macroscopic
translational invariance (see Sec.~\ref{SEC:GeneralProps}), so that
even though it has $\Omega_{{\hat k}}\ne 0$ this is compatible with and
does not disturb the fact that the 1-replica sector remains stable
and that $\Omega_{{\bf k}}^{\alpha}$ remains zero.
\subsection{Free energy}\label{SEC:FrEnSect}
We now set about exploring the nature of the amorphous solid state with
respect to which the liquid state is unstable for $\mu^{2}>1$.
Initially, we do this by following the strategy outlined in
Sec.~\ref{SEC:AppStrat} of making $\calfars$ stationary with respect to
the fields $\Omega_{\bf k}^{\alpha}$ and $\Omega_{\hat k}$. However, we
are unable to parametrise the entire space of possible fields.
Instead, we consider the class of fields for which physical motivation
was presented in Sec.~\ref{SEC:EinsteinMTI} [see
\eqref{EQ:OPhypothesis}],
\bml
\bea
\Omega_{\bf k}^{\alpha}=
&0,
&\quad\mbox{(1-replica sector)},
\label{EQ:WRSopForm}
\\
\Omega_{\hat k}=
&
\DPS
{\LocFr}\,\kdelvecT{k}{0}\dint{\tau}
\exp\left(-\hat{k}^{2}/2\tau\right), 
&\quad\mbox{(higher-replica sector)},
\label{EQ:HRSopForm}
\eea
\eml
evaluate $\calfars$ for such fields, and make the resulting quantity
stationary with respect to the variational quantities, the gel fraction
${\LocFr}$ (a number) and the distribution of (inverse square) localisation
lengths $p(\tau)$ (a normalised function).  This amounts to making a
variational mean-field approximation.  However, as we shall see in the
following subsection, the hypothesis we make for the saddle point
will actually turn out to contain an exact saddle point of
$\calfars$.

By inspecting Eq.~(\ref{EQ:hamnumer}) and employing
Eq.~(\ref{EQ:WRSopForm}) we see that there are two contributions to the
free energy:  a term quadratic in $\Omega_{\hat k}$, and a term that
can be identified as the logarithm of the partition function of a
single replicated macromolecule coupled to $\Omega_{\hat k}$.  The
explicit details of the evaluation of these terms when 
$\Omega_{\hat k}$ is given by Eq.~(\ref{EQ:HRSopForm}) is presented in
App.~\ref{APP:EHEOPH}.  Then the variational mean-field approximation
to the free energy is given by
\beq
\tilde{f}=
\lim_{n\to 0}
\fgenf(\mu^{2},\{0\})
\approx
\lim_{n\to 0}
\min_{{\Omega}_{\bf k}^{\alpha},{\Omega}_{\hat k}}
\calfars\big(\{{\Omega}_{\bf k}^{\alpha},{\Omega}_{\hat k}\}\big)
\approx
\min_{{\LocFr},\,p(\tau)}
\fspa\{{\LocFr},p\}, 
\label{EQ:SeqOfLims}
\eeq
where we have omitted constants independent of the variational 
parameters ${\LocFr}$ and $p(\tau)$, and the variational free energy
$\fspa\{{\LocFr},p\}$ is given by 
\bea
\fspa\{{\LocFr},p\}
&=&
-\frac{1}{2}
\Big(
\exp\big(-\mu^{2}{\LocFr}\big)
-\big(1-\mu^{2}{\LocFr}\big)
-\frac{1}{2}\mu^{2}{\LocFr}^{2}
\Big)
\ln\Big(V^{2/d}/2\pi{\rm e}\Big)
\nonumber
\\
&&
\qquad
+\frac{1}{4}\mu^{2}{\LocFr}^{2}
\dinttau{1}\dinttau{2}
\ln     \Big(
\big(\tau_{1}^{-1}+\tau_{2}^{-1}\big)^{-1}
        \Big)
\nonumber
\\
&&
\qquad\quad
+\frac{1}{2}{\rm e}^{-\mu^{2}{\LocFr}}
\sumr\frac{\mu^{2r}{\LocFr}^{r}}{r!}\sintr
\nonumber
\\
&&
\qquad\qquad\quad
\times
\dintr
\ln\left(\wscar\,\detr\rmatr\right).
\label{EQ:freeSPA}
\eea Here, $\detr$ denotes the determinant of an $r\times r$ matrix,
$\rmatr$ is an ($r\times r$)-matrix--valued function of the $r$
arclength coordinates $\{s_{\nu}\}_{\nu=1}^{r}$ and the $r$ inverse
square localisation lengths $\{\tau_{\nu}\}_{\nu=1}^{r}$, and $\wscar$
is a single such function, $\rmatr$ and $\wscar$ being respectively
defined in Eqs.~(\ref{EQ:rmatdef}) and (\ref{EQ:wscadef}) of
App.~\ref{APP:PELLL}.

As anticipated in Sec.~\ref{SEC:avSym}, in addition to intensive terms
we find a nonintensive term, proportional to $\ln V$, owing to the
omission of the disorder-average of the Gibbs symmetry factor.  The
presence of the $\ln V$ factor signals the fact that the configuration
integral produces additional powers of $V$.  These powers of $V$ can
only be due to degrees of freedom that are allowed to vary over the
entire sample, \ie, to the fraction of macromolecules that are
delocalised.  Thus this term depends on the gel fraction ${\LocFr}$, but it
cannot (and does not) depend on $p(\tau)$, which only describes the
localised degrees of freedom.  In the following subsection we shall
analyse the self-consistency condition for the order parameter directly
and, although no quantity proportional to $\ln V$ will appear, we will
re-obtain the exact same results as in the present subsection. This
approach will be seen to have the additional substantial virtue of
demonstrating that the hypothesised form of the order parameter,
\eqref{EQ:OPhypothesis}, used in the present section as a variational
hypothesis, in fact provides an {\it exact saddle point\/} of the
free energy, not merely a variational approximation.

As a first step towards minimising $\fspa$ we regard the term
proportional to $\ln V$ as dominant, and make it stationary with
respect to the gel fraction ${\LocFr}$.  This leads to the condition 
\cite{REF:ER,REF:AZPGNG,REF:CGZjourEPL}
\beq
\exp\big(-\mu^{2}{\LocFr}\big)=1-{\LocFr},
\label{EQ:erdos}
\eeq
For all values of $\mu^{2}$ this equation has the root ${\LocFr}=0$, corresponding 
to the liquid state.  However, for $\mu^{2}>1$ an additional root 
appears, emerging continuously from ${\LocFr}=0$ at $\mu^{2}=1$, and describing 
the equilibrium amorphous solid state.  In Fig.~\ref{FIG:Qplot} we show 
the dependence of the gel fraction on $\mu^{2}$. 
For $\mu^{2}\gg 1$, \ie, the highly crosslinked regime, ${\LocFr}$ 
approaches unity asymptotically as ${\LocFr}\sim 1-\exp\big(-\mu^{2}\big)$.
In the critical regime, $0\le\mu^{2}-1\ll 1$, it is convenient to 
exchange $\mu^{2}$ for the new control parameter $\epsilon$, defined via 
\beq
\mu^{2}\equiv 1+\epsilon/3, 
\label{EQ:epsdef}
\eeq 
with $0\le\epsilon\ll 1$. 
We may then solve perturbatively for ${\LocFr}$, obtaining 
\beq
{\LocFr}=2\epsilon/3+{\cal O}\big(\epsilon^{2}\big). 
\label{EQ:QforSmallEp}
\eeq

Having determined the condition satisfied by ${\LocFr}$ 
we now turn our attention to the dependence of $\fspa$ on the
distribution $p(\tau)$ of inverse square localisation lengths.  As we
are primarily interested in crosslink densities in the vicinity of the
vulcanisation transition (\ie, $0\le\epsilon\ll 1$), we use the result
that, to order $\epsilon$, we have ${\LocFr}=2\epsilon/3$. This allows us to
retain in the summation over $r$ in \eqref{EQ:freeSPA} only the terms
$r=2,3$ (the $r=1$ term vanishing identically by the construction of
$\wscar$).  Next, we assume that the inverse square localisation
lengths having appreciable statistical weight in $p(\tau)$ are also of
order $\epsilon$, \ie, small compared to unity, in units such that
$\sqrt{\ell L/d}=1$, so that localisation is on length scales much
larger than the size of a free macromolecule.  (We shall confirm the
consistency of this assumption {\it a posteriori\/}.)\thinspace\
Thus, we may use the result from App.~\ref{APP:PELLL} to expand
$\ln\big(\wscar\,\detr\rmatr\big)$ in \eqref{EQ:freeSPA} for small 
$\{\tau_{\rho}\}_{\rho=1}^{r}$, retaining terms to order $\tau_{\rho}$.  
Then we integrate over the arclength variables $\{s_{1},\ldots,s_{r}\}$ 
by using the results of App.~\ref{APP:PELLL}.  Omitting terms that 
are independent of $p(\tau)$ we find that, 
to ${\cal O}(\epsilon^{3})$,
\beq
\fspa=
-\frac{1}{8}\left(\frac{2\epsilon}{3}\right)^{3}
\!\ltave
\ln\left(\frac{\tau_{1}+\tau_{2}}{\tau_{1}\tau_{2}}\right)
\rtave
\!+\frac{1}{12}\left(\frac{2\epsilon}{3}\right)^{3}
\!\ltave
\ln\left(\frac{\tau_{1}+\tau_{2}+\tau_{3}}{\tau_{1}\tau_{2}\tau_{3}}\right)
\rtave
\!+\frac{1}{12}\left(\frac{2\epsilon}{3}\right)^{2}
\!\ltave
\frac{\tau_{1}\tau_{2}}{\tau_{1}+\tau_{2}}
\rtave, 
\label{EQ:freenlt}
\eeq
where the curly brace carrying the subscript $\tau$ 
indicates averaging over the localisation lengths, \ie, 
$\ltave\Psi(\tau_{1},\tau_{2},\dots)\rtave
\equiv
\int_{0}^{\infty}
d\tau_{1}p(\tau_{1})
d\tau_{2}p(\tau_{2})\cdots
\Psi(\tau_{1},\tau_{2},\dots)$, 
as many lengths $\tau_{1},\tau_{2},\dots$ 
as feature as arguments of the arbitrary function $\Psi$.  

For future reference we note that if we suppose that the distribution of 
localisation lengths is sharp, \ie, has no fluctuations, so that 
$p(\tau)=\delta(\tau-\xi^{-2})$, where $\xi$ is the sharp value of the 
localisation lengths, then the expression for $\fspa$ simplifies, becoming
\beq
\fspa=\frac{1}{24}\left(\frac{2\epsilon}{3}\right)^{3}\ln\big(\xi^{2}\big)
 +\frac{1}{24}\left(\frac{2\epsilon}{3}\right)^{2}\frac{1}{\xi^{2}},
\label{EQ:freesharp}
\eeq
correct to ${\cal O}(\epsilon^{3})$.
In this case, demanding that $\fspa$ be stationary with respect to
$\xi^{2}$ yields, to ${\cal O}(\epsilon)$, 
\beq
\frac{1}{\xi^{2}}
=
\cases{0,       &if $\mu^{2}\le 1$;\cr
2\epsilon/3,    &if $\mu^{2}\ge 1$.\cr}
\eeq

We now return to the general situation, in which the distribution of localisation 
lengths is not constrained to be sharp.  Rather than demand that $\fspa$ be 
explicitly stationary with respect to $p(\tau)$ itself, it is convenient to exchange 
its dependence on $p(\tau)$ for dependence on the Laplace transform 
$\hat{p}(\hat{\tau})$ given by 
\beq
\hat{p}(\hat{\tau})
\equiv
\int\nolimits_{0}^{\infty}d\tau p(\tau)\exp(-\hat{\tau}\tau).
\label{EQ:ltdef}
\eeq
The details of this exchange are deferred to App.~\ref{APP:LRFF}; 
what results is the following expression for $\fspa$, correct to 
${\cal O}(\epsilon^{3})$:
\bea
\fspa
&=&
-\frac{1}{8}
\left(
        \frac{2\epsilon}{3}
\right)^{3}
        \int\nolimits_{0}^{\infty}
        \frac{d\hat{\tau}}{\hat{\tau}}
\left(
        - \hat{p}(\hat{\tau})^{2}
        +2\hat{p}(\hat{\tau})
        -{\rm e}^{-\hat{\tau}}
\right)
\nonumber
\\
&&
\qquad\quad
+\frac{1}{12}
\left(
        \frac{2\epsilon}{3}
\right)^{3}
        \int\nolimits_{0}^{\infty}
        \frac{d\hat{\tau}}{\hat{\tau}}
\left(
        - \hat{p}(\hat{\tau})^{3}
        +3\hat{p}(\hat{\tau})
        -2{\rm e}^{-\hat{\tau}}
\right)
\nonumber
\\
&&
\qquad\qquad\qquad
+\frac{1}{12}
\left(
        \frac{2\epsilon}{3}
\right)^{2}
        \int\nolimits_{0}^{\infty}d\hat{\tau}
        \big({d\hat{p}}/{d\hat{\tau}}\big)^{2}.
\label{EQ:fefunction}
\eea This expression has the virtue of being a local functional of 
$\hat{p}(\hat{\tau})$, so that the consequent stationarity 
condition will be a differential equation for $\hat{p}(\hat{\tau})$. 
Moreover, the (global) constraint that $p(\tau)$ be normalised to unity, 
$\int\nolimits_{0}^{\infty}d\tau p(\tau)=1$, is exchanged for the (local) 
boundary condition $\hat{p}(0)=1$. 

We now demand that $\fspa$ be stationary with respect to $\hat{p}(\hat{\tau})$, 
\ie, that $\delta\fspa/\delta \hat{p}(\hat{\tau})=0$.  The details of computing 
the functional derivative of $\fspa$ with respect to $\hat{p}(\hat{\tau})$ are 
deferred to App.~\ref{APP:LRFF}; what results is the following stationarity 
condition, correct to ${\cal O}(\epsilon^{4})$:
\beq
0=\frac{\delta\fspa}{\delta\hat{p}(\hat{\tau})}=
-\left(\frac{2\epsilon}{3}\right)^{3}
 \frac{1}{4\hat{\tau}}\Big(1-\hat{p}(\hat{\tau})\Big)
+\left(\frac{2\epsilon}{3}\right)^{3}
\frac{1}{4\hat{\tau}}\left(1-\hat{p}(\hat{\tau})^{2}\right)
-\left(\frac{2\epsilon}{3}\right)^{2}
\frac{1}{6}\frac{d^{2}\hat{p}}{d\hat{\tau}^{2}},
\eeq 
or, equivalently, 
\beq
{\hat{\tau}}\frac{d^{2}\hat{p}}{d\hat{\tau}^{2}}
=
{\epsilon}\,
\hat{p}(\hat{\tau})\,
\Big(1-\hat{p}(\hat{\tau})\Big), 
\label{EQ:statCondi}
\eeq
correct to ${\cal O}(\epsilon)$.
Normalisation of $p(\tau)$ leads to the boundary condition
$\hat{p}(0)=1$.  As $p(\tau)$ does not contain a $\delta$-function
contribution at $\tau=0$ (see \cite{REF:regularity}), $\hat{p}$ obeys the
additional boundary condition $\hat{p}(\infty)=0$.

Before solving the stationarity condition we note that $p(\tau)$
depends parametrically on the crosslink density, so it would be more
accurate to denote it by $p(\tau;\epsilon)$.  We now introduce the
scaling function $\pi(\theta)$ in terms of which $p(\tau;\epsilon)$ is
given by $p(\tau;\epsilon)=(2/\epsilon)\pi(2\tau/\epsilon)$.  In other
words we transform the dependent and independent variables as follows:
\bml
\bea
{\epsilon}p(\tau;\epsilon)/2
&=&
\pi(\theta),
\label{EQ:pscaledef}
\\
\tau
&=&
{\epsilon}\theta/2.
\label{EQ:tscaledef}.
\eea
\eml In this way, up to an elementary factor, the dependence of
$p(\tau;\epsilon)$ on $\tau$ and $\epsilon$ is combined into a
dependence on a single scaling variable $\theta$; 
see \cite{REF:twice}.  Then the Laplace
transform of the scaling function $\hat{\pi}(\hat{\theta})$ is defined 
via
\beq
\hat{\pi}(\hat{\theta})
\equiv
\int\nolimits_{0}^{\infty}d\theta\pi(\theta)\exp(-\hat{\theta}\theta),
\label{EQ:ltdefscale}
\eeq
so that 
\bml
\bea
\hat{p}(\hat{\tau})
&=&
\hat{\pi}(\hat{\theta})
\\
\epsilon\hat{\tau}/2
&=&\hat{\theta}.
\eea
\eml In terms of $\hat{\pi}(\hat{\theta})$, and neglecting ${\cal
O}(\epsilon)$ contributions, the stationarity condition then becomes
\beq 
{\hat{\theta}}
\frac{d^{2}\hat{\pi}}{d\hat{\theta}^{2}}
=
{2}\hat{\pi}(\hat{\theta})\,
\Big(1-\hat{\pi}(\hat{\theta})\Big),
\label{EQ:PiStatCond}
\eeq 
subject to the boundary conditions $\hat{\pi}(0)=1$ 
and $\hat{\pi}(\infty)=0$.

We have been unable to solve this nonlinear ordinary differential
equation for $\hat{\pi}(\hat{\theta})$ analytically.  One might
consider solving this differential equation numerically, and then
inverting the solution numerically to obtain $\pi(\theta)$ and hence
$p(\tau)$.  Whilst this is possible in principle, the numerical
inversion of Laplace transforms is notoriously unstable.  Instead, we
have found it preferable to take the inverse-transform of the
differential equation analytically, and thus we obtain the following
nonlinear integro-differential equation and constraint for $\pi(\theta)$:
\bml
\bea
\frac{\theta^{2}}{2}
\frac{d\pi}{d\theta}
&=&
(1-\theta)\,\pi(\theta)-
\int_{0}^{\theta}d\theta^{\prime}
\pi(\theta^{\prime})\pi(\theta-\theta^{\prime}),
\label{EQ:scpieq}
\\
\int_{0}^{\infty}d\theta\,\pi(\theta)
&=&
1,
\label{EQ:scpibc}
\eea
\eml the constraint resulting from normalisation.

We shall obtain $\pi(\theta)$ [and hence $p(\tau)$] in
Sec.~\ref{SEC:Character}, and discuss the consequences of the physical
values of ${\LocFr}$ and $p(\tau)$.  Before doing so, we shall adopt a
different point of view, in which we focus not on the variational
extremisation of the free energy but instead on the self-consistent
equation for the order parameter.
\subsection{Self-consistency condition for the order parameter}\label{SEC:SceForOp}
In Sec.~\ref{SEC:FrEnSect} we enforced the stationarity of the
effective hamiltonian only with respect to the parameters ${\LocFr}$ and
$p(\tau)$ of our order parameter hypothesis, and not with respect to
arbitrary variations.  As a consequence, we are not yet in a position
to address whether or not the resulting order parameter is a true
saddle point of the effective hamiltonian.  In the present
subsection we establish that the solution that we have found is indeed
a true saddle point of the effective hamiltonian by directly
analysing the stationarity conditions (\ref{EQ:selfconseqWRS}) and
(\ref{EQ:selfconseqHRS}) themselves.  We emphasise that this approach
allows us to circumvent the difficulties discussed in
Sec.~\ref{SEC:avSym} that arise in the computation of the contribution
to the free energy associated with changes in the indistinguishability
factors introduced by the crosslinks.  The results of this subsection
have been briefly reported in Ref.~\cite{REF:CGZjourEPL}.

We insert the hypothesis given in Eqs.~(\ref{EQ:WRSopForm}) and 
(\ref{EQ:HRSopForm}) into the stationarity conditions (\ref{EQ:selfconseqWRS}) 
and (\ref{EQ:selfconseqHRS}) derived in Sec.~\ref{SEC:AppStrat} to obtain
\bea
&&
(1-{\LocFr})\delta_{\hat{q},\hat{0}}
+
{\LocFr}\,
\delta_{\tilvec{q},{\bf 0}}
\,\,\int_{0}^{\infty}d\tau\,p(\tau)\,
\exp\big(-\hat{q}^{2}/2\tau\big)
\nonumber
\\
\noalign{\medskip}
&&
\qquad
=
{
{\DPS
\Big\langle
\int_{0}^{1}dt\,
\exp\Big(i\hat{q}\cdot\hat{c}(t)\Big)\,
\hfill
\atop{\DPS
\qquad
\hfill
\times
\exp
\Big(
\mu^{2}V^{-n}{\LocFr}{\sum}_{\hat{k}}
\delta_{\tilvec{k},{\bf 0}}
\int_{0}^{\infty}d\tau\,p(\tau)
\exp\big(-\hat{k}^{2}/2\tau\big)
\int_{0}^{1}ds\,{\rm e}^{i\hat{k}\cdot\hat{c}(s)}
\Big)
\Big\rangle_{n+1}^{\rm W}
}}
\over{\DPS
\Big\langle\exp
\Big(
\mu^{2}V^{-n}{\LocFr}{\sum}_{\hat{k}}
\delta_{\tilvec{k},{\bf 0}}
\int_{0}^{\infty}d\tau\,p(\tau)
\exp\big(-\hat{k}^{2}/2\tau\big)
\int_{0}^{1}ds\,{\rm e}^{i\hat{k}\cdot\hat{c}(s)}
\Big)
\Big\rangle_{n+1}^{\rm W}}
}.
\label{EQ:scrho}
\eea 
Here, in both the numerator and the denominator we have relaxed the
constraints on the summations having coefficient $\mu^{2}$ by:
(i)~doubling the range of the summations to include the entire 
higher-replica sector by making use of the property of the hypothesis
$\Omega_{\hat{k}}=\Omega_{-\hat{k}}^{\ast}$; 
(ii)~including the 1-replica sector terms (which vanish by the MTI of
the order parameter hypothesis); and
(iii)~inserting identical factors associated with the 0-replica sector.
It should be noted that \eqref{EQ:scrho} also follows from the direct
application of the Weiss molecular-field method.

As shown at the end of App.~\ref{APP:OPWaverages}, evaluation of the 
numerator and denominator of the right-hand side yields 
\bml
\bea
(1-{\LocFr})\,\delta_{\hat{q},\hat{0}}
&+&
{\LocFr}\,\delta_{\tilvec{q},{\bf 0}}
\,\,\int_{0}^{\infty}d\tau\,p(\tau)\,
{\rm e}^{-\hat{q}^{2}/2\tau}
\nonumber
\\
&=&
{\rm e}^{-\mu^{2}{\LocFr}}\,
\delta_{\hat{q},\hat{0}}
+
{\rm e}^{-\mu^{2}{\LocFr}}\,
\delta_{\tilvec{q},{\bf 0}}
\int_{0}^{\infty}d\tau\,
{\rm e}^{-\hat{q}^{2}/2\tau}
\sum_{r=1}^{\infty}
{\mu^{2r}{\LocFr}^{r}\over{r!}}
\int_{0}^{1}ds_{1}\cdots ds_{r+1}
\nonumber
\\
&&
\qquad\qquad\qquad
\times
\int_{0}^{\infty}
d\tau_{1}\cdots d\tau_{r}\,
p(\tau_{1})\cdots p(\tau_{r})\,
\delta\big(\tau-\Upsilon^{(r)}\big), 
\label{EQ:scone}
\\
\noalign{\medskip}
\frac{1}{\Upsilon^{(r)}}
&\equiv&
\frac{1}{\wscar}
+\smatrCI{r+1,r+1} 
-\frac{2}{\wscar}
\sum_{\rho=1}^{r}
\uvecr\,
\smatrCI{\rho,r+1}      
-\sum_{\rho,\rho^{\prime}=1}^{r}
\smatrCI{r+1,\rho}
\,\cmatr\,
\smatrCI{\rho^{\prime},r+1},
\label{EQ:scmain}
\eea
\eml where the limit $n\rightarrow 0$ has been taken everywhere except
in the dimension of $\hat{q}$, and where $\smatr$, $\uvecrNI$, $\wscar$
and $\cmatrNI$ are, respectively, defined in Eqs.~(\ref{EQ:smatdef}),
(\ref{EQ:uvecdef}), (\ref{EQ:wscadef}) and (\ref{EQ:cmatdef}), and
depend on $\{s_{1},\ldots,s_{r+1}\}$ and
$\{\tau_{1},\ldots,\tau_{r+1}\}$.  It should be emphasised that
\eqref{EQ:scone} is not solved by any sharp distribution of
localisation lengths $p(\tau)=\delta(\tau-\xi^{-2})$.  Thus, a
variational hypothesis involving a sharp distribution gives at best a
variational approximation, whereas a variational hypothesis involving a
non-sharp distribution has the potential to yield an exact saddle
point, and we shall find such an exact saddle point below, at least
in the vicinity of the vulcanisation transition.

We now extract information about ${\LocFr}$ and $p(\tau)$ from
\eqref{EQ:scone}.  First, we take the limit $\hat{q}^{2}\rightarrow 0$,
via a sequence for which $\tilvec{q}={\bf 0}$.  In this limit, the left
hand side becomes ${\LocFr}$, and on the right side each integral gives a
factor of unity, yielding the self-consistency condition for the
gel fraction ${\LocFr}$:
\beq
{\LocFr}=1-{\rm e}^{-\mu^{2}{\LocFr}}, 
\label{EQ:scforQ}
\eeq
\ie, precisely the self-consistency condition for ${\LocFr}$ found from the 
free energy approach is Sec.~\ref{SEC:FrEnSect} and discussed there.

Having decoupled the issue of the gel fraction ${\LocFr}$ from the issue of
the distribution $p(\tau)$ we now return to $p(\tau)$ itself.  By
considering Eq.~(\ref{EQ:scmain}) for a fixed nonzero value of
$\hat{q}^{2}$, and using Lerch's uniqueness theorem for Laplace
transforms \cite{REF:Lerch} we find that indeed the hypothesis solves
the self-consistency condition for 
$\{\Omega_{\bf k}^{\alpha},\Omega_{\hat{k}}\}$ provided that the 
distribution $p(\tau)$ satisfies the condition
\beq
{\LocFr}\,p(\tau)=
{\rm e}^{-\mu^{2}{\LocFr}}
\sum_{r=1}^{\infty}{\mu^{2r}{\LocFr}^{r}\over{r!}}
\int_{0}^{1}ds_{1}\cdots ds_{r+1}
\int_{0}^{\infty}d\tau_{1}\cdots d\tau_{r}\,
p(\tau_{1})\cdots p(\tau_{r})\,
\delta\big(\tau-\Upsilon_{r}\big).
\label{EQ:scforP}
\eeq
This equation for $p(\tau)$ is, for all values of $\mu^{2}$, 
identically satisfied if ${\LocFr}=0$. 

We have not, thus far, made any approximations beyond that of mean
field theory. In order to render Eq.~(\ref{EQ:scforP}) tractable, we
now restrict our attention to the vicinity of the transition regime in
the solid state, \ie, to values of $\epsilon$, as defined in
\eqref{EQ:epsdef}, satisfying $0\le\epsilon\ll 1$.  This restriction
allows us to assume that ${\LocFr}$ is small, and that only localisation
lengths much larger than the free-macromolecule radius of gyration have
an appreciable probability, \ie, $p(\tau)$ only gives appreciable
weight for $0<\tau\ll 1$.  Thus, we need retain in Eq.~(\ref{EQ:scforP}) 
only terms for which $r$ is 1 or 2, and may expand
$\Upsilon^{(1)}$ to ${\cal O}(\tau^{2})$ and 
$\Upsilon^{(2)}$ to ${\cal O}(\tau^{1})$.
As discussed in App.~\ref{APP:DSCE}, in terms of the scaling function
$\pi(\theta)$ introduced in \eqref{EQ:pscaledef}, we recover
\eqref{EQ:scpieq} subject to the normalisation condition
\eqref{EQ:scpibc}, \ie, precisely the stationarity condition for
$p(\tau)$ found from the free energy approach.

Thus, the condition that the order parameter be self-consistent turns
out to be identical to the condition that the effective hamiltonian be
stationary with respect to variations within the subspace spanned by
the hypothesised form of the order parameter.  The form for the order
parameter hypothesised in \eqref{EQ:OPhypothesis} is not merely a
variational form but in fact gives an exact saddle point of the
effective hamiltonian, \eqref{EQ:hamnumer}.

We have obtained the equation for the gel fraction ${\LocFr}$,
Eq.~(\ref{EQ:erdos}), and the equations for the scaled distribution
$\pi(\theta)$, Eqs.~(\ref{EQ:scpieq}) and (\ref{EQ:scpibc}), from two
different points of view.  In the previous subsection we have discussed
the consequences of Eq.~(\ref{EQ:erdos}) for ${\LocFr}$.  In the following
subsection we shall discuss the solution of Eqs.~(\ref{EQ:scpieq}) and
(\ref{EQ:scpibc}) for $\pi(\theta)$, and elaborate on the physical
consequences of our results for ${\LocFr}$ and $p(\tau)$.
\subsection{Characteristics of the amorphous solid state}\label{SEC:Character}
For the sake of completeness we first restate the results concerning
the gel fraction ${\LocFr}$ that were found in Sec.~\ref{SEC:FrEnSect} from
the self-consistency condition on ${\LocFr}$, \eqref{EQ:erdos}.  For all
values of $\mu^{2}$ we find the solution ${\LocFr}=0$, corresponding to the
liquid state.  For $\mu^{2}>1$ an additional solution appears, emerging
continuously from ${\LocFr}=0$ at $\mu^{2}=1$, and describing the equilibrium
amorphous solid state, as shown in  Fig.~\ref{FIG:Qplot}.  For
$\mu^{2}\gg 1$, \ie, the highly crosslinked regime, ${\LocFr}$ approaches
unity asymptotically as ${\LocFr}\sim 1-\exp\big(-\mu^{2}\big)$.  In terms of
the deviation of the crosslink density from criticality, \ie,
$\epsilon$ defined in \eqref{EQ:epsdef}, the critical regime is
$0\le\epsilon\ll 1$.  In this regime we may solve perturbatively for
${\LocFr}$, obtaining ${\LocFr}=2\epsilon/3+{\cal O}\big(\epsilon^{2}\big)$, 
\eqref{EQ:QforSmallEp}.

It should be noted that the stationarity condition on ${\LocFr}$,
\eqref{EQ:erdos}, is precisely the condition obtained by Erd{\H o}s and
R{\'e}nyi in the context of random graph theory \cite{REF:ER}, which 
can also be interpreted as a mean-field treatment of percolation. In
particular, Erd{\H o}s and R{\'e}nyi showed that that for a random
graph of $N$ points and $\mu^{2}N/2$ edges the probability for the
fraction of points in the largest component to differ from the solution
${\LocFr}$ of Eq.~(\ref{EQ:erdos}) vanishes in the $N\rightarrow\infty$ limit.  
A related approach to the theory of macromolecular networks
\cite{REF:AZPGNG} has also led to \eqref{EQ:erdos}.  
This is physically quite reasonable:  one would anticipate that the
transition from liquid to solid would occur when the density of
crosslinks is sufficient to create a macroscopically extended network
of crosslinked macromolecules.

We now address the distribution of localisation lengths via the scaling
form $\pi(\theta)$.  We have solved both the integro-differential
equation (\ref{EQ:scpieq}) and the differential equation
(\ref{EQ:PiStatCond}) numerically \cite{REF:NumSchemeLS,REF:NumSchemeRS}, 
and the solution of \eqref{EQ:scpieq} is shown in Fig.~\ref{FIG:PiTheta}.  
As we see in this figure, the scaling function $\pi(\theta)$ has a
single maximum near $\theta=1$, away from which it decays rapidly. In
fact, states for which $\pi(\theta)$ takes negative values are not
ruled out by the hypothesis \eqref{EQ:OPhypothesis}, but are not found
as solutions of the stationarity condition Eqs.~(\ref{EQ:scpieq}) and
(\ref{EQ:scpibc}).

We are able to obtain asymptotic properties of $\pi(\theta)$
analytically.  The asymptotic form 
$\pi(\theta)\sim a\theta^{-2}\exp\big(-2/\theta\big)$ 
(for $\theta\ll 1$) is obtained
from Eq.~(\ref{EQ:scpieq}) by neglecting the second term on the right
hand side. Notice the essential singularity at the origin:
$\pi(\theta)$ vanishes very rapidly indeed as $\theta\rightarrow 0$.
The coefficient $a\approx 4.554$ cannot be obtained from local
asymptotic analysis.  Instead we have obtained it separately by the
numerical solution of Eq.~(\ref{EQ:PiStatCond}), as discussed in
footnote~\cite{REF:NumSchemeRS}.  The asymptotic form 
$\pi(\theta)\sim 3\big(b\theta-3/5\big)\exp\big(-b\theta\big)$ 
(for $\theta\gg 1$) is obtained by computing the inverse Laplace 
transform of the approximate
analytical solution of Eq.~(\ref{EQ:PiStatCond}) near the point
$\hat{\theta}=-b$, at which $\hat{\pi}(\hat{\theta})$ diverges.  
The coefficient $b\approx 1.678$ was obtained
separately by determining the (negative) value of $\hat{\theta}$ at
which the numerical solution of Eq.~(\ref{EQ:PiStatCond}) diverges. 
Notice the exponential decay of $\pi(\theta)$ for large $\theta$: 
$\pi(\theta)$ goes to zero quickly as $\theta\rightarrow\infty$.  
For the sake of comparison with the numerical results, the 
small- and large-$\theta$ asymptotic forms for $\pi(\theta)$ are 
also shown in Fig.~\ref{FIG:PiTheta}. A distribution of localisation 
lengths also features in Panyukov's approach to the well-crosslinked 
regime; see \cite{REF:PanNew}.

In addition to providing the distribution of localisation lengths,
knowledge of $\pi(\theta)$ allows us to construct the order parameter,
$\Omega_{\hat{k}}$.  By using Eqs.~(\ref{EQ:pscaledef}) and
(\ref{EQ:tscaledef}) in Eq.~(\ref{EQ:OPhypothesis}) we obtain
\bml
\bea
\Omega_{\hat{k}}
&=&
\big(1-2\epsilon/3\big)\,\delta_{\hat{k},\hat{0}}+ 
\big(2\epsilon/3\big)\,\delta_{\tilvec{k},{\bf 0}}\,
\omega\Big(\sqrt{2\hat{k}^{2}/\epsilon}\Big),
\label{EQ:omsc}
\\
\omega(k)
&\equiv&
\int_{0}^{\infty}d\theta\,\pi(\theta)\,
{\rm e}^{-k^{2}/2\theta}.
\label{EQ:omdef}
\end{eqnarray}
\eml Although we do not have an exact analytical expression for $\pi(\theta)$, 
we can compute $\omega(k)$ numerically.  We do this by inserting the 
the numerical values of $\pi(\theta)$ into \eqref{EQ:omdef}, and show 
the result for $\omega(k)$ in Fig.~\ref{FIG:omegaQ}.
We are able to obtain analytical asymptotic expressions for $\omega(k)$.
For $q\ll 1$ the result simply follows from expanding the exponential 
function in \eqref{EQ:omdef} in powers of $q$, thus obtaining
\bml
\bea
\omega(k)
&\sim&
1
-\frac{k^{2}}{2}
\int_{0}^{\infty}d\theta\,\theta^{-1}\pi(\theta)
+\frac{k^{4}}{8}
\int_{0}^{\infty}d\theta\,\theta^{-2}\pi(\theta)
+\cdots
\\
&=&
1-0.4409q^{2}+0.1316q^{4}
+\cdots,
\qquad{\rm for\/}\quad q\ll 1.
\eea
We see that $\omega(k)$ departs quadratically from its absolute 
maximum of unity at the origin.
For $k\gg 1$ one can replace $\pi(\theta)$ by its large-$\theta$ 
asymptotic form in \eqref{EQ:omdef} to obtain 
\beq
\omega(k)\sim
\left(\frac{9\pi k^{3}}{\sqrt{8b}}\right)^{1/2}
\,{\rm e}^{-\sqrt{2bk^{2}}}\,
\left(1+\frac{27}{40\sqrt{2bk^{2}}}+\cdots\right),
\qquad{\rm for\/}\quad k\gg 1.
\eeq
\eml
We see that $\omega(k)$ decays exponentially to zero for large $k$. 
For the sake of comparison with the numerical results, the 
small- and large-$k$ asymptotic forms for $\omega(k)$ are also 
shown in Fig.~\ref{FIG:omegaQ}.

To summarise, as shown in Sec.~\ref{SEC:InstFlPH} the liquid state of a
system of randomly crosslinked macromolecules becomes unstable when the
mean number of crosslinks per macromolecule $[M]/N$ is increased beyond
a certain critical value $M_{\rm c}/N$, corresponding to $\mu^{2}=1$.
At this critical point the system exhibits a continuous phase
transition from the liquid state to the  amorphous solid state.  As
shown in Secs.~\ref{SEC:FrEnSect} and \ref{SEC:SceForOp}, this solid
state is characterised by a gel fraction ${\LocFr}$, which grows from a value
of zero at the critical point with the classical exponent $\beta=1$
(see Ref.~\cite{REF:deg}):  
${\LocFr}\sim\epsilon\sim\mu^{2}-1\sim([M]-M_{\rm c})/N$.  
The amorphous solid state is further characterised by the statistical
distribution of localisation lengths $2\xi^{-3}p\big(\xi^{-2}\big)$.
In the vicinity of the transition the dependence of this distribution
on the control parameter $\epsilon$ and the (inverse square)
localisation length $\tau$ is determined by a universal scaling
function (of a single variable) $\pi(\theta)$, \ie,
$p(\tau)=(2/\epsilon)\pi(2\tau/\epsilon)$. This universality guarantees
that $\pi(\theta)$ need only be computed once for all near-critical
crosslink densities.  As already mentioned in the present subsection,
$\pi(\theta)$ has a single maximum, away from which it decays rapidly.
Hence, the fraction of localised monomers that are localised on
length scales much larger than $\epsilon^{-1/2}$ is exceedingly small.
Our result for $p(\tau)$ also predicts that the fraction of localised
monomers with localisation lengths much smaller than $\epsilon^{-1/2}$
is also exceedingly small.  This provides an {\it a posteriori\/}
confirmation of the internal consistency of the perturbation expansion
in powers of $\xi^{-2}$ upon which our results rely.  However, the
detailed form of the distribution for localisation lengths much smaller
than $\epsilon^{-1/2}$ (\eg, for localisation lengths of the order of
the radius of gyration of a free macromolecule) is unreliable because
such localisation lengths are not within the range of validity of the
perturbation expansion.  The rapid decay of $p(\tau)$ away from its
maximum guarantees that its moments are finite.  This character,
together with the scaling form of $p(\tau)$ ensures that the moments
scale in the following manner:
$\big[\xi^{-2m}\big]\sim\big(([M]-M_{\rm c})/N\big)^{m+1}$.
Furthermore, as the distribution has single maximum it is sensible to
define a typical localisation length $\xi_{\rm typ}$ associated with
the most probable localisation length.  This length $\xi_{\rm typ}$
obeys the scaling relation $\xi_{\rm typ}\sim(\mu^{2}-1)^{-1/2}$.
Thus, a simple, reasonable approximation to the true
distribution $p(\tau)$ would be a sharp distribution, \eg,
$\delta\big(\tau-\epsilon/2\big)$.

We have seen in Sec.~\ref{SEC:SceForOp} that the order parameter
hypothesised in Sec.~\ref{SEC:OPhypoth} and determined in
Secs.~\ref{SEC:FrEnSect} and \ref{SEC:SceForOp} is a solution of the
stationarity condition for the free energy \eqref{EQ:selfconseqHRS}.
This is in contrast with the hypothesis analysed in
Ref.~\cite{REF:PMGandAZprl}, in which it was assumed that ${\LocFr}=1$ and
$p(\xi^{-2})=\delta(\xi^{-2}-\bar{\xi}^{-2})$ (\ie, that all monomers
share a common localisation length).  The hypothesis of
Ref.~\cite{REF:PMGandAZprl} does not satisfy the stationarity
condition, and therefore only provides a variational bound on the free
energy.  That our result for the order parameter is a saddle point
of the free energy, rather than merely a variational bound, is a
feature of considerable significance.  The consequent exact vanishing
of the linear term in the expansion of the effective hamiltonian
\eqref{EQ:hamnumer} in powers of the departure from the known
stationary value streamlines further analysis of, \eg, linear 
stability, fluctuations, correlations, and response to perturbations.
\subsection{Comparison with numerical simulations}\label{SEC:BarPli}
Extensive numerical simulations of both macromolecular melts and
well-crosslinked macromolecular networks have been performed by Grest
and Kremer, and others; for a review see Ref.~\cite{REF:GKreview}.  On
the other hand, until the recent work of Barsky and Plischke
\cite{REF:Barsky} relatively little attention had been devoted to
simulations of the regime in which the number of crosslinks is
comparable to the number of macromolecules (\ie, the vicinity of the
liquid-to-amorphous solid transition).  It should be remarked that from
the computational point of view this is a daunting regime.  In their
simulations, Barsky and Plischke observe a continuous transition from
the liquid to the amorphous solid state, and extract a universal
scaling function describing the distribution of localisation lengths.
Whilst their numerical results are in strong qualitative agreement with
the analytical predictions described in the present article, thus
providing support for the theoretical picture of amorphous
solidification discussed here, there appear to be quantitative
differences.  At present, the precise origin of these differences is
unclear: a possible source is the relatively short length of the
macromolecules used in the simulations, for which mean-field theory is
expected to require substantial fluctuation-corrections over a
moderately wide range of near-critical crosslink densities.
\section{Incorporation of density-sector correlations}\label{SEC:DenSecOV}
Until now we have entirely neglected fluctuations in the fields
$\Omega_{\bf k}^{\alpha}$, $\Omega_{\hat k}$ and $\omega_{\bf k}$ in the
computation of the quotient of functional integrals in the numerator
and denominator of \eqref{EQ:funPartition}.  Instead we have
approximated the functional integrals by the stationary values of their
integrands.  This amounts to making a mean-field approximation, in
which correlations between fluctuations of the fields are neglected.

As a consequence of this strategy, the excluded-volume interaction has
played a subsidiary role: its presence has been required, in order to
maintain the stability of the physical system with respect to the
formation of macroscopically inhomogeneous states, \eg, via
crystallisation or collapse (which would be detected by their nonzero
value of the order parameter in the 1-replica sector), even at
crosslink densities large enough to destabilise the liquid state with
respect to macroscopically translationally invariant states (\ie,
equilibrium amorphous solids).  However, having accomplished this by
guaranteeing the stability of the 1-replica sector, the excluded-volume
interaction has played no further role.  Indeed, the precise value of
the excluded-volume parameter $\lambda$ does not even feature in the
mean-field free energy of either the liquid state or the equilibrium
amorphous solid state (in the same way that the exchange coupling
constant of a magnetic system does not feature in the mean-field free
energy of the paramagnetic state).  Thus, amongst the correlations that
we have entirely neglected are those between fluctuations of the
density of the system, \ie, correlations between the 1-replica sector
fields $\Omega_{\bf k}^{\alpha}$.

There are two reasons why we should seek to improve the theory by 
incorporating density-density correlations, at least at some level. 
(It should be pointed out that Ball and Edwards have undertaken the 
task of incorporating correlations in their approach to crosslinked 
macromolecular systems 
\cite{REF:RCBallPaper,REF:RCBallThesis}.)\thinspace\ 
First, at least at high densities, 
the resulting screening of the excluded-volume interaction gives an 
accurate treatment of interaction effects \cite{REF:SFEdMDoiBook}.
Second, such correlations would reflect the statistical tendency for
the macromolecules to avoid one another.  This would have the effect of
introducing a statistical preference for configurations in which
macromolecules stay apart, and a barrier between topologically 
distinct configurations.  Thus, one might imagine the following
scenario.  In a given realisation of the supercritically crosslinked
system without the correlations associated with self-avoidance there is
no statistical bias against macromolecules passing through one
another.  Thus there is no apparent mechanism for creating distinct
ways for the macromolecules to localise themselves, and the system
would exhibit a single family of (translational- and
rotational-symmetry--related) solid equilibrium states.  If, on the
other hand, correlations are incorporated, then there can be at least
statistical barriers between symmetry-unrelated ways for the
macromolecules to become localised, \eg, topologically distinct
interweavings of the macromolecules.  Thus, one
might anticipate a situation in which the incorporation of correlations
allows the system to exhibit symmetry-unrelated equilibrium states, a
scenario that could be revealed through the mechanism of the
spontaneous breakdown of the permutation symmetry amongst the replicas. 
Certain ideas in this direction have been explored in 
Ref.~\cite{REF:GoZiJoPhyA}.  For a discussion of related issues in 
the context of spin glasses, see Ref.~\cite{REF:MPVbook}.  
\subsection{Free energy}\label{SEC:DenSecFE}
We now set about the task of computing the quotient of functional integrals, 
\eqref{EQ:onePartition}, with improved accuracy, in order to incorporate 
the effect of the excluded-volume interaction. With regard to the functional 
integral in the numerator our strategy will be to treat the 1-replica sector 
in the gaussian approximation, resulting in an improved effective hamiltonian 
for the higher-replica sector field $\Omega_{\hat k}$, which we then treat at 
the mean-field level.  The functional integral in the denominator will also 
be treated in the gaussian approximation. We assume that at the relevant 
saddle points of the improved effective hamiltonian the order parameter 
$\Omega_{\hat k}$ will continue to have the property of being MTI, 
as described in Sec.~\ref{SEC:GeneralProps}.

To implement this strategy, we expand $\calfars$, \eqref{EQ:hamnumer}, to 
quadratic order in the 1-replica sector field $\Omega_{{\bf k}}^{\alpha}$ for 
an arbitrary value of the higher-replica sector field $\Omega_{{\hat k}}$, 
except for the restriction that $\Omega_{{\hat k}}$ be MTI.  Thus we obtain 
\bml
\bea
&&
nd\calfars\big(\{\Omega_{{\bf k}}^{\alpha},\Omega_{{\hat k}}\}\big)
\approx
\frac{\mu^{2}}{V^{n}}
\sumhrsHPNL{k}
\big\lefver
\Omega_{{\hat k}}
\big\rigver^{2}
-\ln
\Big\langle
\exp
\Big(
\frac{2\mu^{2}}{V^{n}}
\sumhrsHPNL{k}
\real\Omega_{\hat{k}}^{\ast}
\int\nolimits_{0}^{1}ds\,
\exp
\big(
i{\hat{k}}\cdot{\hat{c}}(s)
\big)
\Big)
\Big\rangle_{n+1}^{\rm W}
\nonumber
\\
&&
\qquad\qquad\qquad\qquad\quad
+\tlns\denfrac
\sumwrsHP{\alpha,\alpha^{\prime}}{k}
\left(
\delta^{\alpha,\alpha^{\prime}}+
\tlns({N}/{V})\green{k}
\right)
\Omega_{{\bf k}}^{\alpha\ast}\,
\Omega_{{\bf k}}^{\alpha^{\prime}}
+{\cal O}\Big((\Omega_{{\bf k}}^{\alpha})^{3}\Big),
\label{EQ:gausshamnum}
\\
\noalign{\medskip}
&&
\green{q}
\equiv
{\DPS
\Big\langle
\int\nolimits_{0}^{1}dt\,
{\rm e}^{i{\bf q}\cdot{\bf c}^{\alpha}(t)}
\int\nolimits_{0}^{1}dt^{\prime}\,
{\rm e}^{-i{\bf q}\cdot{\bf c}^{\alpha^{\prime}}(t^{\prime})}
\exp
        \Big(
2\mu^{2}V^{-n}
\sumhrsHPNL{k}
\real
\Omega_{\hat{k}}^{\ast}
\int\nolimits_{0}^{1}ds\,
{\rm e}^{i{\hat{k}}\cdot{\hat{c}}(s)}
        \Big)
\Big\rangle_{n+1}^{\rm W}
\over{\DPS
\Big\langle
\exp
        \Big(
2\mu^{2}V^{-n}
\sumhrsHPNL{k}
\real
\Omega_{\hat{k}}^{\ast}
\int\nolimits_{0}^{1}ds\,
{\rm e}^{i{\hat{k}}\cdot{\hat{c}}(s)}
        \Big)
\Big\rangle_{n+1}^{\rm W}
}},
\label{EQ:greendef}
\eea 
\eml in which $\density\equiv N/V$ denotes the number of 
macromolecules per unit volume.  Note that all terms linear in 
$\Omega_{{\bf k}}^{\alpha}$, as well as most terms that are quadratic 
in $\Omega_{{\bf k}}^{\alpha}$ vanish by virtue of the MTI property of 
$\Omega_{\hat{k}}$.  The $\Omega_{\hat{k}}$-dependent correlator 
$\green{q}$ is discussed further, below, and is computed for a specific 
form of $\Omega_{\hat{k}}$ in App.~\ref{APP:FluCo}.  Similarly, we 
expand $\calfzrs$, \eqref{EQ:hamdenom}, to quadratic order in the 
field $\omega_{{\bf k}}$, obtaining
\bml 
\bea
nd\calfzrs\big(\{\omega_{{\bf k}}\}\big)
&\approx&
\tlzs\denfrac
\sumzrsHP{k}
\left(
1+\tlzs\denfrac\debyeZ{k}
\right)
\big\lefver
\omega_{{\bf k}}
\big\rigver^{2},
\label{EQ:gausshamdenom}
\\
\debyeZ{q}
&\equiv&
\Big\langle
\int\nolimits_{0}^{1}dt\,
{\rm e}^{ i{\bf q}\cdot{\bf c}(t)}
\int\nolimits_{0}^{1}dt^{\prime}\,
{\rm e}^{-i{\bf q}\cdot{\bf c}(t^{\prime})}
\Big\rangle_{1}^{\rm W}
\label{EQ:debyeZdef}.
\eea
\eml The basic Debye correlator $\debyeZ{q}$ is computed in
App.~\ref{APP:debye}.  Next, we substitute the approximations to
$\calfars$ and $\calfzrs$, quadratic in $\Omega_{{\bf k}}^{\alpha}$ and
$\omega_{{\bf k}}$ respectively, into \eqref{EQ:onePartition}, and
perform the resulting gaussian integrations over 
$\Omega_{{\bf k}}^{\alpha}$ and $\omega_{{\bf k}}$.  These we do for
each value of ${\bf k}$ in the numerator and denominator, by noting the
normalisations \eqref{EQ:normWRS} and \eqref{EQ:normZRS} and applying
the well-known result,
\beq
{\DPS
\int
\prodaln
d\real z^{\alpha}\,\,
d\imag z^{\alpha}
\exp
        \Big(
-\sum_{\alpha,\alpha^{\prime}=0}^{n}
z^{\alpha\ast}
{\somemat}_{1}^{\alpha\alpha^{\prime}}
z^{\alpha^{\prime}}
        \Big)
\over{\DPS
\int
\prodaln
d\real z^{\alpha}\,\,
d\imag z^{\alpha}
\exp
        \Big(
-\sum_{\alpha,\alpha^{\prime}=0}^{n}
z^{\alpha\ast}
{\somemat}_{2}^{\alpha\alpha^{\prime}}
z^{\alpha^{\prime}}
        \Big)
}}
=
{\detn{\somemat}_{2}\over{\detn{\somemat}_{1}}},
\label{EQ:gaussdet}
\eeq
in which ${\somemat}_{1}$ and ${\somemat}_{2}$ are arbitrary hermitean
complex-valued $(n+1)\times(n+1)$ positive-definite matrices and
$\detn$ denotes a (replica-space) determinant of an $(n+1)\times(n+1)$
matrix.  Hence we obtain
\bea
&&
\exp\big(-ndN\fgenf(\mu^{2},\{0\})\big)
\approx
{\cal B}_{n}\,
\int\dmhrs\Omega
\exp
        \Big(
-\frac{N\mu^{2}}{V^{n}}
\sumhrsHPNL{k}
\big\lefver
\Omega_{{\hat k}}
\big\rigver^{2}
\nonumber
\\
&&
\qquad\qquad
\qquad\qquad
+N\ln
\Big\langle
\exp
\Big(
\frac{2\mu^{2}}{V^{n}}
\sumhrsHPNL{k}
\real
\Omega_{\hat{k}}^{\ast}
\int\nolimits_{0}^{1}ds\,
\exp
\big(
i{\hat{k}}\cdot{\hat{c}}(s)
\big)
\Big)
\Big\rangle_{n+1}^{\rm W}
        \Big)
\nonumber
\\
&&
\qquad\qquad\qquad
\qquad\qquad\qquad
\times
{\prod}_{\bf k}^{\possym}
\Big(
1+\tlzs\density\debyeZ{k}
\Big)
\Big/
{\prod}_{\bf k}^{\possym}
\detn
\Big(
\imatnNI+\tlns\density\greenNI{k}
\Big),
\label{EQ:gaussPartition}
\eea
where $\imatnNI$ is the $(n+1)\times(n+1)$ identity on the replica indices.

Now, our aim is to compute the free energy $\tilde{f}$, \eqref{EQ:logAverage},
and to do this we shall need the difference
$\fgenf(\mu^{2},\{0\})-\fgenf(0,\{0\})$, 
as we see from \eqref{EQ:ReplLim}.
Thus, in addition to $\fgenf(\mu^{2},\{0\})$, which we have
just obtained at the gaussian level of approximation, we shall also
need $\fgenf(0,\{0\})$ at the gaussian level of approximation.
This can readily be obtained from \eqref{EQ:gaussPartition} by setting
$\mu^{2}$ to zero throughout, except in the measure $\dmhrs\Omega$,
\eqref{EQ:measureHRS}, and in the coefficient of the quadratic term in
the exponent:
$-\big({N\mu^{2}}/{V^{n}}\big)
{\overline{\sum}}_{\hat{k}}^{\possym}
\big\lefver\Omega_{{\hat k}}\big\rigver^{2}$. 
Then the logarithmic term vanishes, $\green{k}$ becomes
$\imatn\debyeZ{k}$, and $\tlns$ becomes $\lambda^{2}$, and the functional
integration can be performed by using the normalisation
\eqref{EQ:measureHRS}, all $\mu^{2}$-dependence cancelling from the
result.  Thus, we find
\beq
\exp\big(-ndN\fgenf(0,\{0\})\big)
\approx
{\rm e}^{-n\lambda^{2}N\density/2}
\Big/
{\prod}_{\bf k}^{\possym}
\Big(
1+\lambda^{2}\density\debyeZ{k}
\Big)^{n}
\label{EQ:curly_z_zero}.
\eeq
By forming the difference 
$\fgenf(\mu^{2},\{0\})-\fgenf(0,\{0\})$ 
we thus arrive at the following expression for 
$\big[\tilde{Z}\big(\{i_{e},s_{e};
        i_{e}^{\prime},s_{e}^{\prime}
      \}_{e=1}^{M}
\big)^{n}\big]$, 
in which 1-replica sector density fluctuations are incorporated at 
the gaussian level, as indicated by the superscript 
${\rm gdf}$ on $\cal F$:
\bml
\bea
&&
\left[
\tilde{Z}\big(\{i_{e},s_{e};
        i_{e}^{\prime},s_{e}^{\prime}
      \}_{e=1}^{M}
\big)^{n}
\right]
\approx
\int\dmhrs\Omega
\exp\big(
-ndN\calfgdf\big(\{\Omega_{\hat{k}}\}\big)
\big),
\label{EQ:gdfPF}
\\
\noalign{\medskip}
&&
nd\calfgdf\big(\{\Omega_{\hat{k}}\}\big)
\equiv
\frac{\mu^{2}}{V^{n}}
\sumhrsHPNL{k}
\big\lefver
\Omega_{{\hat k}}
\big\rigver^{2}
-\ln
\Big\langle
\exp
\Big(
\frac{2\mu^{2}}{V^{n}}
\sumhrsHPNL{k}
\real
\Omega_{\hat{k}}^{\ast}
\int\nolimits_{0}^{1}ds\,
\exp
\big(
i{\hat{k}}\cdot{\hat{c}}(s)
\big)
\Big)
\Big\rangle_{n+1}^{\rm W}
\nonumber
\\
&&
\qquad\qquad
+\frac{1}{\deninve V}
\sumzrsHPNL{k}
\ln\detn
\left(
{\imatnNI+\tlns\density\greenNI{k}
\over{1+\lambda^{2}\density\debyeZ{k}}}
\right)
-\frac{1}{\deninve V}
\sumzrsHPNL{k}
\ln\left(
{1+\tlzs\density\debyeZ{k}
\over{1+\lambda^{2}\density\debyeZ{k}}}
\right).
\label{EQ:gdfFE}
\eea
\eml
Following the same line of reasoning made in Sec.~\ref{SEC:FrEnSect}, 
we shall now make a variational mean field approximation for the 
remaining functional integral.  To do this, we evaluate the effective 
hamiltonian (\ref{EQ:gdfFE}) using the hypothesis for $\Omega_{\hat k}$ 
given in \eqref{EQ:HRSopForm}.  As was the case in Sec.~\ref{SEC:FrEnSect}, 
this leads to an expression for the effective hamiltonian in terms of ${\LocFr}$ 
and $p(\tau)$, which we subsequently make stationary with respect to ${\LocFr}$ 
and $p(\tau)$.

The next step in computing 
$\big[
\tilde{Z}\big(\{i_{e},s_{e};
        i_{e}^{\prime},s_{e}^{\prime}
      \}_{e=1}^{M}
\big)^{n}
\big]$ 
concerns the evaluation of the determinant on the replica indices, $\detn$ 
in \eqref{EQ:gdfFE}. This can readily be accomplished provided we make the 
assumption that we only consider values of $\Omega_{\hat{k}}$ that are 
invariant under the permutation of replica indices (\ie, are 
replica-symmetric).  This is indeed the case for the form hypothesised 
in Sec.~\ref{SEC:OPhypoth}.  With this restriction, $\greenNI{k}$ takes 
the form
\beq
\green{k}=\imatn\bigh{k}+\left(1-\imatn\right)\lith{k}
\label{EQ:rs_form}, 
\eeq
which, for a given value of ${\bf k}$, has eigenvalues
\bml
\bea
\bigh{k}+n\lith{k}
&&
\quad
\hbox{with degeneracy }1;
\\
\bigh{k}-\phantom{n}\lith{k}
&&
\quad
\hbox{with degeneracy }n,
\eea
\eml so that $\imatnNI+\tlns\density\greenNI{k}$ has eigenvalues
\bml
\bea
1+\tlns\density(\bigh{k}+n\lith{k})
&&
\quad
\hbox{with degeneracy }1;
\\
1+\tlns\density(\bigh{k}-\phantom{n}\lith{k})
&&
\quad
\hbox{with degeneracy }n.
\eea
\eml Hence, we find that the determinant 
in \eqref{EQ:gdfFE} is given by
\beq
\detn
\left(
{
\imatnNI+\tlns\density\greenNI{k}
\over{
1+\lambda^{2}\density\debyeZ{k}
}}
\right)
=
{
\Big(
1+\tlns\density(\bigh{k}+n\lith{k})
\Big)
\Big(
1+\tlns\density(\bigh{k}-\lith{k})
\Big)^{n}
\over{
\Big(
1+\lambda^{2}\density\debyeZ{k}
\Big)^{n+1}
}}.
\label{EQ:detnzn}
\eeq
It should be noted that this expression has implicit dependence on $n$
through $\bigh{k}$, $\lith{k}$ and $\tlns$ but not through $\debyeZ{k}$.

Our next step involves making use of the fact that we shall ultimately
be taking the replica limit, $n\rightarrow 0$.  With this in mind, we
make the following expansions, valid for small $n$:
\bml
\bea
\bigh{k}&=&\bighz{k}+n\bighw{k}+{\cal O}(n^{2}),
\label{EQ:bighexp}
\\
\lith{k}&=&\lithz{k}+n\lithw{k}+{\cal O}(n^{2}),
\label{EQ:lithexp}
\\
\tlns&=&\tlzs+n(\mu^{2}/\density)\ln V+{\cal O}(n^{2}),
\label{EQ:tlnexp}
\eea
\eml and note that $\bighz{k}=\debyeZ{k}$ (as shown in App.~\ref{APP:FluCo}).
By inserting these expansions into \eqref{EQ:detnzn}, taking the logarithm, and 
expanding for small $n$ we obtain
\bea
&&
\ln\detn
\left(
{
\imatnNI+\tlns\density\greenNI{k}
\over{
1+\lambda^{2}\density\debyeZ{k}
}}
\right)
-
\ln\left(
{1+\tlzs\density\debyeZ{k}
\over{1+\lambda^{2}\density\debyeZ{k}}}
\right)
\nonumber
\\
&&
\qquad
=
n\left(
{\density\tlzs(\lithz{k}+\bighw{k})+\mu^{2}\bighz{k}\ln V
\over{1+\tlzs\density\bighz{k}}}
\right)
+n\ln
\left(
{1+\tlzs\density(\bighz{k}-\lithz{k})
\over{1+\lambda^{2}\density\debyeZ{k}}}
\right)
+{\cal O}(n^{2}).
\label{EQ:RepLogDet}
\eea

Next, we insert \eqref{EQ:RepLogDet} into \eqref{EQ:gdfFE}, 
thus obtaining 
\bea
nd\calfgdf\big(\{\Omega_{\hat{k}}\}\big)
&\approx&
\frac{1}{2}n\mu^{2}\ln V 
+
\frac{\mu^{2}}{V^{n}}
\sumhrsHP{k}
\big\lefver
\Omega_{{\hat k}}
\big\rigver^{2}
\nonumber
\\
&&
\qquad
-\ln
\Big\langle
\exp
\Big(
\frac{2\mu^{2}}{V^{n}}
\sumhrsHPNL{k}
\real
\Omega_{\hat{k}}^{\ast}
\int\nolimits_{0}^{1}ds\,
\exp
\big(
i{\hat{k}}\cdot{\hat{c}}(s)
\big)
\Big)
\Big\rangle_{n+1}^{\rm W}
\nonumber
\\
&&
\qquad\qquad\quad
+\frac{n}{\deninve V}
\sumzrsHP{k}
\left(
{\density\tlzs(\lithz{k}+\bighw{k})+\mu^{2}\debyeZ{k}\ln V
\over{1+\tlzs\density\debyeZ{k}}}
\right)
\nonumber
\\
&&
\qquad\qquad\qquad\qquad
+\frac{n}{\deninve V}
\sumzrsHP{k}
\ln
\left(
{1+\tlzs\density(\debyeZ{k}-\lithz{k})
\over{1+\lambda^{2}\density\debyeZ{k}}}
\right)
+{\cal O}(n^{2}),
\label{EQ:spFEgdf}
\eea
which is to be made stationary with respect to $\Omega_{{\hat k}}$ of
the form (\ref{EQ:HRSopForm}) on which $\calfgdf$ depends both
explicitly, and also implicitly, through $\lithz{k}$ and $\bighw{k}$.
It should be noted that in obtaining this expression for
$\calfgdf\big(\{\Omega_{\hat{k}}\}\big)$ we have made use of the
result, established in App.~\ref{APP:FluCo}, that
$\bighz{k}=\debyeZ{k}$, without which the limit $n\rightarrow 0$ would
not exist.

We now evaluate $\calfgdf\big(\{\Omega_{\hat{k}}\}\big)$ for
$\Omega_{\hat{k}}$ given by \eqref{EQ:HRSopForm}.  The second and third
contributions to the right side of \eqref{EQ:spFEgdf} are mean-field
contributions, which have already been discussed in
Sec.~\ref{SEC:StatPtCrit}, and which are computed in
App.~\ref{APP:EHEOPH}. The fourth and fifth contributions are
fluctuation corrections, resulting from the incorporation of 1-replica
sector density fluctuations: it is upon these that we now focus.  By
reorganising \eqref{EQ:spFEgdf} and omitting contributions that are
independent of both the gel fraction ${\LocFr}$ and the distribution $p(\tau)$
we obtain
\bml
\bea
\fgdf\{{\LocFr},p\}
&\equiv&
\lim_{n\to 0}
\calfgdf\big(\{\Omega_{\hat{k}}\}\big)
\\
&=&
\fspa\{{\LocFr},p\}
+\frac{1}{d\denfrac V}
\sumzrsHP{k}
{\bighw{k}\over{1+\sigz\debyeZ{k}}}
+\frac{\sigz}{d\denfrac V}
\sumzrsHP{k}
{\lithz{k}\over{1+\sigz\debyeZ{k}}}
\nonumber
\\
&&
\qquad\qquad
+\frac{1}{d\denfrac V}
\sumzrsHP{k}\ln
\left(
1-{\sigz\lithz{k}\over{1+\sigz\debyeZ{k}}}
\right), 
\label{EQ:gdfFnoexp}
\eea
\eml where $\sigz\equiv\tlzs N/V\equiv\tlzs\density$.  Next, we recall 
that from \eqref{EQ:erdos} we know that ${\LocFr}$ is small in the vicinity of 
the transition. This, together with the observation that terms for which 
${\bf k}={\bf 0}$ are to be omitted from the fluctuation contributions, 
and that $\lithz{k}\sim {\LocFr}$ (from \eqref{EQ:litHZacc} of App.~\ref{APP:FluCo}), 
shows that we may expand the last term in \eqref{EQ:gdfFnoexp} as a power 
series in $\lithz{k}$, which yields
\bea
\fgdf\{{\LocFr},p\}
&=&
\fspa\{{\LocFr},p\}
+\frac{1}{d\denfrac V}
\sumzrsHP{k}
{\bighw{k}\over{1+\sigz\debyeZ{k}}}
-\frac{1}{2d\denfrac V}
\sumzrsHP{k}
\left({\sigz\lithz{k}\over{1+\sigz\debyeZ{k}}}\right)^{2}
\nonumber
\\
&&
\qquad\qquad\qquad
-\frac{1}{3d\denfrac V}
\sumzrsHP{k}
\left({\sigz\lithz{k}\over{1+\sigz\debyeZ{k}}}\right)^{3}
+{\cal O}({\LocFr}^{4}).
\label{EQ:gdfFexp}
\eea

We now examine the fluctuation corrections in \eqref{EQ:gdfFexp} in detail. 
Equation~(\ref{EQ:litHZacc}) shows that $\lithz{k}$ contains factors of the 
form $\exp\big({-k^{2}/\tau}\big)$.  For localisation length distributions 
$p(\tau)$ that give appreciable weight only to $\tau\sim\epsilon$, we can 
assert that, where present, such factors effectively eliminate contributions 
from wave vectors larger in magnitude than $\sqrt{\epsilon}$.  Thus, the two 
terms in \eqref{EQ:gdfFexp} that involve $\lithz{k}$ respectively scale 
with $\epsilon$ in 
$d$ dimensions as $\epsilon^{2+d/2}$ and $\epsilon^{3+d/2}$, and thus are 
negligible for $d>2$ and $d>0$, respectively.  On the other hand, $\bighw{k}$ 
contains no such factors.  In fact, the factorisation of $\bighw{k}$ into 
$\tau$- and ${\bf k}$-dependent pieces exhibited in \eqref{EQ:bigHWacc} 
enables us to identify the contribution to \eqref{EQ:gdfFexp} associated 
with $\bighw{k}$ as a renormalisation of the coefficient of one particular 
term in $\fspa\{{\LocFr},p\}$:
\bml
\beq
\frac{1}{12}
\mu^{4}{\LocFr}^{2}
\ltave
\frac{\tau_{1}\tau_{2}}{\tau_{1}+\tau_{2}}
\rtave
\rightarrow
(1+{\cal E})
\frac{1}{12}
\mu^{4}{\LocFr}^{2}
\ltave
\frac{\tau_{1}\tau_{2}}{\tau_{1}+\tau_{2}}
\rtave,
\label{EQ:fgdf_renorm}
\eeq
where the renormalisation parameter ${\cal E}$ is given by 
\beq
{\cal E}
\equiv
\frac{6\sigz}{d\denfrac}
\frac{1}{V}
\sumzrsHP{k}
k^{2}{\debyeW{k}-\debyeT{k}
\over{1+\sigz\debyeZ{k}}}.
\label{EQ:coeff_renorm}
\eeq
\eml Now, in App.~\ref{APP:debye}, the functions $\debyeZ{k}$, $\debyeW{k}$ 
and $\debyeT{k}$ are computed.  From the asymptotic properties of these 
functions we find that the renormalisation of the coefficient ${\cal E}$ is 
finite for $-2<d<4$, it being infra-red divergent for $d<-2$ and ultra-violet 
divergent for $d>4$. Thus, we find that in the physically relevant 
three-dimensional case the 1-replica-sector gaussian density fluctuations 
simply give a finite renormalisation of a coefficient in the variational  
free energy. 

This renormalisation has no effect on the behaviour of ${\LocFr}$ as a function
of $\mu^{2}$, which undergoes precisely the transition at $\mu^{2}=1$
that it undergoes at the mean-field level.  Moreover, the only effect
of this renormalisation on $p(\tau)$ as a function of $\mu^{2}$, in the
vicinity of the transition at $\mu^{2}=1$, is the finite rescaling:
$p(\tau)=(2/\epsilon)\pi(2\tau/\epsilon)\rightarrow
(2/\epsilon^{\prime})\pi(2\tau/\epsilon^{\prime})$, where
$\epsilon^{\prime}\equiv\epsilon/(1+{\cal E})$. 
However, it must also be recognised that the relationship between the
control parameter $\mu^{2}$ and the mean number of crosslinks per
macromolecule $[M]$ is modified by the fluctuation corrections. Thus,
once the typical localisation length is rescaled, the properties of the
system are as found in Sec.~\ref{SEC:Character}.  Note, however, that
although the location of the transition as measured by $\mu^{2}$ is
unaltered, the relationship between $\mu^{2}$ and $[M]/N$ is changed by
the incorporation of fluctuations; therefore, the location of the
transition as measured by $[M]$ is altered.
\section{Concluding remarks}\label{SEC:Conclude}
As Charles Goodyear discovered in 1839 \cite{REF:Good}, the
introduction of a sufficient density of permanent, random crosslinks
causes a macromolecular liquid to be transformed into an amorphous
solid, in which a nonzero fraction of macromolecules have spontaneously
become localised.  This system is a disordered system, in the sense
that it comprises both thermally equilibrating  as well as quenched
random variables (\ie, the crosslink locations).  In the present
article we have presented a theoretical description of the physical
properties of systems of macromolecules that have been randomly
crosslinked.  Our focus has been on the equilibrium properties of such
systems, especially in the regime of the vulcanisation transition.

The qualitative picture of randomly crosslinked macromolecular systems
that emerges has the following primary features.  For sufficiently few
crosslinks, the equilibrium state of the system is liquid, thermal
fluctuations causing all macromolecules to wander throughout the entire
container given sufficient time (\ie, all macromolecules are
delocalised).  For sufficiently many crosslinks, the equilibrium state
of the system is an amorphous solid state. In this state a nonzero
fraction of macromolecules are self-consistently localised, exhibiting
thermal fluctuations in location only over a certain distance scale,
which we refer to as a localisation length and which varies randomly
from monomer to monomer.  At a critical density of crosslinks, of order
one per macromolecule, there is a continuous thermodynamic phase
transition from the liquid state to the equilibrium amorphous solid
state.  The amorphous solid state is most unusual: the mean positions
of the localised monomers are homogeneously random, exhibiting no
periodicity whatsoever.  The state is characterised by the fraction of
monomers that are localised, together with the statistical
distribution of their localisation lengths. The symmetry properties of
the amorphous solid state are striking. Microscopically, the amorphous
solid state is one in which translational symmetry is spontaneously
broken, certain monomers becoming localised about fixed mean positions,
in contrast with the liquid state.  Macroscopically, however, the
amorphous solid state retains the translational symmetry of the liquid
state, owing to the homogeneous randomness of the mean positions of the
localised monomers.  (Technically, this shows up as the macroscopic
translational invariance of the value of the order parameter in the
amorphous solid state.)\thinspace\  This state bears the same
relationship to the liquid and crystalline states as the spin glass
state of certain magnetic systems bears to the paramagnetic and
ferromagnetic states, in the following sense:  the local static density
fluctuations in the amorphous solid state correspond to local static
magnetisation fluctuations in the spin glass state.  In both cases
these fluctuations vanish, if averaged over the entire sample, and
hence cannot serve as a global order parameter. Thus, moments of the
static fluctuations higher than the first must be considered.  In fully
connected long-range spin glasses the distribution of these static
fluctuations is gaussian (in the replica symmetric theory), so that the
second moment---the Edwards-Anderson order parameter---characterises
the state of the system completely.  In contrast, in the amorphous
solid state under consideration here, we find that all moments are
equally important. Such non-gaussian statistics are also encountered in
strongly diluted long-range spin glasses \cite{REF:Viana}, which show a
percolation transition as described by \eqref{EQ:erdos}. 

To construct our picture of their physical properties, we have
developed a field-theoretic representation of the statistical mechanics
of randomly crosslinked macromolecular systems.  The order parameter
capable of distinguishing between the various candidate states (liquid,
amorphous, crystalline solid and globular) naturally emerges from this
representation.  The presence of quenched as well as annealed variables
has been addressed by invoking the replica technique.  We have derived
the saddle-point equation from the effective hamiltonian of this
field-theoretic representation, this equation being equivalent to the
self-consistent mean-field equation satisfied by the order parameter.
Whilst it is not apparent how one might obtain the most general
solution for the order parameter, we have proposed a
physically-motivated form for it, which allows for the possibilities of
a liquid state and an amorphous solid state.  This form is parametrised
by the fraction of localised monomers, together with the statistical
distribution of localisation lengths.  In fact, this form turns out to
yield an exact solution of the saddle-point equation.  It should be
noted that we are only able to proceed with the calculation in the
vicinity of the amorphous solidification transition, where the typical
localisation length is substantially larger than the radius of gyration
of an isolated non--self-interacting macromolecule.  In a refinement of
this approach, we have incorporated a certain physically relevant class
of correlations, associated with macromolecular repulsion, at the
gaussian level of approximation.  No qualitative changes in our results
stem from this refinement.

The quantitative picture of randomly crosslinked macromolecular systems
that emerges from our field-theoretic representation has the following
primary elements.  At the mean-field level of approximation there is,
for any crosslink density, a saddle point of the effective hamiltonian
that corresponds to the liquid state.  However, for crosslink densities
greater than a certain critical value, this liquid state is unstable.
At this critical crosslink density a new saddle point of the effective
hamiltonian bifurcates continuously from the liquid-state saddle
point.  This new saddle point, which is characterised by a nonzero
gel-fraction and a specific distribution of localisation lengths,
corresponds to the amorphous solid state. The transition between the
liquid and amorphous solid states is therefore continuous: in
particular, the gel fraction and the inverse-square of the typical
localisation length both increase from zero linearly with the excess of
the crosslink density from its critical value.  Moreover, the entire
distribution of (inverse-square) localisation lengths has a scaling
form determined by a universal function of a single variable (which
only need be computed once for all near-critical crosslink densities).
Detailed results for the gel fraction and the distribution of
localisation lengths have been given.  Thus, we see that what emerges
is a quantitative picture of the amorphous solid state that confirms
the qualitative picture proposed roughly a decade ago.

There are several other contexts in which one can make use of the
circle of ideas that we have been using to explore the physical
properties of randomly crosslinked macromolecular systems.  First, one
can apply them to a wide class of random-network--forming systems.
Indeed, a straightforward extension \cite{REF:Roos} of the present work
yields a theory of randomly crosslinked manifolds (\ie,
higher-dimensional analogues of linear macromolecules
\cite{REF:KaKaNe,REF:Jeru}).  Similarly, one can address macromolecular
networks formed via a random {\it endlinking\/} (rather than
crosslinking) process, in which one end from each of several randomly
selected macromolecules are linked to one other \cite{REF:EndLink}.
One can also consider networks formed via the (freely-jointed)
endlinking of rigid or semi-flexible rods \cite{REF:EndLink}, subjects
that are of particular relevance to certain biological structures.

This circle of ideas has also been used to develop a
statistical-mechanical theory of continuous random (atomic or molecular)
networks, and thus to develop a view of the structural glass
transition \cite{REF:StrGla}.  In this case, what emerges is a picture
of glass-formation in which atomic or molecular units are (permanently
chemically) bonded together at random, so as to develop an infinite
network.  Not only do the translational freedoms of the units become
localised but also do the orientational freedoms.

A particularly interesting problem arises in the context of NMR data on
protein-folding. The basic question is this: How much phase space is
reduced due to a given number of constraints on the relative positions
of certain monomer-pairs \cite{REF:SGSnmr}?  We also mention the
formal analogy between such questions and Gardner-type calculations in
the field of neural networks \cite{REF:Hertz}.  Here, again, the basic
issue is the reduction of phase space by quenched random constraints,
even the distribution of randomness being generated in close analogy
with the Deam-Edwards distribution, which leads to the same formal
construction of $(n+1)$-fold replicated systems.

Finally, we raise the fascinating and difficult issue of the dynamical
properties of randomly crosslinked macromolecular systems.  In
particular, it would be interesting to develop a semi-microscopic
theory of the viscosity of the liquid state, and its divergence at the
onset of solidification, as well as of the dynamics of the solid state
itself.  Our semi-microscopic approach is well-suited to the
application of the dynamical lagrangian methods \cite{REF:GZDynUn} that
have been developed to address the dynamics of spin glasses
\cite{REF:SoZi}.

\acknowledgments
It is a pleasure for us to thank 
Nigel Goldenfeld and Grzegorz Szamel
for many stimulating and useful conversations. 
We gratefully acknowledge support from 
the U.S.~National Science Foundation through grants 
DMR91-57018 (PMG, HEC), and	
DMR94-24511 (PMG, HEC), 	
from NATO through Collaborative Research Grant 940909 (PMG, AZ), and 
from the Deutsche Forschungsgemeinschaft through the 
Sonderforschungsbereich 345 (AZ).
HEC gratefully acknowledges support from a Graduate Fellowship
at the University of Illinois at Urbana-Champaign.
PMG thanks for its hospitality 
the Instit{\"u}t fur Theoretische Physik at
Georg August Universit{\"a}t, G{\"o}ttingen, 
where parts of the work reported here were undertaken.
AZ thanks for its hospitality 
the Department of Physics at the 
University of Illinois at Urbana-Champaign
where other parts of the work reported here were undertaken. 
PMG wishes to use this occasion to record his sincere gratitude to 
Sir Sam Edwards, Phil Pincus and David Sherrington for the crucial 
opportunities that they have graciously provided for him.

\appendix               
\section{Replica representation of disorder-averaged observables}\label{APP:RepAndObs}
In this appendix we demonstrate how to evaluate the disorder average of
products of equilibrium expectation values by using the replica
technique. This complements the demonstration, given in
Sec.~\ref{SEC:ReThPhiFE}, of the connection between the
disorder-averaged free energy $\tilde{f}$ and the replica-Helmholtz free
energy $\fgenf$.

Consider the disorder average of the product of $g$ equilibrium
expectation values of physical observables, \ie, 
$[\langle{\cal O}_{1}\rangle_{\disfac}\ldots\langle{\cal O}_{g}\rangle_{\disfac}]$.
(The meaning of the expectation value $\langle\cdots\rangle_{\disfac}$
is discussed in Sec.~\ref{SEC:GeneralProps}, immediately after
\eqref{EQ:OPdefNDA}.)\thinspace\ In terms of sums over disorder
realisations and microscopic configurations this quantity is given by
\beq
[\langle{\cal O}_{1}\rangle_{\disfac}\ldots\langle{\cal O}_{g}\rangle_{\disfac}]
=\left[
{
{\syfac}^{-1}\trfac{1}{\rm e}^{-\achfac{1}}\delfac{1}{\cal O}_{1}
\cdots
{\syfac}^{-1}\trfac{g}{\rm e}^{-\achfac{g}}\delfac{g}{\cal O}_{g}
\over{
{\syfac}^{-1}\trfac{1}{\rm e}^{-\achfac{1}}\delfac{1}
\cdots
{\syfac}^{-1}\trfac{g}{\rm e}^{-\achfac{g}}\delfac{g}
}}
\right].
\eeq
Here,
$\trfac{1}$ denotes the trace (\ie, sum) over the configurations one
copy (or replica) of the system, without regard for distinguishability
(\ie, without dividing by any symmetry factor), and $\trfac{2}$ denotes
a similar trace over another copy, \etc.
The factor $\syfac(\equiv\prod_{a}\nu_{a}!)$ denotes the exact 
indistinguishability factor appropriate for the specific realisation of 
the disorder and, accordingly, takes on values between $N!$ and unity.
$\delfac{1}$ implements the constraints on the first copy of the system,
being a factor that is unity for configurations of the first copy that
satisfy the constraints $\disfac$, and zero otherwise.  $\delfac{2}$
indicates a similar factor for the second copy, \etc.
The factor $\exp\big(-\achfac{1}\big)$ indicates the weight for
configurations of the first copy, and similarly the factor
$\exp\big(-\achfac{2}\big)$ indicates the weight for the second copy,
\etc. We now introduce a further $(n-g)$ factors
${\syfac}^{-1}\trfac{}{\rm e}^{-\achfac{}}\delfac{}$ in both the
numerator and denominator, which gives
\beq
[\langle{\cal O}_{1}\rangle_{\disfac}\ldots\langle{\cal O}_{g}\rangle_{\disfac}]
=
\left[
{\DPS
{\DPS
{\syfac}^{-1}\trfac{1}{\rm e}^{-\achfac{1}}\delfac{1}{\cal O}_{1}
\cdots
{\syfac}^{-1}\trfac{g}{\rm e}^{-\achfac{g}}\delfac{g}{\cal O}_{g}
\hfill\atop{\DPS\qquad
\hfill\times
{\syfac}^{-1}\trfac{g+1}{\rm e}^{-\achfac{g+1}}\delfac{g+1}
\cdots
{\syfac}^{-1}\trfac{n}{\rm e}^{-\achfac{n}}\delfac{n}
}}
\over{\DPS
{\DPS
\phantom{x}
\atop{\DPS
{\syfac}^{-1}\trfac{1}{\rm e}^{-\achfac{1}}\delfac{1}
\cdots
{\syfac}^{-1}\trfac{n}{\rm e}^{-\achfac{n}}\delfac{n}
}}
}}
\right].
\eeq
In principle, this quantity is defined only for integral $n\ge g$, for
which values of $n$ it is constant. However, in the replica approach we
regard $n$ as a continuous variable, and formally determine this
constant value by computing its limit as $n\rightarrow 0$.  Thus, we
obtain
\bea
&&
[\langle{\cal O}_{1}\rangle_{\disfac}\ldots\langle{\cal O}_{g}\rangle_{\disfac}]
=
\\
\noalign{\medskip}
&&
\quad
\lim_{n\to 0}
\left[
\trfac{1}{\rm e}^{-\achfac{1}}\delfac{1}{\cal O}_{1}
\cdots
\trfac{g}{\rm e}^{-\achfac{g}}\delfac{g}{\cal O}_{g}
\trfac{g+1}{\rm e}^{-\achfac{g+1}}\delfac{g+1}
\cdots
\trfac{n}{\rm e}^{-\achfac{n}}\delfac{n}
\right], 
\nonumber
\eea
where we have used the fact that 
$\lim_{n\to 0}(\trfac{}{\rm e}^{-\achfac{}}\delfac{})^{n}=1$. 
As we have seen in Sec.~\ref{SEC:DECrossDist}, in the context of macromolecular
networks the disorder distribution is generated by the partition
function itself, which is represented via the inclusion of an
additional replica.  In fact, it turns out that the naturally emerging
order parameter involves the expectation value of variables associated
with this additional replica, as we see in Sec.~\ref{SEC:OPFields}.
Such expectation values may be regarded as introducing additional
$\disfac$-dependent factors into our disorder averages, \eg,
\bea
&&
[
\Theta(\disfac)
\langle{\cal O}_{1}\rangle_{\disfac}\ldots\langle{\cal O}_{g}\rangle_{\disfac}]
=
\label{EQ:withzero}
\\
\noalign{\medskip}
&&
\quad
\lim_{n\to 0}
\left[
\Theta(\disfac)
\trfac{1}{\rm e}^{-\achfac{1}}\delfac{1}{\cal O}_{1}
\cdots
\trfac{g}{\rm e}^{-\achfac{g}}\delfac{g}{\cal O}_{g}
\trfac{g+1}{\rm e}^{-\achfac{g+1}}\delfac{g+1}
\cdots
\trfac{n}{\rm e}^{-\achfac{n}}\delfac{n}
\right].
\nonumber
\eea
By using Eq.~(\ref{EQ:withzero}) and the Deam-Edwards distribution, 
Eq.~(\ref{EQ:distribute}), in a manner analogous to that used in the 
context of the free energy in Sec.~\ref{SEC:ReThPhiFE}, 
we arrive at the result
\FL
\beq
[\Theta(\disfac)
\langle{\cal O}_{1}\rangle_{\disfac}\ldots\langle{\cal O}_{g}\rangle_{\disfac}]
=
\lim_{n\to 0}
{\DPS
{\DPS
\Big\langle
{\cal O}_{0}\{{\bf c}_{i}^{0}(s)\}
{\cal O}_{1}\{{\bf c}_{i}^{1}(s)\}\cdots
{\cal O}_{g}\{{\bf c}_{i}^{g}(s)\}
\hfill
\atop{\DPS
\quad
\hfill
\times
\exp
        \Big( 
\frac{\mu^{2}V}{2N}
\sum_{i,i^{\prime}=1}^{N}
\int\nolimits_{0}^{1}ds
\int\nolimits_{0}^{1}ds^{\prime}
\prod_{\alpha=0}^{n}
\delta^{(d)}
\big(
     {\bf c}_{i         }^{\alpha}(s         )
    -{\bf c}_{i^{\prime}}^{\alpha}(s^{\prime})
\big)
        \Big)
\Big\rangle_{n+1}^{\rm E}
}}
\over{\DPS
\Big\langle
\exp
        \Big( 
\frac{\mu^{2}V}{2N}
\sum_{i,i^{\prime}=1}^{N}
\int\nolimits_{0}^{1}ds
\int\nolimits_{0}^{1}ds^{\prime}
\prod_{\alpha=0}^{n}
\delta^{(d)}
\big(
     {\bf c}_{i         }^{\alpha}(s         )
    -{\bf c}_{i^{\prime}}^{\alpha}(s^{\prime})
\big)
        \Big)
\Big\rangle_{n+1}^{\rm E}
}},
\label{EQ:newsource}
\eeq
where $\langle{\cal O}_{0}\rangle_{\disfac}=\Theta(\disfac)$.

We now focus on a particularly important example of a disorder-averaged
observable, \viz, the amorphous solid order
parameter~\eqref{EQ:opDefinition}.  This quantity is obtained as the
special case of \eqref{EQ:newsource} for the choices:
\beq
\matrix{
\Theta(\disfac)
&=&
\langle\exp\big(i{\bf k}^{0}\cdot{\bf c}_{i}(s)\big)\rangle_{\disfac},
\cr
\noalign{\medskip}
{\cal O}_{1}&=&
\exp\big(i{\bf k}^{1}\cdot{\bf c}_{i}(s)\big),
\cr
\vdots&\vdots&\vdots\cr
{\cal O}_{g}&=&
\exp\big(i{\bf k}^{g}\cdot{\bf c}_{i}(s)\big).
\cr
}
\eeq
By using \eqref{EQ:newsource} for this case we see that 
\bml
\bea
&&
\left[
\frac{1}{N}\sumin\int_{0}^{1}ds\,
\langle\exp\big(i{\bf k}^{0}\cdot{\bf c}_{i}(s)\big)\rangle_{\disfac}
\langle\exp\big(i{\bf k}^{1}\cdot{\bf c}_{i}(s)\big)\rangle_{\disfac}
\cdots
\langle\exp\big(i{\bf k}^{g}\cdot{\bf c}_{i}(s)\big)\rangle_{\disfac}
\right]
\nonumber
\\
&&
\qquad
=
\lim_{n\to 0}
{\DPS
\Big\langle
{\FTDen}_{\hat{l}}\,
\exp
        \Big( 
\frac{\mu^{2}V}{2N}
\sum_{i,i^{\prime}=1}^{N}
\int\nolimits_{0}^{1}ds
\int\nolimits_{0}^{1}ds^{\prime}
\prod_{\alpha=0}^{n}
\delta^{(d)}
\big(
     {\bf c}_{i         }^{\alpha}(s         )
    -{\bf c}_{i^{\prime}}^{\alpha}(s^{\prime})
\big)
        \Big)
\Big\rangle_{n+1}^{\rm E}
\over{\DPS
\Big\langle
\exp
        \Big( 
\frac{\mu^{2}V}{2N}
\sum_{i,i^{\prime}=1}^{N}
\int\nolimits_{0}^{1}ds
\int\nolimits_{0}^{1}ds^{\prime}
\prod_{\alpha=0}^{n}
\delta^{(d)}
\big(
     {\bf c}_{i         }^{\alpha}(s         )
    -{\bf c}_{i^{\prime}}^{\alpha}(s^{\prime})
\big)
        \Big)
\Big\rangle_{n+1}^{\rm E}
}},
\label{EQ:ManyOpsDis}
\eea
\eml where ${\FTDen}_{\hat{k}}$ is defined in \eqref{EQ:density}, and 
$\hat{l}=\{
{\bf k}^{0},{\bf k}^{1},\ldots,
{\bf k}^{g},{\bf 0},    \ldots,{\bf 0}\}$. 
As was anticipated in Sec.~\ref{SEC:OPhypoth},
the order parameter probes not only the $g$ physical copies of the
system but also the additional copy used to generate the crosslink
distribution.  
\section{Wiener correlator}\label{APP:WMC}
In this appendix we derive the basic correlator associated 
with the Wiener measure:
\beq
\Big\langle
\exp
\Big(
-i\sumrhor{\bf k}_{\rho}\cdot{\bf c}(s_{\rho})
\Big)
\Big\rangle_{1}^{\rm W}
=
\delta_{{\bf 0},\sum\nolimits_{\rho=1}^{r}{\bf k}_{\rho}}^{(d)}
\exp
\Big(
-\frac{1}{2}
\sumrhorhopr
\smatr\,
{\bf k}_{\rho}\cdot{\bf k}_{{\rho}^{\prime}}
\Big),
\label{EQ:WMCresult}
\eeq
where $\smatr$ is a function of the pair of arclength 
coordinates $s_{\rho}$ and $s_{\rho^{\prime}}$ defined via
\beq
\smatr
\equiv
\min\left(
s_{\rho         },
s_{\rho^{\prime}}\right).
\label{EQ:smatdef}
\eeq
In terms of the Wiener measure, the correlator is given, up to 
normalisation, by
\beq
\Big\langle
\exp\Big(-i\sumrhor{\bf k}_{\rho}\cdot{\bf c}(s_{\rho})\Big)
\Big\rangle_{1}^{\rm W}
\propto
\int{\cal D}{\bf c}\,
\exp\Big({-\frac{1}{2}\int_{0}^{1}ds\vert{\dot{\bf c}}(s)\vert^{2}}\Big)
\exp\Big({-i\sumrhor{\bf k}_{\rho}\cdot{\bf c}(s_{\rho})}\Big),
\eeq
where the overdot denotes a derivative with respect to $s$.  We express 
the configuration of the macromolecule in terms of the position of the 
end at $s=0$, together with the tangent vector $\dot{\bf c}(s)$ via: 
${\bf c}(s)=
{\bf c}(0)+
\int_{0}^{1}ds^{\prime}
\dot{\bf c}(s^{\prime})
\theta(s-s^{\prime})$, 
where $\theta(s)$ is the Heaviside $\theta$-function. 
Then the measure ${\cal D}{\bf c}$ is given by 
${\cal D}\dot{\bf c}\,d{\bf c}(0)$, and the correlator becomes 
proportional to 
\beq
\int\!d{\bf c}(0)\,
\exp\Big({-i{\bf c}(s_{0})\cdot\!\sumrhor{\bf k}_{\rho}}\Big)
\int\!{\cal D}\dot{\bf c}\,
\exp\Big({-\frac{1}{2}\int\limits_{0}^{1}\!ds\vert{\dot{\bf c}}(s)\vert^{2}}\Big)
\exp\Big({-i\int\limits_{0}^{1}\!ds\,\dot{\bf c}(s)\cdot\!
\sumrhor{\bf k}_{\rho}\theta(s_{\rho}-s)}\Big).
\eeq
By performing the integral over ${\bf c}(0)$ we obtain the 
Kronecker $\delta$-function factor: 
$\delta_{{\bf 0},\sum\nolimits_{\rho=1}^{r}{\bf k}_{\rho}}^{(d)}$. 
By performing the integrals over the tangent vectors $\dot{\bf c}(s)$ we 
obtain the gaussian factor: 
\beq
\exp
\Big(
-\frac{1}{2}\sumrhorhopr
{\bf k}_{\rho}\cdot{\bf k}_{{\rho}^{\prime}}
\int_{0}^{1}ds\,ds^{\prime}
\theta(s_{\rho}-s)
\theta(s_{{\rho}^{\prime}}-s^{\prime})
\delta(s-s^{\prime})
\Big).
\eeq
By performing the arclength integrals, and by setting to zero the 
wave vectors $\{{\bf k}_{\rho}\}_{\rho=1}^{r}$ in order to establish 
the correct normalisation factor, we obtain the Wiener measure 
correlator \eqref{EQ:WMCresult}.  It should be noted that because
$\min\left(s_{\rho},s_{\rho^{\prime}}\right)=
 \frac{1}{2}(s_{\rho}+s_{\rho^{\prime}})
-\frac{1}{2}\vert{s_{\rho}-s_{{\rho}^{\prime}}}\vert$, 
and because of the Kronecker 
$\delta$-function factor, the exponent of the Wiener measure correlator
can also be expressed as 
$\sumrhorhopr
{\bf k}_{\rho}\cdot{\bf k}_{{\rho}^{\prime}}
\vert{s_{\rho}-s_{\rho^{\prime}}}\vert/4$.
\section{Debye function and related functions}\label{APP:debye}
In this appendix we give the basic Debye function $\debyeZ{k}$, 
as well as the two related functions $\debyeW{k}$ and $\debyeT{k}$.
The Debye function $\debyeZ{k}$ is defined as the integral over arclength 
variables of the Wiener correlator \eqref{EQ:WMCresult} for the case 
$r=2$ and $-{\bf k}_{1}={\bf k}_{2}={\bf k}$:
\bea
\debyeZ{k}
&\equiv&
\int_{0}^{1}ds_{1}ds_{2}\,
\Big\langle
{\rm e}^{i{\bf k}\cdot
\big({\bf c}(s_{1})-{\bf c}(s_{2})\big)}
\Big\rangle_{1}^{\rm W}
=\int_{0}^{1}ds_{1}ds_{2}\,
{\rm e}^{-k^{2}\vert{s_{1}-s_{2}}\vert/2}
\nonumber
\\
\noalign{\medskip}
&=&
{
{\rm e}^{-k^{2}/2}-\Big(1-\frac{1}{2}k^{2}\Big)
\over{
\frac{1}{2}\left(\frac{1}{2}k^{2}\right)^{2}
}}
\sim
\cases{1 - k^{2}/6,     &if $k^{2}\ll 1$;\cr
       4/k^{2}&if $k^{2}\gg 1$.\cr}
\eea The function $\debyeW{k}$ is defined as 
\beq
\debyeW{k}
\equiv
\int_{0}^{1}ds_{1}ds_{2}ds_{3}\,
{\rm e}^{-k^{2}\vert{s_{1}-s_{2}}\vert/2}
\big(
-\smatrCI{1,3}
+\smatrCI{2,3}
\big)^{2}
\sim
\cases{1/15,    &if $k^{2}\ll 1$;\cr
       16/k^{6}&if $k^{2}\gg 1$.\cr}
\eeq 
The function $\debyeT{k}$ is defined as 
\bea
\debyeT{k}
&\equiv&
\int_{0}^{1}ds_{1}ds_{2}ds_{3}\,
{\rm e}^{-k^{2}\vert{s_{1}-s_{2}}\vert/2}
\big(
-\smatrCI{1,3}
+\smatrCI{2,3}
\big)
\big(
-\smatrCI{1,4}
+\smatrCI{2,4}
\big)
\nonumber
\\
\noalign{\medskip}
&\sim&
\cases{2/45,    &if $k^{2}\ll 1$;\cr
       32/3k^{6}&if $k^{2}\gg 1$.\cr}
\eea In $d=3$  the relevant correction due to gaussian density fluctuations 
involves the quantity $\debyeW{k}-\debyeT{k}$, the asymptotic behaviour of 
which is given by
\beq
\debyeW{k}-\debyeT{k}
\sim
\cases{1/45,    &if $k^{2}\ll 1$;\cr
       16/3k^{6}&if $k^{2}\gg 1$.\cr}
\eeq
\section{Effective hamiltonian evaluated for the order 
	parameter hypothesis}\label{APP:EHEOPH}
\subsection{Quadratic contribution}\label{APP:QuadContr}
We now compute the contribution in the $n\rightarrow 0$ limit to the
effective hamiltonian, \eqref{EQ:hamnumer}, that is quadratic in
$\Omega_{\hat k}$, for the specific form of $\Omega_{\hat k}$ given in
\eqref{EQ:OPhypothesis}. No approximations will be made. To this end we
focus on the quantity
\beq
\lim_{n\to 0}
\frac{1}{n}
\frac{2}{V^{n}}
\sumhrsHP{k}\lefver\Ohat{k}\rigver^{2}.
\eeq
Inserting $\Omega_{\hat k}$ from 
\eqref{EQ:OPhypothesis}, we obtain
\bml
\bea
\frac{2}{V^{n}}
\sumhrsHP{k}\lefver\Ohat{k}\rigver^{2}
&=&
\frac{1}{V^{n}}
\sumhrsNP{k}\lefver\Ohat{k}\rigver^{2}
\nonumber
\\
&=&
\frac{1}{V^{n}}
\sumhrsNP{k}
{\LocFr}^{2}
\dinttau{1}\dinttau{2}\,
\kdelvecT{k}{0}
\exp
\Big(
-{\hat{k}^{2}}
\big(\tau_{1}^{-1}+\tau_{2}^{-1}\big)/2
\Big)
\nonumber
\\
&=&
{\LocFr}^{2}
\int\limits_{0}^{\infty}d\tau
\dinttau{1}\dinttau{2}\,
\delta
\Big(
\tau-
\big(
 \tau_{1}^{-1}
+\tau_{2}^{-1}
\big)^{-1}
\Big)
\nonumber
\\
&&
\qquad\qquad\qquad\qquad
\qquad\qquad\qquad
\times
\frac{1}{V^{n}}
\sumhrsNP{k}\,\kdelvecT{k}{0}\,
\exp\left(
-{\hat{k}^{2}}/{2\tau}
\right).
\eea 
We now add and subtract the terms 
in the $0$- and $1$-replica sectors 
in order to relax the constraint on the summation over $\hat{k}$ 
indicated by the overbar on the summation symbol. 
In fact, owing to the factor of $\kdelvecT{k}{0}$, the summand 
vanishes for $\hat{k}$ in the $1$-replica sector, so that we obtain 
\bea
\frac{2}{V^{n}}
\sumhrsHP{k}
\lefver
\Ohat{k}
\rigver^{2}
&=&
-\frac{{\LocFr}^{2}}{V^{n}}
+{\LocFr}^{2}
\int\limits_{0}^{\infty}
d\tau
\dinttau{1}\dinttau{2}\,
\delta
\Big(
\tau-
\big(
 \tau_{1}^{-1}
+\tau_{2}^{-1}
\big)^{-1}
\Big)
\nonumber
\\
&&
\qquad\qquad\qquad\qquad\qquad
\qquad\qquad
\times
\frac{1}{V^{n}}
\sumhat{k}\kdelvecT{k}{0}
\exp
\left(
-\hat{k}^{2}/2\tau
\right).
\label{EQ:relax}
\eea
\eml We now focus on the remaining, unconstrained summation on 
the right hand side of \eqref{EQ:relax}, which we compute 
via the following steps. First, we introduce an integral 
representation of $\kdelvecT{k}{0}$, \viz.,
\beq
\kdelvecT{k}{0}=
\frac{1}{V}
\int\nolimits_{V}d\lamvec
\exp\Big(i\lamvec\cdot\sum\nolimits_{\alpha=0}^{n}{\bf k}^{\alpha}\Big),
\eeq
where the integral is taken over the volume $V$.  Then we convert 
the summation over $\hat{k}$ into an integral by using 
\beq
\frac{1}{V^{n+1}}
\sumhat{k}\cdots\rightarrow 
\int\dbar\hat{k}\cdots,
\label{EQ:sumtoint}
\eeq
where $\dbar\hat{k}\equiv\prod_{\alpha=0}^{n}\dbar{\bf k}^{\alpha}$ 
and $\dbar{\bf k}\equiv (2\pi)^{-d}d{\bf k}$, and the integral is 
taken over the entire $\hat{k}$ space.  Thus we obtain
\bea
\frac{1}{V^{n}}
\sumhat{k}\kdelvecT{k}{0}
\exp\Big(
-{\hat{k}^{2}}/{2\tau}
    \Big)
&=&
\int\dbar\hat{k}
\exp\Big(
-{\hat{k}^{2}}/{2\tau}
    \Big)
\int\nolimits_{V}d\lamvec
\exp\Big(
i\lamvec\cdot\sum\nolimits_{\alpha=0}^{n}{\bf k}^{\alpha}
     \Big)
\nonumber
\\
&=&
\int\nolimits_{V}d\lamvec
        \left\{
\int\dbar{\bf k}
\exp\Big(
-{\bf k}\cdot{\bf k}/{2\tau}
     \Big)
\exp\Big(
i{\bf k}\cdot\lamvec
     \Big)
        \right\}^{n+1}
\nonumber
\\
&=&
\big(\tau/2\pi\big)^{(n+1)d/2}
\int\nolimits_{V}d\lamvec
\exp\Big(
-(n+1)\tau\lamvec\cdot\lamvec/2
    \Big)
\nonumber
\\
&=&
\big(\tau/2\pi\big)^{nd/2}
(1+n)^{-d/2},
\eea where we have used the gaussian integral and, in the 
last step, have assumed that $\tau^{-1/2}\ll V^{1/d}$ 
for inverse square localisation lengths that are given significant 
weight by the distribution $p(\tau)$, \eqref{EQ:OPhypothesis}. 
Thus we find 
\bea
\frac{2}{V^{n}}
\sumhrsHP{k}\lefver\Ohat{k}\rigver^{2}
&=&
-\frac{{\LocFr}^{2}}{V^{n}}
+
{\LocFr}^{2}\int_{0}^{\infty}d\tau
\dinttau{1}\dinttau{2}\,
\delta
        \Big(
\tau-
                \big(
 \tau_{1}^{-1}
+\tau_{2}^{-1}
                \big)^{-1}
        \Big)
\nonumber
\\
&&
\qquad\qquad\qquad\qquad\qquad\qquad\qquad
\times
\big(\tau/2\pi\big)^{nd/2}
(1+n)^{-d/2}.
\label{EQ:quad_any_n}
\eea As we shall need this quadratic term only in the vicinity of 
$n=0$, we expand for small $n$ using \eqref{EQ:limitA}, 
and by taking the $n\rightarrow 0$ limit we obtain
\beq
\lim_{n\to 0}
\frac{1}{n}
\frac{2}{V^{n}}
\sumhrsHP{k}\lefver\Ohat{k}\rigver^{2}
={d\over{2}}{\LocFr}^{2} 
\dinttau{1}\dinttau{2}
\ln
\left(
\frac{V^{2/d}}{2\pi{\rm e}}
\left(\tau_{1}^{-1}+\tau_{2}^{-1}\right)^{-1}
\right).
\eeq
\subsection{Logarithmic contribution}\label{SEC:logtrace}
We now compute the contribution in the $n\rightarrow 0$ limit to the
effective hamiltonian, \eqref{EQ:hamnumer}, that can be identified and
the partition function of a single macromolecule coupled to
$\Omega_{\hat k}$ for the specific form of $\Omega_{\hat k}$ given in
\eqref{EQ:OPhypothesis}.  The calculation will be undertaken as a
perturbation expansion in the typical inverse square localisation
length, to first order in this quantity. Thus, we focus on the quantity
\beq
\Big\langle
\exp
        \Big(
2\mu^{2}V^{-n}
{\overline{\sum}}_{\hat{k}}^{\possym}
\real
\Omega_{\hat{k}}^{\ast}
\int\nolimits_{0}^{1}ds\,
{\rm e}^{i{\hat{k}}\cdot{\hat{c}}(s)}
        \Big)
\Big\rangle_{n+1}^{\rm W}.
\label{EQ:traceA}
\eeq 
First, we observe that $\Ohat{k}$ has only been introduced for 
$\hat{k}\cdot\hat{n}>0$, as follows from the discussion after 
\eqref{EQ:symmPartition}.  We are therefore free to introduce 
$\Ohat{k}$ for $\hat{k}\cdot\hat{n}<0$ at our 
convenience, and we do so via the definition
\beq
\Ohat{k}\equiv
\Omega_{-{\hat k}}^{\ast}
\qquad{\rm for\/}\quad\hat{k}\cdot\hat{n}<0.
\eeq 
This allows us to eliminate the $\real$ operation and to extend the 
range of the summation in \eqref{EQ:traceA}, which becomes
\beq
\Big\langle
\exp
                \Big(
\mu^{2}V^{-n}
{\overline{\sum}}_{\hat{k}}
\Omega_{\hat{k}}
\int\nolimits_{0}^{1}ds\,
{\rm e}^{-i{\hat{k}}\cdot{\hat{c}}(s)}
                \Big)
\Big\rangle_{n+1}^{\rm W}.
\label{EQ:trace_B}
\eeq
Next, we insert $\Omega_{\hat k}$ from 
\eqref{EQ:OPhypothesis}, which gives 
\beq
\Big\langle
\exp
\Big(
\mu^{2}{\LocFr} V^{-n}
{\overline{\sum}}_{\hat{k}}
\kdelvecT{k}{0}
\dint{\tau}
{\rm e}^{-\hat{k}^{2}/2\tau}
\int\nolimits_{0}^{1}ds\,
{\rm e}^{-i{\hat{k}}\cdot{\hat{c}}(s)}
\Big)
\Big\rangle_{n+1}^{\rm W}.
\label{EQ:trace_C}
\eeq
We now add and subtract the terms in the $0$- and $1$-replica 
sectors to the summation over $\hat{k}$ in order to relax the 
constraint indicated by the overbar on the summation symbol. 
In fact, owing to the factor of $\kdelvecT{k}{0}$, the summand 
vanishes for $\hat{k}$ in the $1$-replica sector, so that we obtain 
\beq
\exp\Big(-\mu^{2}{\LocFr} V^{-n}\Big)
\Big\langle
\exp
\Big(
\mu^{2}{\LocFr} V^{-n}
\sum\nolimits_{\hat{k}}
\kdelvecT{k}{0}
\dint{\tau}\,
{\rm e}^{-\hat{k}^{2}/2\tau}
\int\nolimits_{0}^{1}ds\,
{\rm e}^{-i{\hat{k}}\cdot{\hat{c}}(s)}
\Big)
\Big\rangle_{n+1}^{\rm W}.
\label{EQ:trace_D}
\eeq
The next step is to make the power series expansion of the 
exponential in the expectation value, and make $r$-fold use 
of the integral representation of the Kronecker $\delta$-function,
\beq
\kdelvec{k}{0}
=
\frac{1}{V}
\int\nolimits_{V}
d\lamvec
\exp\Big(i\lamvec\cdot{\bf k}\Big),
\label{EQ:kronHype}
\eeq
in which the $\lamvec$-integral is taken over the 
volume $V$, to obtain
\bea
&&
\exp\Big(\mu^{2}{\LocFr} V^{-n}\Big)
\Big\langle
\exp
\Big(
2\mu^{2}V^{-n}
\sumhrsHP{k}
\real\Omega_{\hat{k}}^{\ast}
\int\nolimits_{0}^{1}ds\,
{\rm e}^{i{\hat{k}}\cdot{\hat{c}}(s)}
\Big)
\Big\rangle_{n+1}^{\rm W}
\nonumber
\\
&&
\qquad
=1+
\frac{1}{V}
\sumr
\frac{\mu^{2r}{\LocFr}^{r}}{V^{nr}\,r!}
\sumhatr 
\int\nolimits_{V}
\frac{d{\lamvec}_{1}}{V}\dots
\frac{d{\lamvec}_{r}}{V}
\exp\Big(
i\sumrhor
{\lamvec}_{\rho}\cdot
\sumaln
{\bf k}_{\rho}^{\alpha}
\Big)
\nonumber
\\
&&
\qquad\qquad\quad
\times
\dintr
\exp
\Big(
-\frac{1}{2}
\sumrhor
\frac{1}{\tau_{\rho}}
\sumaln
\big\lefver
{\bf k}_{\rho}^{\alpha}
\big\rigver^{2}
\Big)
\nonumber
\\
&&
\qquad\qquad\qquad\qquad\qquad
\times
\sintr
\Big\langle
\exp
\Big(
-i\sumrhor\sumaln{\bf k}_{\rho}^{\alpha}
\cdot{\bf c}^{\alpha}(s_{\rho})
\Big)
\Big\rangle_{n+1}^{\rm W}. 
\label{EQ:traceE}
\eea The remaining expectation value in \eqref{EQ:traceE},
$\big\langle
\exp
\big(
-i\sumrhor\sumaln{\bf k}_{\rho}^{\alpha}
\cdot{\bf c}^{\alpha}(s_{\rho})
\big)
\big\rangle_{n+1}^{\rm W}$,
factorises on the replica index, giving 
$\prod_{\alpha=0}^{n}
\big\langle
\exp\big(
-i\sumrhor{\bf k}_{\rho}^{\alpha}
\cdot{\bf c}(s_{\rho})
\big)
\big\rangle_{1}^{\rm W}$. 
Each factor in this product is of the form of the expectation value 
computed in App.~\ref{APP:WMC}, in terms of which we express the 
remaining expectation value in \eqref{EQ:traceE}.  
The result contains an $(n+1)$-fold product of Kronecker 
$\delta$-functions, which we replace with the 
integral representation 
\beq
\prodaln
\delta_{{\bf 0},\sum\nolimits_{\rho=1}^{r}{\bf k}_{\rho}^{\alpha}}^{(d)}
=
\frac{1}{V^{n+1}}
\int\nolimits_{V}
d\muvec^{0}
\dots
d\muvec^{n}
\exp
\Big(
-i\sumaln
\muvec^{\alpha}\cdot
\sumrhor{\bf k}_{\rho}^{\alpha}
\Big),
\label{EQ:kronCorrel}
\eeq
in which each of the $(n+1)$ $\muvec$-integrals is taken over the volume
$V$.  Next, we convert the summations over
$\{{\hat{k}}_{1},\cdots,{\hat{k}}_{r}\}$ to integrations by using
\eqref{EQ:sumtoint} $r$ times, after which the summation on the right
hand side of \eqref{EQ:traceE} becomes
\bea
&&
\frac{1}{V^{n}}
\sumr\frac{\mu^{2r}{\LocFr}^{r}}{r!}
\sintr\dintr
\frac{1}{V}
\int\nolimits_{V}
\frac{d{\lamvec}_{1}}{V}
\cdots
\frac{d{\lamvec}_{r}}{V}
\int\nolimits_{V}
d\muvec^{0}
\cdots
d\muvec^{n}
\nonumber
\\
&&
\quad
\times
\inthatr 
\exp
\Big(
i\sumaln\sumrhor
{\bf k}_{\rho}^{\alpha}
\cdot
\big(
\lamvec_{\rho}-\muvec^{\alpha}
\big)
\Big)
\,
\exp
\Big(
-\frac{1}{2}\sumrhorhopr\rmatr\sumaln
{\bf k}_{\rho}^{\alpha}
\cdot
{\bf k}_{\rho^{\prime}}^{\alpha}
\Big)
\nonumber
\\
&&
\qquad
=
\frac{1}{V^{n}}
\sumr\frac{\mu^{2r}{\LocFr}^{r}}{r!}
\sintr\dintr
\frac{1}{V}
\int\nolimits_{V}
\frac{d{\lamvec}_{1}}{V}
\cdots
\frac{d{\lamvec}_{r}}{V}
\nonumber
\\
&&
\qquad\quad
\times
\bigg(
\int\limits_{V}
d\muvec
\int
\dbar{\bf k}_{1}
\cdots
\dbar{\bf k}_{r}
\exp
\Big(
i\sumrhor
{\bf k}_{\rho}
\cdot
\big(
\lamvec_{\rho}-\muvec
\big)
\Big)
\nonumber
\\
&&
\qquad\quad
\qquad\quad
\times
\exp
\Big(
-\frac{1}{2}\sumrhorhopr\rmatr\,
{\bf k}_{\rho}
\cdot
{\bf k}_{\rho^{\prime}}
\Big)
\bigg)^{n+1}.
\label{EQ:traceF}
\eea Here, $\rmatr$ is an ($r\times r$)-matrix--valued function of the
$r$ arclength coordinates $\{s_{\nu}\}_{\nu=1}^{r}$ and the $r$ inverse
square localisation lengths $\{\tau_{\nu}\}_{\nu=1}^{r}$, defined in
\eqref{EQ:rmatdef}.  We now focus on the quantity in this expression
that is raised to the $(n+1)^{\rm th}$ power.

By performing the gaussian integrations, first over 
$\{{\bf k}_{1},\dots,{\bf k}_{r}\}$ and then over $\muvec$, we obtain
\bea
&&
\int\limits_{V}
d\muvec
\int
\dbar{\bf k}_{1}
\cdots\,
\dbar{\bf k}_{r}
\exp
\Big(
i\sumrhor
{\bf k}_{\rho}
\cdot
\big(
\lamvec_{\rho}-\muvec
\big)
\Big)
\,
\exp
\Big(
-\frac{1}{2}\sumrhorhopr\rmatr\,
{\bf k}_{\rho}
\cdot
{\bf k}_{\rho^{\prime}}
\Big)
\nonumber
\\
&&
\qquad\qquad\qquad\qquad
=
(2\pi)^{-(r-1)d/2}
\left(\wscar\,\detr\rmatrNI\right)^{-d/2}
\exp
\Big(
-\frac{1}{2}\sumrhorhopr\cmatr\,
\lamvec_{\rho}
\cdot
\lamvec_{\rho^{\prime}}
\Big),
\label{EQ:kmuInt}
\eea where we have introduced the ($r\times r$)-matrix--valued function 
$\cmatrNI$ of the $r$ arclength coordinates $\{s_{\nu}\}_{\nu=1}^{r}$ 
and the $r$ inverse square localisation lengths $\{\tau_{\nu}\}_{\nu=1}^{r}$, 
which is built from $\rmatrNI$, \eqref{EQ:rmatdef}, in a manner described in 
App.~\ref{APP:PELLL}.  The gaussian integrals in \eqref{EQ:kmuInt} are 
convergent, owing to the positive-definiteness of the eigenvalue spectrum of 
$\rmatrNI$ and of $\wscar$ for finite $\tau$.  By inserting the 
result~(\ref{EQ:kmuInt}) into expression~(\ref{EQ:traceF}) we obtain
\bea
&&
\frac{1}{V^{n}}
\sumr\frac{\mu^{2r}{\LocFr}^{r}}{r!}\sintr\dintr
\frac{1}{V}
\int\limits_{V}
\frac{d{\lamvec}_{1}}{V}\dots
\frac{d{\lamvec}_{r}}{V}
\nonumber
\\
&&
\qquad
(2\pi)^{-(r-1)(n+1)d/2}
\left(\wscar\,\detr\rmatrNI\right)^{-(n+1)d/2}
\exp
\Big(
-\frac{n+1}{2}\sumrhorhopr\cmatr\,
\lamvec_{\rho}
\cdot
\lamvec_{\rho^{\prime}}
\Big).
\label{EQ:traceG}
\eea The quantity $\wscar$ is built from $\rmatrNI$, \eqref{EQ:rmatdef},
in a manner described in App.~\ref{APP:PELLL}.

Next we perform the integration over
$\{\lamvec_{1},\dots,\lamvec_{r}\}$.  This integration is not quite
gaussian, instead being quasi-gaussian, owing to the presence of a
single {\it zero-mode\/}, the eigenvalue spectrum of $\cmatrNI$
containing a single zero eigenvalue, the remaining $r-1$ eigenvalues
being positive-definite. The presence of this zero-mode can readily be
ascertained by observing that from the definition of $\cmatrNI$,
\eqref{EQ:cmatdef}, we identically have
$\sum_{\rho^{\prime}=1}^{r}\cmatr=0$, \ie, the normalised $r$-vector
$(1,1,\ldots,1)/\sqrt{r}$ is an eigenvector of $\cmatr$ with zero
eigenvalue.  The necessary quasi-gaussian integral is computed in
App.~\ref{APP:QGI}.  By using it, \eqref{EQ:traceG} becomes
\bea
&&
\frac{1}{V^{n}}
\sumr\frac{\mu^{2r}{\LocFr}^{r}}{r!}\sintr\dintr
(2\pi)^{-(r-1)nd/2}
\nonumber
\\
&&
\qquad
\times
\left(\wscar\,\detr\rmatrNI\right)^{-nd/2}
\left(n+1\right)^{-(r-1)d/2}
\left(r^{-1}\wscar\,\detr\rmatrNI\,\qdetr\cmatrNI\right)^{-d/2}
\label{EQ:traceH}, 
\eea where $\qdetr\cmatrNI$ indicates the {\it quasi-determinant\/} of
$\cmatrNI$, \ie, the product of all the nonzero eigenvalues of
$\cmatrNI$; see \cite{REF:quasigeometry}.  We now make use of the
result, established in App.~\ref{APP:PUI}, that the factor
$r^{-1}\,\wscar\,\detr\rmatrNI\,\qdetr\cmatrNI$ is identically unity.
Thus, \eqref{EQ:traceH} is simplified, becoming
\bea
&&
\frac{1}{V^{n}}
\sumr\frac{\mu^{2r}{\LocFr}^{r}}{r!}\sintr\dintr
\nonumber
\\
&&
\qquad\qquad\qquad
\times
(2\pi)^{-(r-1)nd/2}
\left(\wscar\,\detr\rmatrNI\,\right)^{-nd/2}
\left(n+1\right)^{-(r-1)d/2}.
\label{EQ:trace_I}
\eea We take this expression and insert it into 
\eqref{EQ:traceE} to obtain an expression for
$\Big\langle
\exp
\Big(
2\mu^{2}V^{-n}
{\overline{\sum}}_{\hat{k}}^{\possym}
\real\Omega_{\hat{k}}^{\ast}
\int\nolimits_{0}^{1}ds\,
{\rm e}^{i{\hat{k}}\cdot{\hat{c}}(s)}
\Big)
\Big\rangle_{n+1}^{\rm W}$ 
that is valid for $n>-1$. By expanding this result 
for small $n$, we finally obtain
the desired quantity:
\bea
&&
\lim_{n\to 0} 
\frac{2}{nd}
\ln
\Big\langle
\exp
        \Big(
2\mu^{2}V^{-n}
{\overline{\sum}}_{\hat{k}}^{\possym}
\real\Omega_{\hat{k}}^{\ast}
\int\nolimits_{0}^{1}ds\,
{\rm e}^{i{\hat{k}}\cdot{\hat{c}}(s)}
        \Big)
\Big\rangle_{n+1}^{\rm W}
\nonumber
\\
&&
\qquad\quad
=
        \Big(
{\rm e}^{-\mu^{2}{\LocFr}}-(1-\mu^{2}{\LocFr})
        \Big)
\ln
        \Big(
V^{2/d}/2\pi{\rm e}
        \Big)
-{\rm e}^{-\mu^{2}{\LocFr}}
\sumr\frac{\mu^{2r}{\LocFr}^{r}}{r!}
\nonumber
\\
&&
\qquad\qquad
\qquad\qquad
\times
\sintr\dintr
\ln\Big(\wscar\,\detr\rmatrNI\Big).
\eea 
\section{Quasi-gaussian integration}\label{APP:QGI}
Consider the following integral,
\beq
\frac{1}{V}
\int\nolimits_{V}
d\lamvec_{1}\cdots
d\lamvec_{r}
\exp
\Big(
-\frac{1}{2}
\sumrhorhopr
\cmatr\,
\lamvec_{\rho}
\cdot
\lamvec_{\rho^{\prime}}
\Big)
\exp
\Big(
-i\sumrhor
\lamvec_{\rho}
\cdot
{\bf J}_{\rho}
\Big), 
\eeq
taken over $r$ copies of the volume $V$, in which $\cmatrNI$ has as its
sole non-positive definite eigenvalue the zero eigenvalue corresponding
to the $r$-eigenvector $(1,1,\ldots,1)/\sqrt{r}$.  We shall need both
the general case, in which the sources $\{{\bf J}_{\rho}\}_{\rho=1}^{r}$
are arbitrary, and also the special case, in which the sources 
$\{{\bf J}_{\rho}\}_{\rho=1}^{r}$ all vanish.  By working in a basis in
which $\cmatrNI$ is diagonal and assuming that $V$ is sufficiently large
(or, equivalently that no nonzero eigenvalue of $\cmatrNI$ is
arbitrarily small) one finds that this {\it quasi-gaussian\/} integral
is given by
\beq
r^{d/2}
\,
\big(2\pi\big)^{(r-1)d/2}
\Big(
\qdetr\cmatrNI
\Big)^{-d/2}
\exp
\Big(
-\frac{1}{2}
\sumrhorhopr
\cmatrQi\,
{\bf J}_{\rho}
\cdot
{\bf J}_{\rho^{\prime}}
\Big)
\,
{\delta_{{\bf 0},\sumrhor{\bf J}_{\rho}}^{(d)}}\,\,,
\eeq
where $\cmatrQiNI$ is the {\it quasi-inverse\/} of $\cmatrNI$, \ie, the
eigenvector expansion of the inverse of $\cmatrNI$ from which the term
corresponding to the zero eigenvalue has been omitted; see
\cite{REF:quasigeometry}. Similarly, as mentioned in
Sec.~\ref{SEC:logtrace},  $\qdetr\cmatrNI$ indicates the {\it
quasi-determinant\/} of $\cmatrNI$; see \cite{REF:quasigeometry}.
The factor of $r^{d/2}$ is subtle, but is familiar from the context of
so-called collective coordinates.  It arises from the fact that owing to
the presence of the zero eigenvalue there are $d$ integration directions
for which convergence is not controlled by a gaussian integrand. Not
only do the corresponding integrations yield a volume factor; they also
each yield a factor of $r^{1/2}$ by virtue of the limits on them
determined by the form of the corresponding eigenvector. For the special
case in which the sources $\{{\bf J}_{\rho}\}_{\rho=1}^{r}$ all vanish
we have:
\beq
\frac{1}{V}
\int\nolimits_{V}
d\lamvec_{1}\cdots
d\lamvec_{r}
\exp
\Big(
-\frac{1}{2}
\sumrhorhopr
\cmatr\,
\lamvec_{\rho}
\cdot
\lamvec_{\rho^{\prime}}
\Big)
=
r^{d/2}
\,
\big(2\pi\big)^{(r-1)d/2}
\Big(
\qdetr\cmatrNI
\Big)^{-d/2}.
\label{EQ:QGInormal}
\eeq
\section{Perturbation expansion at long localisation lengths: 
	free energy}\label{APP:PELLL}
Consider the quantity $\rmatrNI$, an ($r\times r$)-matrix--valued
function of the $r$ arclength coordinates $\{s_{\nu}\}_{\nu=1}^{r}$ and
the $r$ inverse square localisation lengths
$\{\tau_{\nu}\}_{\nu=1}^{r}$, given by
\beq
\rmatr
\equiv
\tau_{\rho}^{-1}\,
\delta_{\rho\rho^{\prime}}
+\smatr,
\label{EQ:rmatdef}
\eeq
where $\smatrNI$ is given by \eqref{EQ:smatdef}.  We have found it
useful to construct from $\rmatrNI$ several auxiliary quantities:
\bml
\bea
\uvecr
&\equiv&
\sum_{\rho^{\prime}=1}^{r}
\Big({\rmatrNI}\Big)^{-1}\Big\vert_{\rho\rho^{\prime}},
\label{EQ:uvecdef}
\\
\wscar
&\equiv&
\sum_{\rho=1}^{r}
\uvecr,
\label{EQ:wscadef}
\\
\cmatr
&\equiv&
\Big({\rmatrNI}\Big)^{-1}\Big\vert_{\rho\rho^{\prime}}
-
\uvecr\,{\cal U}_{\rho^{\prime}}^{(r)}{\Big/}\wscar.
\label{EQ:cmatdef}
\eea
\eml 

We shall need to develop perturbative expansions in
$\{\tau_{\nu}\}_{\nu=1}^{r}$ of $\ln\left(\wscar\,\detr\rmatrNI\right)$
to linear order and of $\cmatrNI$ to quadratic order.  To this end, we
introduce the ($r\times r$) identity matrix
$\imatrNI\equiv\delta_{\rho\rho^{\prime}}$, and make the definition
\beq
\rmatrzero
\equiv
\tau_{\rho}^{-1}\,
\imatr,
\label{EQ:rmatzerodef}
\eeq
so that 
$\rmatrzeroinvtxt
\equiv
\tau_{\rho}\,
    \imatr
$.
First, we consider $\ln\detr\rmatrNI$:
\bea
\ln\detr\rmatrNI
&=&
\ln\detr\Big(\rmatrzeroNI+\smatrNI\Big)
\nonumber
\\
&=&
\ln\detr
\rmatrzeroNI\cdot\Big(\imatrNI+\rmatrzeroinvNI\cdot\smatrNI\Big)
\nonumber
\\
&=&
\ln\detr\rmatrzeroNI
+\ln\detr\Big(\imatrNI+\rmatrzeroinvNI\cdot\smatrNI\Big)
\nonumber
\\
&=&
\ln\detr\rmatrzeroNI
+\trar\ln\Big(\imatrNI+\rmatrzeroinvNI\cdot\smatrNI\Big)
\nonumber
\\
&=&
\ln
\Big(
\prod\nolimits_{\rho=1}^{r}\tau_{\rho}^{-1}
\Big)
+\trar
\Big(
\rmatrzeroinvNI\cdot\smatrNI+\cdots
\Big)
\nonumber
\\
&=&
-\sum\nolimits_{\rho=1}^{r}\ln\tau_{\rho}
+\sum\nolimits_{\rho=1}^{r}\tau_{\rho}\,\smatrNI_{\rho\rho}
+{\cal O}(\tau^{2}),
\label{EQ:lndetrpert}
\eea where $\trar$ denotes the trace of an $r\times r$ matrix.  Second, 
we consider $\rmatrinvNItxt$. From \eqref{EQ:rmatdef} we have 
$\rmatrNI=\rmatrzeroNI+\smatrNI$, from which follows the Dyson-type equation 
\beq
\rmatrinvNI=\rmatrzeroinvNI-\rmatrzeroinvNI\cdot\smatrNI\cdot\rmatrinvNI.
\eeq
Iterating the Dyson-type equation once and then truncating gives
\beq
\rmatrinv=\tau_{\rho}\delta_{\rho\rho^{\prime}}
-\tau_{\rho}\smatr\tau_{\rho^{\prime}}
+{\cal O}(\tau^{3}).
\label{EQ:rinvpert}
\eeq 
By using this result in Eqs.~(\ref{EQ:uvecdef}) and (\ref{EQ:wscadef}) we obtain
\bml
\bea
\uvecr
&=&
\tau_{\rho}-\tau_{\rho}\sum_{\rho^{\prime}=1}^{r}\smatr\tau_{\rho^{\prime}},
\label{EQ:uvecpert}
\\
\wscar
&=&
\sum_{\rho=1}^{r}\tau_{\rho}-
\sum_{\rho,\rho^{\prime}=1}^{r}\tau_{\rho}\smatr\tau_{\rho^{\prime}}
+{\cal O}(\tau^{3}).
\label{EQ:wscapert}
\eea
\eml Third, we use Eqs.~(\ref{EQ:lndetrpert}) and 
(\ref{EQ:wscapert}) to obtain
\beq
\ln\left(\wscar\,\detr\rmatrNI\right)
=
\ln
\left(
\frac{\sum_{\rho=1}^{r}\tau_{\rho}}{\prod_{\rho=1}^{r}\tau_{\rho}}
\right)
+
\left(
\sum_{\rho=1}^{r}\tau_{\rho}\smatrNI_{\rho\rho}
-{\sum_{\rho,\rho^{\prime}=1}^{r}\tau_{\rho}\,\smatr\,\tau_{\rho^{\prime}}
\over{\sum_{\sigma=1}^{r}\tau_{\sigma}}}
\right)
+
{\cal O}(\tau^{2}).
\label{EQ:lnwdetrpert}
\eeq Finally, by using Eqs.~(\ref{EQ:rinvpert}), (\ref{EQ:uvecpert}) and 
(\ref{EQ:wscapert}) in \eqref{EQ:cmatdef} we obtain
\bea
\cmatr
&=&
\bigg(
  \tau_{\rho}\delta_{\rho\rho^{\prime}}
-
{
\tau_{\rho}
  \tau_{\rho^{\prime}}
\over{
\sum_{\sigma=1}^{r}
\tau_{\sigma}
}}
\bigg)
\nonumber
\\
&&
+
\left(
-\tau_{\rho}\,
 \smatr\,
 \tau_{\rho^{\prime}}
+\tau_{\rho}
 \tau_{\rho^{\prime}}
{
\sum_{\nu=1}^{r}
 \tau_{\nu}(
 \smatrNI_{\nu\rho^{\prime}}
+\smatrNI_{\rho\nu}
           )
\over{
\sum_{\sigma=1}^{r}\tau_{\sigma}
}}
-\tau_{\rho}\tau_{\rho^{\prime}}
{
\sum_{\nu,\nu^{\prime}=1}^{r}\tau_{\nu}\tau_{\nu^{\prime}}
\,\smatrNI_{\nu\nu^{\prime}}
\over{
(\sum_{\sigma=1}^{r}\tau_{\sigma})^{2}
}}
\right)
+{\cal O}(\tau^{3}).
\label{EQ:cmatpert}
\eea

In reducing the free energy (\ref{EQ:freeSPA}) to the form
(\ref{EQ:freenlt}) we have used, {\it inter alia\/}, the perturbation
expansion (\ref{EQ:lnwdetrpert}).  In this way, the free energy reduces
to an assembly of terms each being a functional of $p(\tau)$.  Each
term has a coefficient determined by integration over the arclength
variables $\{s_{1},\ldots,s_{r}\}$ of integrands arising from factors
of $\smatrNI_{\rho\rho^{\prime}}$, defined in \eqref{EQ:smatdef}, which
depend on the arclength variables.  The necessary integrals are
readily expressed in terms of the following ones:
\bml
\bea
\int_{0}^{1}ds_{1}\,
\smatrNI_{11}
&=&
\int_{0}^{1}ds_{1}\,
\min\left(s_{1},s_{1}\right)
=
\int_{0}^{1}ds_{1}\,
s_{1}
=1/2,
\\
\int_{0}^{1}ds_{1}
\int_{0}^{1}ds_{2}\,
\smatrNI_{12}
&=&
\int_{0}^{1}ds_{1}
\int_{0}^{1}ds_{2}\,
\min\left(s_{1},s_{2}\right)
=
2\int_{0}^{1}ds_{1}
\int_{0}^{s_{1}}ds_{2}\,
s_{2}
=1/3.
\eea
\eml
\section{A useful identity}\label{APP:PUI}
We make repeated use of the identity
\beq
r^{-1}\,\wscar\,\detr\rmatrNI\,\,\qdetr\cmatrNI=1,
\label{EQ:repeatID}
\eeq
where $\rmatrNI$, $\wscar$ and $\cmatrNI$ are, respectively, defined in
Eqs.~(\ref{EQ:rmatdef}), (\ref{EQ:wscadef}) and (\ref{EQ:cmatdef}) of
App.~\ref{APP:PELLL}.  To derive this identity, we evaluate the
following quantity [which arises, \eg, in \eqref{EQ:traceF}] in two
ways:
\beq
V^{-1}
\int\limits_{V}
d\muvec
\int
\dbar{\bf k}_{1}
\cdots
\dbar{\bf k}_{r}
\exp
\Big(
i\sumrhor
{\bf k}_{\rho}
\cdot
\big(
\lamvec_{\rho}-\muvec
\big)
\Big)
\,
\exp
\Big(
-\frac{1}{2}\sumrhorhopr\rmatr
{\bf k}_{\rho}
\cdot
{\bf k}_{\rho^{\prime}}
\Big).
\label{EQ:startUSiden}
\eeq
First, by integrating over $\{\lamvec_{\rho}\}_{\rho=1}^{r}$, then over
$\{{\bf k}_{\rho}\}_{\rho=1}^{r}$, and then over $\muvec$ we obtain the
result: unity.  Second, by integrating over 
$\{{\bf k}_{\rho}\}_{\rho=1}^{r}$,then over $\muvec$, and then, by using 
the quasi-gaussian integral \eqref{EQ:QGInormal}, over
$\{\lamvec_{\rho}\}_{\rho=1}^{r}$, we obtain the result:
\beq
\Big(r^{-1}\wscar\,\detr\rmatrNI\,\qdetr\cmatrNI\Big)^{-d/2};
\label{EQ:end_iden}
\eeq
hence the identity \eqref{EQ:repeatID}.
\section{Laplace representation of free energy}\label{APP:LRFF}
In this appendix we describe in detail how to exchange the 
dependence of the three contributions to $\fspa$, \eqref{EQ:freenlt}, 
from $p(\tau)$ to its Laplace transform $\hat{p}(\hat{\tau})$.
We begin by noting two integrals:
\bml
\bea
\ln\tau
&=&
\int\nolimits_{0}^{\infty}\frac{d\hat{\tau}}{\hat{\tau}}
\left(
{\rm e\/}^{-\hat{\tau}}-{\rm e\/}^{-\hat{\tau}\tau}
\right),
\label{EQ:frul}
\\
\frac{1}{\tau}
&=&
\int\nolimits_{0}^{\infty}d\hat{\tau}{\rm e\/}^{-\hat{\tau}\tau}.
\label{EQ:inlt}
\eea
\eml The latter integral is elementary; the former is an example of a
Frullanian integral; see Ref.~\cite{REF:Zwill}.  We use \eqref{EQ:frul}
to express the first contribution to $\fspa$ as
\bml
\bea
\ltave
\ln\left(\frac{\tau_{1}+\tau_{2}}{\tau_{1}\tau_{2}}\right)
\rtave
&=&
\ltave
 \ln\left(\tau_{1}+\tau_{2}\right)
-\ln\left(\tau_{1}\right)
-\ln\left(\tau_{2}\right)
\rtave
\nonumber
\\
&=&
\int\nolimits_{0}^{\infty}d\tau_{1}p(\tau_{1})
\int\nolimits_{0}^{\infty}d\tau_{2}p(\tau_{2})
\int\nolimits_{0}^{\infty}
\frac{d\hat{\tau}}{\hat{\tau}}
\left(
-        {\rm e\/}^{-\hat{\tau}}
-{\rm e\/}^{-\hat{\tau}(\tau_{1}+\tau_{2})}
+{\rm e\/}^{-\hat{\tau}\tau_{1}}
+{\rm e\/}^{-\hat{\tau}\tau_{2}}
\right)
\nonumber
\\
&=&
\int\nolimits_{0}^{\infty}
\frac{d\hat{\tau}}{\hat{\tau}}
\left(
- \hat{p}(\hat{\tau})^{2}
+2\hat{p}(\hat{\tau})
-{\rm e\/}^{-\hat{\tau}}
\right), 
\eea where the curly braces $\ltave\cdots\rtave$ were defined immediately 
following \eqref{EQ:freenlt}.  By following the identical strategy, we use 
\eqref{EQ:frul} to express the second contribution to $\fspa$ as
\bea
\ltave
\ln
\left(
\frac{\tau_{1}+\tau_{2}+\tau_{3}}{\tau_{1}\tau_{2}\tau_{3}}
\right)
\rtave
&=&
\ltave
 \ln\left(\tau_{1}+\tau_{2}+\tau_{3}\right)
-\ln\left(\tau_{1}\right)
-\ln\left(\tau_{2}\right)
-\ln\left(\tau_{3}\right)
\rtave
\nonumber
\\
&=&
\int\nolimits_{0}^{\infty}
\frac{d\hat{\tau}}{\hat{\tau}}
\left(
- \hat{p}(\hat{\tau})^{3}
+3\hat{p}(\hat{\tau})
-2{\rm e\/}^{-\hat{\tau}}
\right).
\eea To express the third contribution to $\fspa$ in terms of
$\hat{p}(\hat{\tau})$ we make use of \eqref{EQ:inlt}.  This yields
\bea
\ltave
\frac{\tau_{1}\tau_{2}}{\tau_{1}+\tau_{2}}
\rtave
&=&
\int\nolimits_{0}^{\infty}d\tau_{1}p(\tau_{1})
\int\nolimits_{0}^{\infty}d\tau_{2}p(\tau_{2})
\frac{\tau_{1}\tau_{2}}{\tau_{1}+\tau_{2}}
\nonumber
\\
&=&
\int\nolimits_{0}^{\infty}d\tau_{1}p(\tau_{1})\tau_{1}
\int\nolimits_{0}^{\infty}d\tau_{2}p(\tau_{2})\tau_{2}
\int\nolimits_{0}^{\infty}d\hat{\tau}
{\rm e}^{-\hat{\tau}(\tau_{1}+\tau_{2})}
=
\int\nolimits_{0}^{\infty}d\hat{\tau}
\big(
{d\hat{p}}/{d\hat{\tau}}
\big)^{2}.
\eea
\eml

We now take the functional derivative with respect to $\hat{p}(\hat{\tau})$ 
of these three contributions to $\fspa$.  Being local, the first two are 
straightforward to compute, respectively yielding 
\bml
\bea
\frac{\delta}{\delta\hat{p}(\hat{\tau})}
\ltave
\ln\left(\frac{\tau_{1}+\tau_{2}}{\tau_{1}\tau_{2}}\right)
\rtave
&=&
\frac{2}{\hat{\tau}}\Big(1-\hat{p}(\hat{\tau})\Big),
\\
\frac{\delta}{\delta\hat{p}(\hat{\tau})}
\ltave
\ln\left(\frac{\tau_{1}+\tau_{2}+\tau_{3}}{\tau_{1}\tau_{2}\tau_{3}}\right)
\rtave
&=&
\frac{3}{\hat{\tau}}\Big(1-\hat{p}(\hat{\tau})^{2}\Big).
\eea To evaluate the functional derivative with respect to 
$\hat{p}(\hat{\tau})$ of the third contribution to $\fspa$ requires 
an integration by parts.  The integrated piece vanishes because 
$\hat{p}(\hat{\tau})$ is to be varied at neither $\hat{\tau}=0$  
nor $\hat{\tau}=\infty$, due to the boundary conditions.
Thus we obtain for the third contribution to $\fspa$:
\beq
\frac{\delta}{\delta\hat{p}(\hat{\tau})}
\ltave
\frac{\tau_{1}\tau_{2}}{\tau_{1}+\tau_{2}}
\rtave
=-2\frac{d^{2}\hat{p}}{d\hat{\tau}^{2}}.
\eeq
\eml
\section{Order-parameter weighted averages}\label{APP:OPWaverages}
In this appendix we focus on the computation of the following quantity, 
defined for arbitrary $\hat{l}$ and $\hat{l}^{\prime}$:
\beq    
{\DPS
\Big\langle
\int\nolimits_{0}^{1}dt\,
{\rm e}^{-i\hat{l}\cdot\hat{c}(t)}
\int\nolimits_{0}^{1}dt^{\prime}\,
{\rm e}^{-i\hat{l}^{\prime}\cdot\hat{c}(t^{\prime})}
\hfill
\atop{\DPS
\qquad
\hfill
\times
\exp
        \Big(
\mu^{2}{\LocFr} V^{-n}
\sumhat{k}
\int\nolimits_{0}^{1}ds\,
{\rm e}^{-i{\hat{k}}\cdot{\hat{c}}(s)}\,
\kdelvecT{k}{0}\dint{\tau}
\exp
\big(
-\hat{k}^{2}/2\tau
\big)
        \Big)
\Big\rangle_{n+1}^{\rm W}\,\,.
}}
\label{EQ:gencorr}
\eeq 
In addition, we make two applications of it. 
We proceed by expanding the exponential, which yields
\bea
&&
\Big\langle
\int\nolimits_{0}^{1}dt\,
{\rm e}^{-i\hat{l}\cdot\hat{c}(t)}
\int\nolimits_{0}^{1}dt^{\prime}\,
{\rm e}^{-i\hat{l}^{\prime}\cdot\hat{c}(t^{\prime})}
\Big\rangle_{n+1}^{\rm W}
+
\sumr\frac{\mu^{2r}{\LocFr}^{r}}{V^{nr}r!}
\sintrPT\dintr
\nonumber
\\
&&
\qquad\quad
\times
\sumhatr\prodrhor
{\delta_{{\tilvec{k}}_{\rho},{\bf 0}}^{(d)}}
\exp
\Big(
-\frac{1}{2}
\sumrhor
\frac{1}{\tau_{\rho}}
{\hat{k}_{\rho}}^{2}
\Big)
\Big\langle
{\rm e}^{-i\hat{l}\cdot\hat{c}(s_{r+1})}
{\rm e}^{-i\hat{l}^{\prime}\cdot\hat{c}(s_{r+2})}
{\rm e}^{-i\sumrhor\hat{k}_{\rho}\cdot\hat{c}(s_{\rho})}
\Big\rangle_{n+1}^{\rm W}.
\eea
Next, we observe the factorisation of the Wiener measure correlators 
on the replica index, 
use the explicit result for this correlator given in App.~\ref{APP:WMC}, 
use integral representations for the Kronecker $\delta$-functions 
[Eqs.~(\ref{EQ:kronCorrel}) and (\ref{EQ:kronHype})], and 
convert summations to integrals by using \eqref{EQ:sumtoint},
thus obtaining
\bea
&&
\kdelhatNS_{\hat{l}+\hat{l}^{\prime},\hat{0}}
\int\nolimits_{0}^{1}dt\,
\int\nolimits_{0}^{1}dt^{\prime}\,
{\rm e}^{-\vert{t-t^{\prime}}\vert\hat{l}^{2}}
\nonumber
\\
&&
\quad
+
\frac{1}{V^{n}}
\sumr\frac{\mu^{2r}{\LocFr}^{r}}{r!}
\sintrPT\dintr
\frac{1}{V}
\int\nolimits_{V}
d\lamvec_{1}\dots
d\lamvec_{r}
\int\nolimits_{V}
d\muvec^{0}
\cdots
d\muvec^{n}
\nonumber
\\
&&
\quad
\times
\!\inthatr 
\exp
\Big(
i\sumrhor\lamvec_{\rho}
\!\cdot\!
\sumaln
{\bf k}_{\rho}^{\alpha}
\Big)
\exp
\Big(
\!
-i\sumaln
\muvec^{\alpha}
\!
\cdot
\big(
 {\bf l}^{\alpha}
+{\bf l}^{\prime\alpha}
+\sumrhor
{\bf k}_{\rho}^{\alpha}
\big)
\Big)
\exp
\Big(
\!-\!
\sumrhor
{\hat{k}}_{\rho}^{2}/2\tau_{\rho}
\Big)
\nonumber
\\
&&
\quad
\times
\exp
\Big(
\!
-\frac{1}{2}
\sumaln
\sumrhorhopr\!
{\bf k}_{\rho}^{\alpha}
\cdot\!
{\bf k}_{\rho^{\prime}}^{\alpha}
\,\smatr
\Big)
\exp
\Big(
\!
-\sumaln
\sumrhor
{\bf l}^{\alpha}
\!\cdot
{\bf k}_{\rho}^{\alpha}
\,\smatrCI{r+1,\rho}
\Big)
\exp
\Big(
\!
-\sumaln
\sumrhor
{\bf l}^{\prime\alpha}
\cdot
{\bf k}_{\rho}^{\alpha}
\,\smatrCI{r+2,\rho}
\Big)
\nonumber
\\
&&
\quad
\times
\exp
\Big(
-\frac{1}{2}
\hat{l}^{2}
\,\smatrCI{r+1,r+1}
\Big)
\exp
\Big(
-\frac{1}{2}
{\hat{l}}^{\prime 2}
\,\smatrCI{r+2,r+2}
\Big)
\exp
\Big(
-\hat{l}\cdot{\hat{l}}^{\prime}
\,\smatrCI{r+1,r+2}
\Big).
\eea
We now recognise that the integrals over 
$\{\muvec^{\alpha}\}_{\alpha=0}^{n}$ and 
$\{{\bf k}_{1}^{\alpha},\dots,{\bf k}_{r}^{\alpha}\}_{\alpha=0}^{n}$ 
factorise on the replica index, giving 
\bea
&&
\kdelhatNS_{\hat{l}+\hat{l}^{\prime},\hat{0}}
\int\nolimits_{0}^{1}dt\,
\int\nolimits_{0}^{1}dt^{\prime}\,
{\rm e}^{-\vert{t-t^{\prime}}\vert\hat{l}^{2}}
+\frac{1}{V^{n}}
\sumr\frac{\mu^{2r}{\LocFr}^{r}}{r!}
\sintrPT\dintr
\nonumber
\\
&&
\quad
\times
\exp
\Big(
-\frac{1}{2}
\hat{l}^{2}
\,\smatrCI{r+1,r+1}
-\frac{1}{2}
{\hat{l}}^{\prime 2}
\,\smatrCI{r+2,r+2}
-\hat{l}\cdot{\hat{l}}^{\prime}
\,\smatrCI{r+1,r+2}
\Big)
\frac{1}{V}
\int\nolimits_{V}
d\lamvec_{1}\cdots
d\lamvec_{r}
\nonumber
\\
&&
\quad
\times
\prodaln
\bigg(
\int\nolimits_{V}
d\muvec
\int
\dbar{\bf k}_{1}
\cdots
\dbar{\bf k}_{r}
\exp
\Big(
i\sumrhor\lamvec_{\rho}
\cdot
{\bf k}_{\rho}
\Big)
\exp
\Big(
-i\muvec
\cdot
\big(
 {\bf l}^{\alpha}
+{\bf l}^{\prime\alpha}
+\sumrhor
{\bf k}_{\rho}
\big)
\Big)
\nonumber
\\
&&
\quad
\times
\exp
\Big(-\sumrhor
{\bf k}_{\rho}^{2}/2\tau_{\rho}
\Big)
\exp
\Big(
-\frac{1}{2}
\sumrhorhopr
{\bf k}_{\rho}
\cdot
{\bf k}_{\rho^{\prime}}
\,\smatr
\Big)
\nonumber
\\
&&
\quad
\times
\exp
\Big(
-\sumrhor
{\bf l}^{\alpha}
\cdot
{\bf k}_{\rho}
\,\smatrCI{r+1,\rho}
\Big)
\exp
\Big(
-\sumrhor
{\bf l}^{\prime\alpha}
\cdot
{\bf k}_{\rho}
\,\smatrCI{r+2,\rho}
\Big)
\bigg).
\label{EQ:factInt}
\eea
At this stage we focus on the factorised integrals over 
$\{{\bf k}_{\rho}^{\alpha}\}_{\rho}^{r}$ and $\muvec^{\alpha}$ occurring under 
the product.  As they are gaussian integrals they can straightforwardly be 
performed (we prefer to integrate over 
$\{{\bf k}_{\rho}^{\alpha}\}_{\rho=1}^{r}$ 
first and $\muvec^{\alpha}$ second), yielding
\bea
&&
\big(2\pi\big)^{-(r-1)d/2}\big(\wscar\,\det\rmatrNI\big)^{-d/2} 
\exp
\Big(
-\big\vert{\bf l}^{\alpha}+{\bf l}^{\prime\alpha}\big\vert^{2}
/2\wscar
\Big)
\nonumber
\\
&&
\qquad
\times
\exp
\Big(
-\frac{1}{2}\sumrhorhopr
\cmatr
\big(
\lamvec_{\rho}
+i{\bf l}^{\alpha}\,\smatrCI{r+1,\rho}
+i{\bf l}^{\prime\alpha}\,\smatrCI{r+2,\rho}
\big)\cdot\big(
\lamvec_{\rho^{\prime}}
+i{\bf l}^{\alpha}\,\smatrCI{r+1,\rho^{\prime}}
+i{\bf l}^{\prime\alpha}\,\smatrCI{r+2,\rho^{\prime}}
\big)
\Big)
\nonumber
\\
&&
\qquad\qquad
\times
\exp
\Big(
-\frac{i}{\wscar}
\big(
{\bf l}^{\alpha}+{\bf l}^{\prime\alpha}
\big)\cdot\sumrhor\uvecr\,\big(
\lamvec_{\rho}
+i{\bf l}^{\alpha}\,\smatrCI{r+1,\rho}
+i{\bf l}^{\prime\alpha}\,\smatrCI{r+2,\rho}
\big)\Big),
\eea
where $\uvecrNI$, $\wscar$ and $\cmatrNI$ are respectively defined in 
Eqs.~(\ref{EQ:uvecdef}), (\ref{EQ:wscadef}) and (\ref{EQ:cmatdef}).  
Next, we insert 
the result of integrating over $\{{\bf k}_{\rho}^{\alpha}\}_{\rho=1}^{r}$ and 
$\muvec^{\alpha}$ into \eqref{EQ:factInt} and focus on the remaining integrals 
over $\{\lamvec_{\rho}\}_{\rho=1}^{r}$.  Just as we encountered in 
App.~\ref{APP:EHEOPH}, this integration is not quite gaussian, instead being 
quasi-gaussian, owing to the presence a single zero in the eigenvalue 
spectrum of $\cmatrNI$.  The necessary quasi-gaussian integral is computed in 
App.~\ref{APP:QGI}.  By using it, \eqref{EQ:factInt} becomes
\bea
&&
\kdelhatNS_{\hat{l}+\hat{l}^{\prime},\hat{0}}
\int\nolimits_{0}^{1}dt\,
\int\nolimits_{0}^{1}dt^{\prime}\,
{\rm e}^{-\vert{t-t^{\prime}}\vert\hat{l}^{2}}
\nonumber
\\
&&
\quad
+\delta_{{\tilvec{l}}+
        {\tilvec{l}}^{\prime},{\bf 0}}^{(d)}
\frac{1}{V^{n}}
\sumr\frac{\mu^{2r}{\LocFr}^{r}}{r!}
\sintrPT\dintr
\nonumber
\\
&&
\quad
\times
\big(2\pi\big)^{-(r-1)nd/2}
\big(\wscar\,\detr\rmatrNI\big)^{-nd/2}
\big(n+1\big)^{-(r-1)d/2}
\nonumber
\\
&&
\quad
\times
\big(r^{-1}\,\wscar\,
\detr\rmatrNI\,\qdetr\cmatrNI 
\big)^{-d/2}            
\nonumber
\\
&&
\quad
\times
\exp
\bigg(
\big(\wscar\big)^{-1}\sumrhor\uvecr
\Big(
{\hat{l}}^{2}\,\smatrCI{r+1,\rho}+
{\hat{l}^{\prime\,{2}}}\,\smatrCI{r+2,\rho}+
\hat{l}\cdot\hat{l}^{\prime}\,
(\smatrCI{r+1,\rho}+\smatrCI{r+2,\rho})
\Big)
\bigg)
\nonumber
\\
&&
\quad
\times
\exp\Big(
-\big(
{\hat{l}}^{2}\,\smatrCI{r+1,r+1}+
{\hat{l}^{\prime\,{2}}}\,\smatrCI{r+2,r+2}+
2\hat{l}\cdot\hat{l}^{\prime}\,
\smatrCI{r+1,r+2}
\big)/2
\Big)
\nonumber
\\
&&
\quad
\times
\exp
\Big(
\frac{1}{2}{\hat{l}}^{2}
\!
\sumrhorhopr
\cmatr\,\smatrCI{r+1,\rho}\,\smatrCI{r+1,\rho^{\prime}}+
\frac{1}{2}{\hat{l}^{\prime\,{2}}}
\!\!
\sumrhorhopr
\cmatr\,\smatrCI{r+2,\rho}\,\smatrCI{r+2,\rho^{\prime}}+
\hat{l}\cdot\hat{l}^{\prime}
\!\!
\sumrhorhopr
\cmatr\,\smatrCI{r+1,\rho}\,\smatrCI{r+2,\rho^{\prime}}
\Big)
\nonumber
\\
&&
\quad
\times
\exp
\Big(
-
\big(
{\hat{l}}^{2}+
{\hat{l}^{\prime\,{2}}}+
2\hat{l}\cdot\hat{l}^{\prime}
\big)/2\wscar
\Big)
\nonumber
\\
&&
\quad
\times
\exp\Big(-
(n+1)^{-1}\sumrhorhopr\cmatr
\big(
\tilvec{l}\,\smatrCI{r+1,\rho}+
\tilvec{l}^{\prime}\,\smatrCI{r+2,\rho}
\big)
\cdot
\big(\tilvec{l}\,\smatrCI{r+1,\rho^{\prime}}+
     \tilvec{l}^{\prime}\,\smatrCI{r+2,\rho^{\prime}}\big)/2
\Big).
\label{EQ:NumerCorr}
\eea
We have simplified this result by making use of the identity given in
\eqref{EQ:repeatID} of App.~\ref{APP:PUI}.  We have further simplified
it by observing that as $\cmatrQiNI$ is the quasi-inverse
\cite{REF:quasigeometry} of $\cmatrNI$ (see App.~\ref{APP:PELLL} and
footnote~\cite{REF:quasigeometry}), the relevant zero-mode being
$(1,1,\ldots,1)/\sqrt{r}$, we have
\bml
\bea
\cmatrQiNI\cdot\cmatrNI\big\vert_{\rho\rho^{\prime}}
&=&
\delta_{\rho\rho^{\prime}}-r^{-1},
\\
\cmatrNI\cdot\cmatrQiNI\cdot\cmatrNI
&=&
\cmatrNI.
\eea
\eml In addition, we have used the fact that there is a single 
zero-mode to ascertain that 
$\qdetr\big((n+1)\,\cmatrNI\big)=(n+1)^{r-1}\qdetr\cmatrNI$.

As our first application of \eqref{EQ:NumerCorr}, we set
$\hat{l}=\hat{l}^{\prime}=\hat{0}$, thus obtaining a normalisation
factor that we shall use subsequently:
\bea
&&
1+\frac{1}{V^{n}}
\sumr\frac{\mu^{2r}{\LocFr}^{r}}{r!}
\sintrPT\dintr
\nonumber
\\
&&
\qquad\qquad\qquad\qquad\quad
\times
\big(2\pi\big)^{-(r-1)nd/2}
\big(\wscar\,\detr\rmatrNI\big)^{-nd/2}
\big(n+1\big)^{-(r-1)d/2}.
\label{EQ:GreenDenom}
\eea
As our second application of \eqref{EQ:NumerCorr} we compute the right
hand side of \eqref{EQ:scrho} by forming the quotient of
\eqref{EQ:NumerCorr} with $\hat{l}^{\prime}=\hat{0}$ and
\eqref{EQ:GreenDenom}.  By making use of the indentity
\eqref{EQ:repeatID} and taking the limit $n\rightarrow 0$ we obtain
Eqs.~(\ref{EQ:scone}) and (\ref{EQ:scmain}).
\section{Perturbation expansion at long localisation lengths: 
	order parameter}\label{APP:DSCE}
In this appendix we compute the perturbative expansions of
$\Upsilon^{(1)}$ and $\Upsilon^{(2)}$ needed to compute the right hand
side of \eqref{EQ:scforP}.  By using the definitions (\ref{EQ:smatdef}),
(\ref{EQ:uvecdef}), (\ref{EQ:wscadef}) and (\ref{EQ:cmatdef}), we find
\bml
\bea
\Upsilon^{(1)}
&=&
\Big(
	\tau_{1}^{-1}+\vert{s_{1}-s_{2}}\vert
\Big)^{-1}
=
\tau_{1}
-\vert{s_{1}-s_{2}}\vert\tau_{1}^{2}
+{\cal O}\big(\tau^{3}\big),
\\
\Upsilon^{(2)}
&=&
\tau_{1}+\tau_{2}
+{\cal O}\big(\tau^{2}\big).
\eea
\eml By inserting these results into \eqref{EQ:scforP}, 
and using \eqref{EQ:scforQ}, we obtain
\bea
{\LocFr} p(\tau)
&=&
(1-{\LocFr})\mu^{2}{\LocFr}
\int_{0}^{1}ds_{1}ds_{2}           
\int_{0}^{\infty}d\tau_{1}p(\tau_{1})\,
\delta
\big(
\tau-\tau_{1}+\tau_{1}^{2}\vert{s_{1}-s_{2}}\vert
\big)
\nonumber
\\
&&
\qquad
\qquad
+\frac{1}{2}\mu^{4}{\LocFr}^{2}
\int_{0}^{1}ds_{1}ds_{2}ds_{3}     
\int_{0}^{\infty}d\tau_{1}p(\tau_{1})\,d\tau_{2}p(\tau_{2})\,
\delta\big(\tau-\tau_{1}-\tau_{2}\big)
+{\cal O}\big(\epsilon^{2}\big).
\label{EQ:SCpartial}
\eea
Next, we expand the Dirac $\delta$-function, 
$\delta\big(\tau-\tau_{1}
+\tau_{1}^{2}\vert{s_{1}-s_{2}}\vert\big)
\approx
\delta\big(\tau-\tau_{1}\big)
+\tau_{1}^{2}\vert{s_{1}-s_{2}}\vert
\delta^{\prime}\big(\tau-\tau_{1}\big)$, 
and perform the $\tau$ and $s$ integrals.  (Equivalently, we take the Laplace 
transform of \eqref{EQ:SCpartial}, expand the resulting exponential function, 
perform the $\tau$ and $s$ integrals, and back-transform the resulting nonlinear 
ordinary differential equation.)\thinspace\  Finally, we transform to the scaling 
form $\pi(\theta)$ defined in \eqref{EQ:pscaledef}, and observe that it satisfies 
\eqref{EQ:scpieq}.  
\section{Correlator for one-replica sector fluctuations}\label{APP:FluCo}
In this appendix we compute the quantity $\greenNI{q}$ defined in 
\eqref{EQ:greendef} for $\Omega_{\hat{k}}$ taking the form given in 
\eqref{EQ:OPhypothesis}.  Ultimately, we shall be concerned with 
the behaviour at small $n$ of the diagonal and off-diagonal elements 
of $\greenNI{q}$ via the quantities $\bighw{k}$ and $\lithz{k}$ 
defined in Eqs.~(\ref{EQ:bighexp}) and (\ref{EQ:lithexp}).

First, we note that by using the invariance under 
$\hat{k}\rightarrow -\hat{k}$ of the hypothesis for $\Omega_{\hat{k}}$,
\eqref{EQ:OPhypothesis}, along with the adding and subtracting of terms
in the 0-replica sector, so as to relax the constraints on the summations
${\overline{\sum}}_{\hat{k}}^{\possym}$ in the exponents in the
numerator and denominator of \eqref{EQ:greendef}, we arrive at the form:
\beq
\green{q}=
{
{\DPS
\Big\langle
\int\nolimits_{0}^{1}dt\,
{\rm e}^{i{\bf q}\cdot{\bf c}^{\alpha}(t)}
\int\nolimits_{0}^{1}dt^{\prime}\,
{\rm e}^{-i{\bf q}\cdot{\bf c}^{\alpha^{\prime}}(t^{\prime})}
\hfill
\atop{\DPS
\qquad
\hfill
\times
\exp
        \Big(
\mu^{2}{\LocFr} V^{-n}
\sumhat{k}
\int\nolimits_{0}^{1}ds\,
{\rm e}^{-i{\hat{k}}\cdot{\hat{c}}(s)}
\kdelvecT{k}{0}\dint{\tau}
\exp
\big(
-\hat{k}^{2}/2\tau
\big)
        \Big)
\Big\rangle_{n+1}^{\rm W}
}}
\over{\DPS
\Big\langle
\exp
        \Big(
\mu^{2}{\LocFr} V^{-n}
\sumhat{k}
\int\nolimits_{0}^{1}ds\,
{\rm e}^{-i{\hat{k}}\cdot{\hat{c}}(s)}
\kdelvecT{k}{0}\dint{\tau}
\exp
\big(
-\hat{k}^{2}/2\tau
\big)
        \Big)
\Big\rangle_{n+1}^{\rm W}
}}.
\label{EQ:greenQuo}
\eeq 

Second, we set 
$\hat{l}=         \{-{\bf q},{\bf 0},\dots,{\bf 0}\}$ and 
$\hat{l}^{\prime}=\{ {\bf q},{\bf 0},\dots,{\bf 0}\}$ 
to obtain the replica--diagonal elements of the numerator 
correlator in \eqref{EQ:greenQuo}:
\bea
&&
\debyeZ{q}+
\frac{1}{V^{n}}
\sumr\frac{\mu^{2r}{\LocFr}^{r}}{r!}
\sintrPT\dintr
\nonumber
\\
&&
\qquad\quad
\times
\big(2\pi\big)^{-(r-1)nd/2}
\big(\wscar\,\detr\rmatrNI\big)^{-nd/2}
\big(n+1\big)^{-(r-1)d/2}
\nonumber
\\
&&
\qquad\qquad\qquad
\times
\exp
\bigg(
-\frac{1}{2}q^{2}
                \Big(
\big(\smatrCI{r+1,r+1}+
\smatrCI{r+2,r+2}-
2\smatrCI{r+1,r+2}\big)
\nonumber
\\
&&
\qquad\qquad\qquad\qquad\quad
-\frac{n}{n+1}
\sumrhorhopr\cmatr\,
\big(-\smatrCI{r+1,\rho}
     +\smatrCI{r+2,\rho}
\big)
\big(-\smatrCI{r+1,\rho^{\prime}}
     +\smatrCI{r+2,\rho^{\prime}}
\big)
                \Big)
\bigg).
\label{EQ:GreenNumDiag}
\eea
Third, we set 
$\hat{l}         =\{-{\bf q},{\bf 0},{\bf 0},\dots,{\bf 0}\}$ and 
$\hat{l}^{\prime}=\{ {\bf 0},{\bf q},{\bf 0},\dots,{\bf 0}\}$ 
to obtain the replica--off-diagonal elements of the numerator 
correlator in \eqref{EQ:greenQuo}:
\bea
&&
\kdelvec{q}{0}+
\frac{1}{V^{n}}
\sumr\frac{\mu^{2r}{\LocFr}^{r}}{r!}
\sintrPT\dintr
\nonumber
\\
&&
\qquad
\times
\big(2\pi\big)^{-(r-1)nd/2}
\big(\wscar\,\detr\rmatrNI\big)^{-nd/2}
\big(n+1\big)^{-(r-1)d/2}
\nonumber
\\
&&
\qquad
\times
\exp
\bigg(
-\frac{1}{2}q^{2}
\Big(
\big(
\smatrCI{r+1,r+1}+
\smatrCI{r+2,r+2}
\big)
+\frac{2}{\wscar}
-\frac{2}{\wscar}
\sumrhor\uvecr
\big(
\smatrCI{r+1,\rho}+
\smatrCI{r+2,\rho}
\big)
\nonumber
\\
&&
\qquad
-\frac{n}{n+1}
\sumrhorhopr\cmatr
\big( 
 \smatrCI{r+1,\rho}\smatrCI{r+1,\rho^{\prime}}
+\smatrCI{r+2,\rho}\smatrCI{r+2,\rho^{\prime}}
\big)
-\frac{2}{n+1}
\sumrhorhopr\cmatr
\smatrCI{r+1,\rho}\smatrCI{r+2,\rho^{\prime}}
\Big)
\bigg).
\label{EQ:GreenNumOff}
\eea

Next, by using the denominator, \eqref{EQ:GreenDenom}, and respective 
numerators, Eqs.~(\ref{EQ:GreenNumDiag}) and (\ref{EQ:GreenNumOff}), we 
build the diagonal and off-diagonal elements, $\bighw{k}$ and $\lithz{k}$, 
of $\greenNI{q}$, \eqref{EQ:greenQuo}.  At this stage, all dependence on 
$n$ is explicit.  Thus, by making series expansions in $n$ about $n=0$ of 
the denominator and respective numerators we are able to identify 
$\bighz{k}$, $\bighw{k}$ and $\lithz{k}$ of Eqs.~(\ref{EQ:bighexp})
and (\ref{EQ:lithexp}) (for ${\bf q}\ne{\bf 0}$):
\bml
\bea
\bighz{q}
&=&
\debyeZ{q},
\label{EQ:bighz_term}
\\
\bighw{q}
&=&
{\rm e}^{-\mu^{2}{\LocFr}}
\sum_{r=2}^{\infty}
\frac{\mu^{2r}{\LocFr}^{r}}{r!}
\sintrPT
\exp
\Big(-q^{2}\big\vert{s_{r+1}-s_{r+2}}\big\vert/2\Big)
\dintr
\nonumber
\\
&&
\times
(q^{2}/2)
\sumrhorhopr
\cmatr
\big( 
-\smatrCI{r+1,\rho}
+\smatrCI{r+2,\rho}
\big)
\big(
-\smatrCI{r+1,\rho^{\prime}}
+\smatrCI{r+2,\rho^{\prime}}
\big),
\label{EQ:bighwTerm}
\\
\lithz{q}
&=&
{\rm e}^{-\mu^{2}{\LocFr}}
\sumr\frac{\mu^{2r}{\LocFr}^{r}}{r!}
\sintrPT\dintr
\nonumber
\\
&&
\qquad
\times
\exp
\bigg(
-q^{2}
\Big(
\frac{1}{2}
\big(
\smatrCI{r+1,r+1}+\smatrCI{r+2,r+2}
\big)
+\frac{1}{\wscar}
-\frac{1}{\wscar}\sumrhor
\uvecr
\big(\smatrCI{r+1,\rho}+\smatrCI{r+2,\rho}\big)
\nonumber
\\
&&
\qquad\qquad
-\sumrhorhopr\cmatr\,\smatrCI{r+1,\rho}\,\smatrCI{r+2,\rho^{\prime}}
\Big)
\bigg),
\label{EQ:lithz_term}
\eea
\eml where we have restricted our attention to ${\bf q}\ne{\bf 0}$, 
which is all that is necessary, and have used the fact that 
${\cal C}^{(1)}$ vanishes identically. 

As a final component of this appendix, we expand $\bighw{q}$ and $\lithz{q}$, 
assuming that the gel fraction ${\LocFr}$ and the inverse square localisation 
lengths to which $p(\tau)$ gives appreciable weight are both asymptotically of 
order $\epsilon(\ll 1)$, \ie, which is appropriate in the vicinity on the 
transition, as we can verify {\it a posteriori\/}.  The accuracy to which 
$\bighw{q}$ and $\lithz{q}$ must be computed can be established by observing 
the manner in which they appear in \eqref{EQ:gdfFexp}, from which we see that 
$\fgdf\{{\LocFr},p\}$ can be computed to ${\cal O}(\epsilon^{3})$ 
by computing 
$\bighw{q}$ to ${\cal O}(\epsilon^{3})$ and 
$\lithz{q}$ to ${\cal O}(\epsilon^{2})$. 

We first focus on $\bighw{q}$.  As we see from \eqref{EQ:cmatpert}, $\cmatr$ is 
of order $\tau$, so that we need retain only the $r=2$ term in \eqref{EQ:bighwTerm}.  
Thus, to sufficient accuracy we obtain
\bea
\bighw{q}
&=&
\frac{1}{4}q^{2}\mu^{4}{\LocFr}^{2}
\int_{0}^{1}ds_{3}ds_{4}
\exp\big(-q^{2}\vert{s_{3}-s_{4}}\vert/2\big)
\nonumber
\\
&&
\qquad
\times
\int_{0}^{1}ds_{1}ds_{2}
\sum_{\rho,\rho^{\prime}=1}^{2}
\big( 
-\smatrCI{3,\rho}
+\smatrCI{4,\rho}
\big)
\big(
-\smatrCI{3,\rho^{\prime}}
+\smatrCI{4,\rho^{\prime}}
\big)
\bigg\{
  \tau_{\rho}\delta_{\rho\rho^{\prime}}
-{\tau_{\rho}
  \tau_{\rho^{\prime}}
\over{
\sum_{\sigma=1}^{r}
\tau_{\sigma}
}}
\bigg\}_{\tau}
\nonumber
\\
&=&
\frac{1}{2}
\mu^{4}{\LocFr}^{2}
\bigg\{
{\tau_{1}\tau_{2}
\over
{\tau_{1}+\tau_{2}}}
\bigg\}_{\tau}\,\,
q^{2}
\Big(\debyeW{q}-\debyeT{q}\Big),
\label{EQ:bigHWacc}
\eea where the curly braces $\ltave\cdots\rtave$ were defined immediately 
following \eqref{EQ:freenlt}, and the functions $\debyeW{q}$ and $\debyeT{q}$ 
are given by elementary integrals defined in App.~\ref{APP:debye}.  Note the 
convenient factorisation into ${\bf q}$-dependent and $p(\tau)$-dependent terms.

Last, we focus on $\lithz{q}$, for which we also need only retain terms 
with $r\le 2$.  However, the $r=1$ term is not identically zero, and as it 
carries only a single power of ${\LocFr}$ care must be taken to compute it with 
appropriate accuracy.  (In practice, the absence of matrix algebra renders 
the perturbative calculation of $\big({\cal R}^{(1)}\big)^{-1}$ and quantities 
built from it straightforward.)\thinspace\ Thus, to sufficient accuracy, we obtain
\bea
\lithz{q}
&=&
\big(1-\mu^{2}{\LocFr}\big)\mu^{2}{\LocFr}
\ltave{\rm e}^{-q^{2}/\tau_{1}}\rtave
\int_{0}^{1}ds_{1}ds_{2}ds_{3}
{\rm e}^{
-q^{2}
\big(
\vert{s_{2}-s_{1}}\vert+
\vert{s_{3}-s_{1}}\vert
\big)/2}
\nonumber
\\
&&
\quad
+\frac{1}{2}\mu^{4}{\LocFr}^{2}
\Big\{
{\rm e}^{-q^{2}/(\tau_{1}+\tau_{2})}
\int_{0}^{1}ds_{1}ds_{2}ds_{3}ds_{4}\,
{\rm e}^{-q^{2}(s_{3}+s_{4})/2}\,
{\rm e}^{-q^{2}(\tau_{1}+\tau_{2})^{-2}
\sum_{\rho,\rho^{\prime}=1}^{2}
\tau_{\rho}\tau_{\rho^{\prime}}\,\smatr}\,
\nonumber
\\
&&
\qquad\qquad\qquad
\times
{\rm e}^{q^{2}(\tau_{1}+\tau_{2})^{-1}
\sum_{\rho=1}^{2}
\tau_{\rho}(\smatrCI{3,\rho}+\smatrCI{4,\rho})}
\Big\}_{\tau}.
\label{EQ:litHZacc}
\eea

\begin{figure}
\caption{Dependence of the gel fraction ${\LocFr}$ on $\mu^{2}$: 
stable solution (full line); unstable solution (dotted line).
\label{FIG:Qplot}}
\end{figure}
\begin{figure}
\caption{Scaling function $\pi(\theta)$ (full line) for the probability
distribution of localisation lengths; 
asymptotic form for $\theta\rightarrow 0$ (dotted line); 
asymptotic form for $\theta\rightarrow\infty$ (broken line). 
\label{FIG:PiTheta}}
\end{figure}
\begin{figure}
\caption{Scaling function $\omega(k)$ for the order parameter 
(full line); 
asymptotic form for $k\rightarrow 0$ (dotted line);
asymptotic form for $k\rightarrow\infty$ (broken line).
\label{FIG:omegaQ}}
\end{figure}
\begin{table}
\caption{Values of the order parameter in the liquid and amorphous 
solid states.}
\begin{tabular}{cccc}
Sector	&Order parameter	&Liquid state  	&Amorphous solid state
\\
      	&	       		&$(\mu^{2}\le 1)$	&($\mu^{2}>1$)	    
\\
\tableline
one-replica&
	$\langle {\FTDen}_{\bf k}^{\alpha}\rangle_{n+1}^{\rm P}$
				&$=0$	&$=0$
\\
higher-replica&
	$\langle {\FTDen}_{\hat{k}}       \rangle_{n+1}^{\rm P}$
				&$=0$	&$\ne0$
\\
\end{tabular}
\label{TAB:phases}
\end{table}

\begin{references}
\bibitem{REF:DimensionForm}
It is straightforward to see that in terms of dimensionful variables 
the Edwards measure takes the form 
\[
\exp
        \left(
-\frac{d}{2}
\sum_{i=1}^{N}
\int\limits_{0}^{L}\frac{d\sigma}{\ell}\,
\Big\lefver
  {d\over{d\sigma}}{\bf R}_{i}(\sigma)
\Big\rigver^{2}
-\frac{u_{0}}{2}
\sum_{i,i^{\prime}=1}^{N}
\int\limits_{0}^{L}\frac{d\sigma}{\ell}
\int\limits_{0}^{L}\frac{d\sigma^{\prime}}{\ell}\,
\delta^{(d)}
\big( 
 {\bf R}_{i}         (\sigma         )
-{\bf R}_{i^{\prime}}(\sigma^{\prime})
\big)
        \right), 
\]
where the excluded-volume parameter $u_{0}$ is given by  
$u_{0}=
\left(
\ell L/d
\right)^{d/2}
\left(
\ell/L
\right)^{2}
\lambda^{2}$, the units of which are those of volume.
\bibitem{REF:NoDelta}
At the microscopic level, a pair of monomers participating in a crosslink 
are not located at precisely the same position, being instead separated 
by the length of the bond that connects them. At the semi-microscopic level, 
however, positions are not resolved beyond a scale of order $\ell$. 
Thus, it is adequate to model the constraints via the Dirac $\delta$-function.
A more detailed description, involving the modelling of the constraint via a 
function that varies on the scale of the bond-length, would yield a theory that 
differs only at scales on the order of $\ell$: the results of the present 
semi-microscopic approach would thus be unchanged.
\bibitem{REF:discrete}
Two remarks are in order here.  As a point of practice, only crosslink
locations that differ by more than $\ell$ can be distinguished in our
semi-microscopic approach.  Besides, as a point of principle, the set
of possible crosslink locations is discrete, owing to the underlying
quantum-mechanical nature of the chemical bonds.  Thus, the cluster
index $a$ is discrete.  If it were continuous, one would reasonably
expect that for all cluster species (other than the uncrosslinked
species) the likelihood of having more that one specimen would be
zero.
\bibitem{REF:TDlimit}
The thermodynamic limit 
($N\rightarrow\infty$, 
 $V\rightarrow\infty$, 
 $N/V$ fixed, 
 $\mu^{2}$ fixed)
will eventually be taken, in a manner that is discussed in detail 
in Sec.~\ref{SEC:InstFlPH}.
\bibitem{REF:XLkinetics}
This can be achieved, \eg, by some suitable photo-chemical process.
Such schemes have the virtue that they are accurately determined by the
equilibrium properties of the system prior to crosslinking. To describe
crosslinking processes that are not simultaneous and instantaneous would
require information pertaining to the kinetics of the system as the
number of crosslinks grows.
\bibitem{REF:CGZforth}
In a forthcoming article \cite{REF:CGZelastic} we shall report in detail
on the extension of the present work to the investigation of the elastic
properties of the amorphous solid state of randomly crosslinked
macromolecular networks.
\bibitem{REF:PanNew}
A different field-theoretic treatment, intended for the
well-crosslinked regime (in which the many-macromolecule system is
modelled by a single, linear, self-crosslinked macromolecule) can be
found in Ref.~\cite{REF:PanPapers}.  For a review of this work, see
Ref.~\cite{REF:PanReview}.
\bibitem{REF:cautionA}
We believe that this is the appropriate order in which to take the
thermodynamic and replica limits, although this order is opposite to
that commonly used in the context of spin glass theory; see, \eg,
{\S}3C of the second citation in Ref.~\cite{REF:MPVbook}.
\bibitem{REF:cautionB}
We have not explored the curious feature of the 
non--positive-definiteness of $\hat{k}^{2}$, which arises for modes,
say, in the $(n-2)$-replica sector.  In principle, this feature could
modify the notion that the least stable modes are the long wavelength
modes.
\bibitem{REF:HighNumber}
In Ref.~\cite{REF:PMGandAZprl} the instability of the liquid state was
found to occur at a higher density of crosslinks, corresponding to
$\mu^{2}\approx 1.59$.  This was due to the variational hypothesis
adopted there, which did not allow for a fraction of delocalised
monomers to be present in the amorphous solid state.
\bibitem{REF:regularity}
If we had chosen to include the delocalised monomers along with  
the localised monomers in the distribution $p(\tau)$ then 
$p(\tau)$ would have a term of the form $(1-{\LocFr})\delta(\tau)$. 
Such a singularity at $\tau=0$ would lead to $\hat{p}(\infty)=(1-{\LocFr})$,
which would in general be nonzero.  As this would be a technical 
inconvenience, we have chosen to exclude the delocalised monomers 
from $p(\tau)$, separating them from the localised monomers, as 
was done explicitly in Sec.~\ref{SEC:OPFields}.
\bibitem{REF:twice}
It is noteworthy that a divergent length scale emerges already at the
level of mean-field theory. The analysis of fluctuations around the
saddle-point solution would presumably reveal a second length scale,
the conventional correlation length, which characterises the decay of
order-parameter correlations.  It is not clear at present whether these
two length scales are related.
\bibitem{REF:NumSchemeLS}
Equation~(\ref{EQ:PiStatCond}) was integrated numerically using the
Bulirsch-Stoer method \cite{REF:NumRecip}.  Choosing the solution that
satisfies the boundary conditions is a delicate matter because the
differential equation is unstable in the following sense. Assuming that
$\hat{\pi}(0)=1$, there is a unique value of
$d\hat{\pi}/d\hat{\theta}\vert_{\hat{\theta}=0}$ that gives rise to
$\hat{\pi}(\infty)=0$: if
$d\hat{\pi}/d\hat{\theta}\vert_{\hat{\theta}=0}$ is larger then
$\hat{\pi}(\hat{\theta})$ will oscillate indefinitely about unity; if
$d\hat{\pi}/d\hat{\theta}\vert_{\hat{\theta}=0}$ is smaller then
$\hat{\pi}(\hat{\theta})$ will become negative and diverge to $-\infty$.
We determined $d\hat{\pi}/d\hat{\theta}\vert_{\hat{\theta}=0}$ by using
a variant of the shooting method for which we defined a monotonic
function $t(x)$ via
(i)~$t(x)=1/\theta^{\ast}$ 
if integrating \eqref{EQ:PiStatCond} with $\hat{\pi}(0)=1$
and $d\hat{\pi}/d\hat{\theta}\vert_{\hat{\theta}=0}=x$ gives 
a solution for which $\hat{\pi}(\hat{\theta}^{\ast})=1$ and 
$0<\hat{\pi}(\hat{\theta})<1$ for 
$0<\hat{\theta}<\hat{\theta}^{\ast}$; and 
(ii)~$t(x)=-1/\theta^{\ast}$ 
if integrating \eqref{EQ:PiStatCond} with $\hat{\pi}(0)=1$
and $d\hat{\pi}/d\hat{\theta}\vert_{\hat{\theta}=0}=x$ gives 
a solution for which $\hat{\pi}(\hat{\theta}^{\ast})=0$ and 
$0<\hat{\pi}(\hat{\theta})<1$ for 
$0<\hat{\theta}<\hat{\theta}^{\ast}$.
We identified the desired value of 
$d\hat{\pi}/d\hat{\theta}\vert_{\hat{\theta}=0}$ 
as the root of $t(x)=0$.
\bibitem{REF:NumSchemeRS}
Equation~(\ref{EQ:scpieq}) was also integrated numerically using the
Bulirsch-Stoer method \cite{REF:NumRecip}.  To satisfy the normalisation
condition we first analytically obtained the asymptotic solution of
\eqref{EQ:PiStatCond} for $\hat{\theta}\gg 1$.  We identified the
numerical prefactor by comparison with the numerical solution, and then
inverse-Laplace-transformed this asymptotic form to obtain
$\pi(\theta)\sim 4.554\theta^{-2}\exp(-2/\theta)$  (for $\theta\ll 1$).
We used this form to start the integration of \eqref{EQ:scpieq} at a
small but nonzero value of $\theta$, thus avoiding the singularity at
$\theta=0$. After integrating to obtain the full function $\pi(\theta)$
we computed $\int_{0}^{\infty}d\theta\,\pi(\theta)$ numerically, and
verified (to within numerical accuracy) that the value was unity.  As a
second check, we computed $\hat{\pi}(\hat{\theta})$ as the (numerical)
Laplace transform of our (numerically obtained) solution of
\eqref{EQ:scpieq}, and established that it coincided (to within
numerical accuracy) with the  numerical solution of
\eqref{EQ:PiStatCond}.
\end{references}
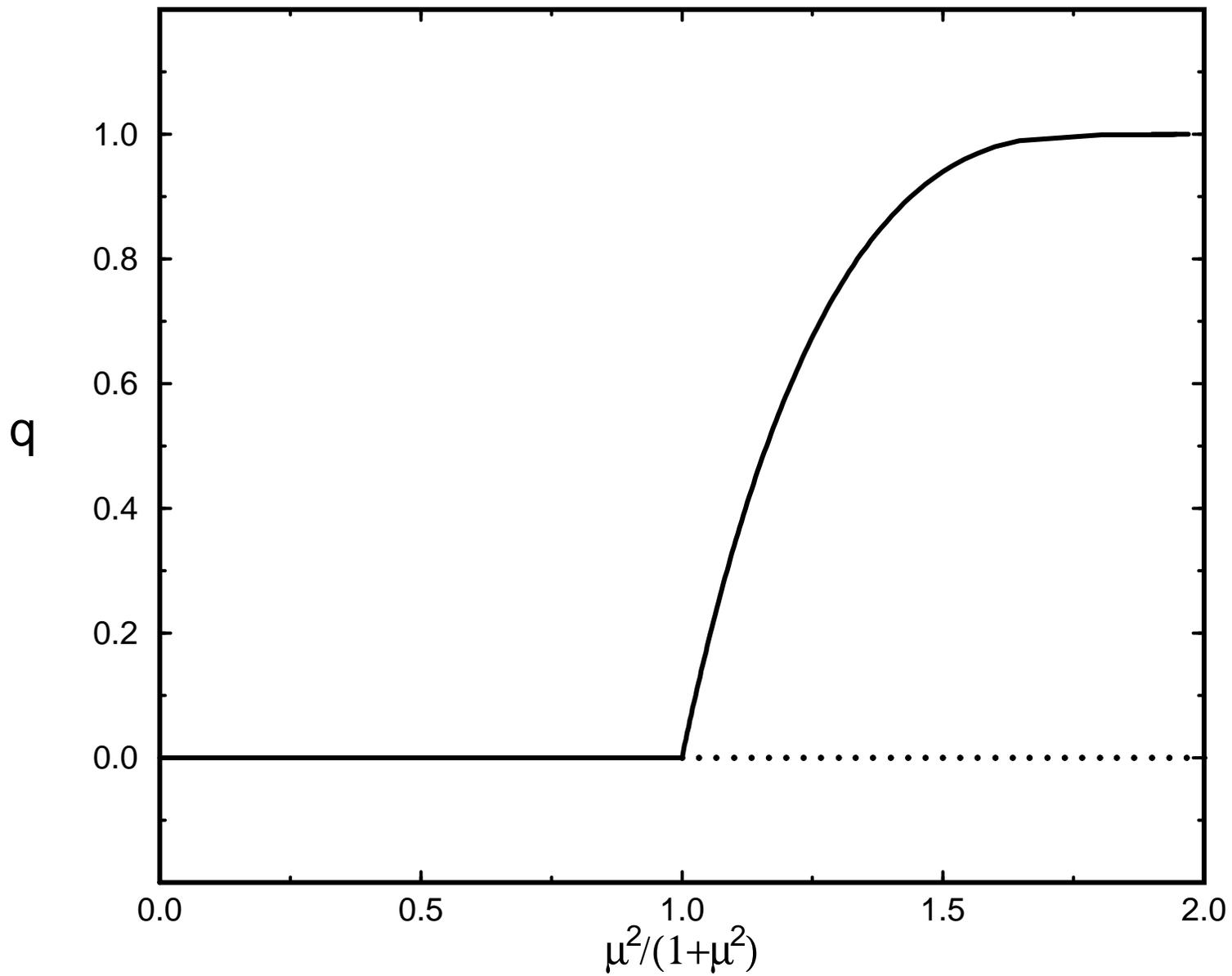
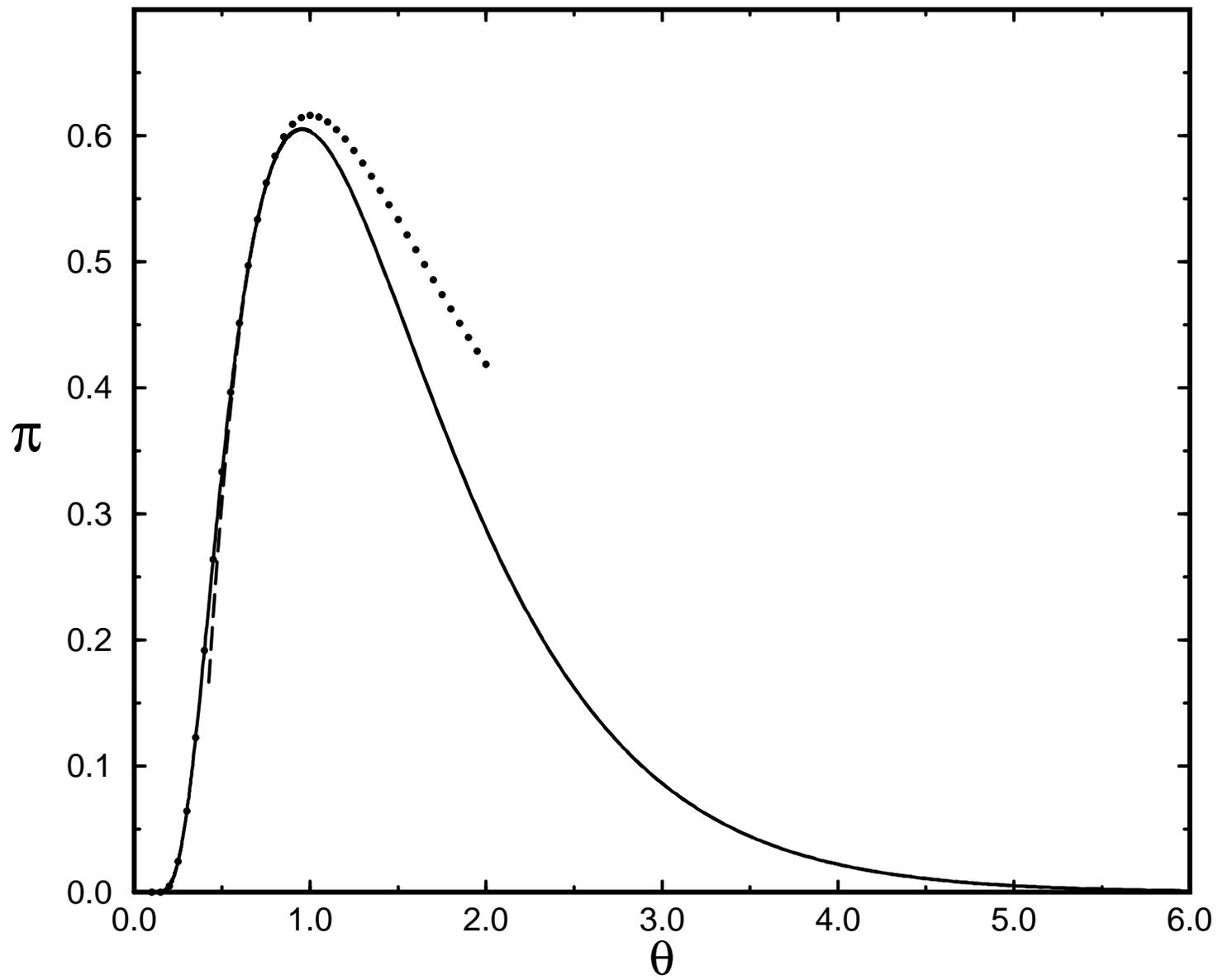
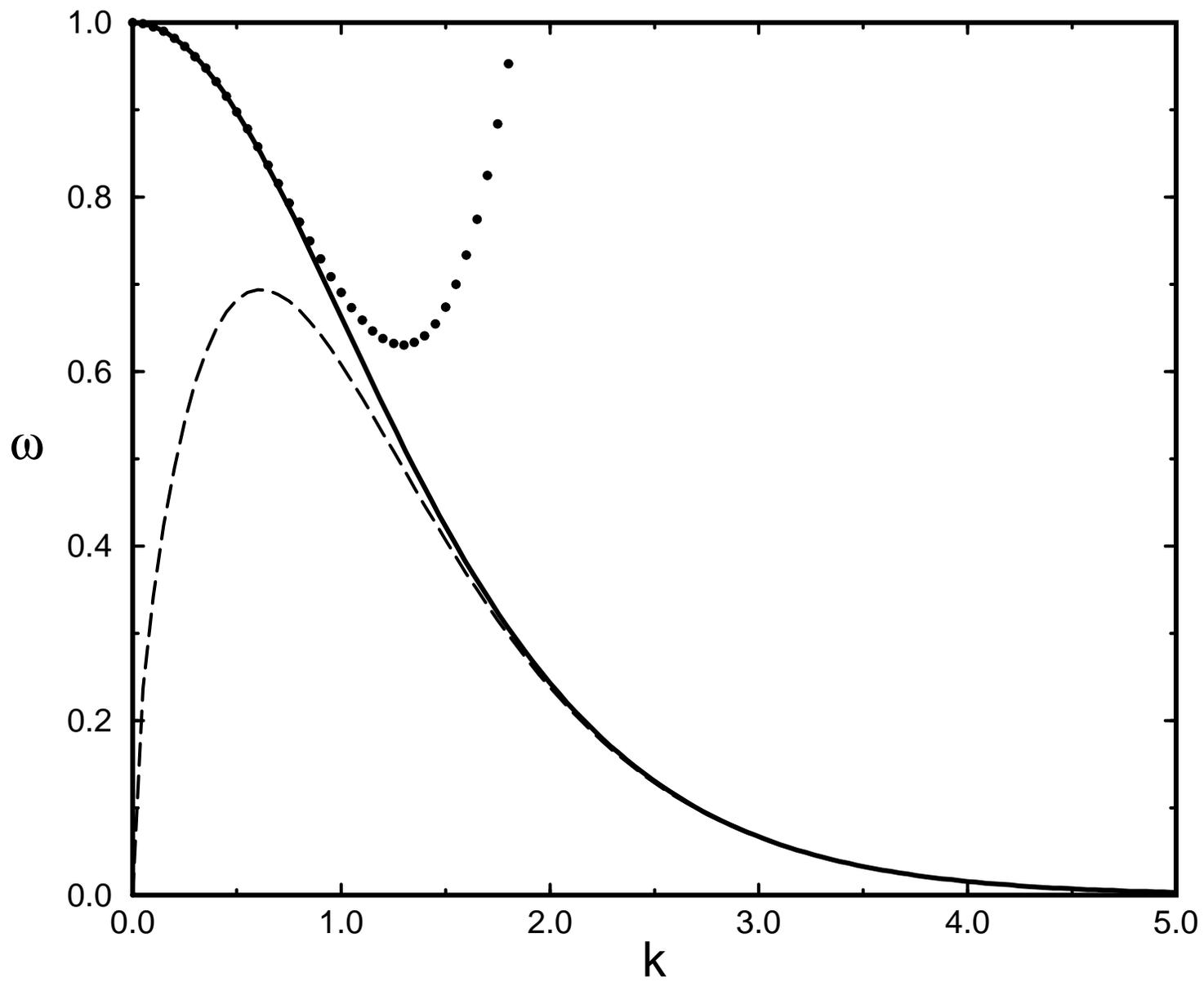
\end{document}